\documentclass[12pt,preprint]{aastex}

\usepackage{color}

\shorttitle{Photospheric response to EB-like event}
\shortauthors{Danilovic et al.}

\begin{document}

\renewcommand{\thefootnote}{\fnsymbol{footnote}}
\renewcommand{\thempfootnote}{\fnsymbol{mpfootnote}}

\newcommand{\sunrise}{\textsc{Sunrise}}
\newcommand{\carcsec}{$\mbox{.\hspace{-0.5ex}}^{\prime\prime}$}

\title{Photospheric response to Ellerman Bomb-like event - analogy of \sunrise{}/IM\lowercase{a}X observations and MHD simulations}
\author{\textsc{
S.~Danilovic,$^{1}$
S.~K.~Solanki,$^{1,2}$
P.~Barthol,$^{1}$
A.~Gandorfer,$^{1}$
L.~Gizon,$^{1,3}$
J.~Hirzberger,$^{1}$
T.~L.~Riethm\"uller,$^{1}$
M.~van~Noort,$^{1}$
J.~Blanco~Rodr\'{\i}guez,$^{4}$
J.~C.~Del~Toro~Iniesta,$^{5}$
D.~Orozco~Su\'arez,$^{5}$
W.~Schmidt$^{6}$
V.~Mart\'{i}nez Pillet,$^{7}$
\& M.~Kn\"olker,$^{8}$
}}
\affil{
$^{1}$Max-Planck-Institut f\"ur Sonnensystemforschung, Justus-von-Liebig-Weg 3, 37077 G\"ottingen, Germany; danilovic@mps.mpg.de\\
$^{2}$School of Space Research, Kyung Hee University, Yongin, Gyeonggi, 446-701, Republic of Korea\\
$^{3}$ Institut f\"ur Astrophysik, Georg-August-Universit\"at G\"ottingen,
Friedrich-Hund-Platz 1, 37077 G\"ottingen, Germany \\
$^{4}$Grupo de Astronom\'{i}a y Ciencias del Espacio, Universidad de Valencia, 46980 Paterna, Valencia, Spain\\
$^{5}$Instituto de Astrof\'{i}sica de Andaluc\'{i}a (CSIC), Apartado de Correos 3004, 18080 Granada, Spain\\
$^{6}$Kiepenheuer-Institut f\"ur Sonnenphysik, Sch\"oneckstr. 6, 79104 Freiburg, Germany\\
$^{7}$National Solar Observatory, 3665 Discovery Drive, Boulder, CO 80303, USA\\
$^{8}$High Altitude Observatory, National Center for Atmospheric Research,\footnote{The National Center for Atmospheric Research is sponsored by the National Science Foundation.} P.O. Box 3000, Boulder, CO 80307-3000, USA\\
}

\begin{abstract}
Ellerman Bombs are signatures of magnetic reconnection, which is an important physical process in the solar atmosphere. How and where they occur is a subject of debate. In this paper we analyse \sunrise{}/IMaX data together with 3D MHD simulations that aim to reproduce the exact scenario proposed for the formation of these features. Although the observed event seems to be more dynamic and violent than the simulated one, simulations clearly confirm the basic scenario for the production of EBs. The simulations also reveal the full complexity of the underlying process. The simulated observations show that the Fe~I~525.02 nm line gives no information on the height where reconnection takes place. It can only give clues about the heating in the aftermath of the reconnection. The information on the magnetic field vector and velocity at this spatial resolution is, however, extremely valuable because it shows what numerical models miss and how they can be improved.
\end{abstract}


\section{Introduction}

One of the important physical processes that has a major effect on the energy budget in the solar atmosphere is magnetic reconnection. It is a mechanism behind a myriad of dynamic atmospheric phenomena, starting from small scale flux cancellation in the solar photosphere to the largest solar disruptions -  flares. 

Recently, a number of studies concentrated on explaining one of these phenomena called Ellerman Bombs \citep[EBs;][]{Ellerman1917} that pose a considerable challenge to  models because of their specific characteristics. They are defined as transient brightenings of the extended wings of the H$\alpha$ line, but
leave signatures also in Ca~II~H and Ca~II~IR~854~nm \citep{Gregal:2013,Reza2015} and sometimes even in observables which should sample orders of magnitude higher temperatures than the former \citep{Vissers2015,Tian2016}. Observations show that EBs are almost exclusively formed in young emerging active regions, usually between spots/pores where the emergence of serpentine field lines takes place. Series of EBs usually appear  aligned along the orientation of the active region in so-called Bald Patches (BPs) - dips in magnetic field lines \citep{Pariat:etal:2004,Pariat:2006}. This scenario is also supported by MHD simulations \citep{Isobe2007,Archontis2009}.

Because EBs assume the shape of a flame that seems to be rooted in the intergranular lanes \citep{Matsumoto2008,Watanabe2011}, it is suggested that their formation begins very low, near the surface. However, one- and two-dimensional modelling strongly suggests that only a temperature increase starting at heights of a few hundred km above the solar surface can produce the observed H$\alpha$ line profile without continuum brightening \citep{Kitai1983,Fang:2006,Bello:etal:2013,Berlicki:2014,Fang:2006,Hong:2014}. 
\cite{Nelson2013} analysed observations together with similar events in 3D MHD simulations and inferred that EBs occur co-spatially with regions of strong opposite-polarity magnetic field at locations where the Fe~I~630.25~nm line core intensity increases. According to them this gives enough evidence that EBs are in fact signatures of photospheric magnetic reconnection. Later it was established that most of the features observed by \cite{Nelson2013} were actually 'pseudo-EBs', just strong-field magnetic concentrations \citep{Rutten2013}. \cite{Reid2016} imposed a more discriminating threshold for detecting EBs and corrected the previously made mistake. Their detailed analysis of a large number of these features goes beyond just looking at the  Fe~I~630.25~nm line's core intensity. Using inversions of the spectropolarimetric data they retrieved temperature enhancements of 200~K at the EBs' footpoints. This value is below the lowest number required by H$\alpha$ line modelling. \cite{Reid2016} attributed this to the low formation height of Fe~I~630.25~nm line.

In this paper we analyse the photospheric response to an EB observed by the Imaging Magnetograph eXperiment
\citep[IMaX,][]{MartinezPillet2011} on board of \sunrise{}, a
balloon-borne solar observatory \citep{Solanki2010,Barthol2011,Berkefeld2011,Gandorfer2011}. Similar to \cite{Nelson2013} we compare the observations with MHD simulations, but unlike them we carry out an appropriate numerical experiment. While they used a run where cancellation of weak network was simulated, we reproduce the exact scenario proposed for the formation of EBs, i.e. the emergence of serpentine magnetic flux. Furthermore, we take into account the spectral and spatial resolution of our instrument, as well as its polarimetric sensitivity and apply the same inversion strategy as in the case of the observations. We pinpoint possible errors that this kind of analysis could produce and finally show which characteristics of EBs numerical simulations fail to reproduce and why.

\section{An event observed by \sunrise~II}

\begin{figure*}
  \centering
  \includegraphics[angle=0,width=0.5\linewidth ,trim=1.5cm 3.5cm 0cm 3cm,clip=true]{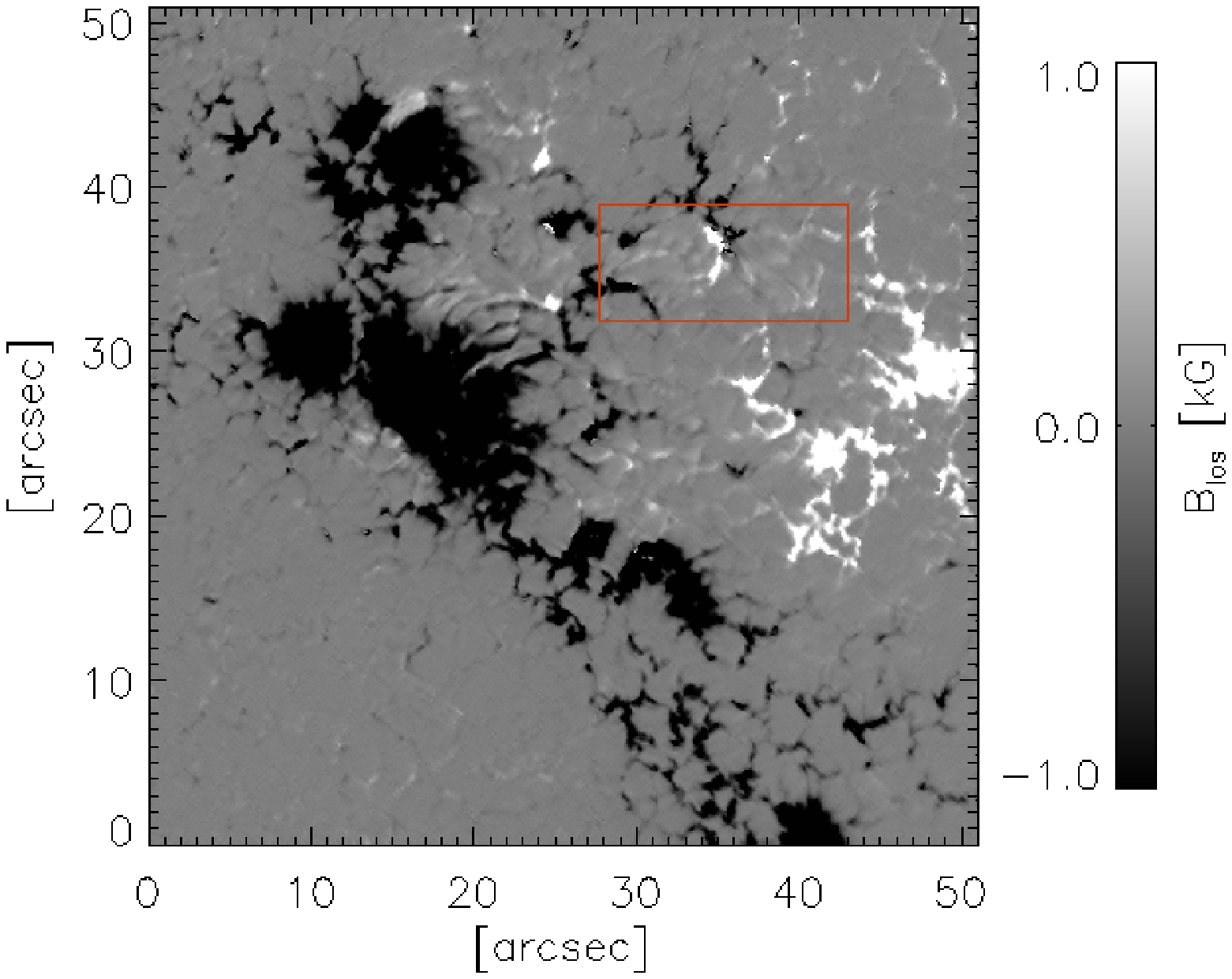} \\
\includegraphics[angle=0,width=0.5\linewidth ,trim=1.5cm 1cm 0cm 3cm,clip=true]{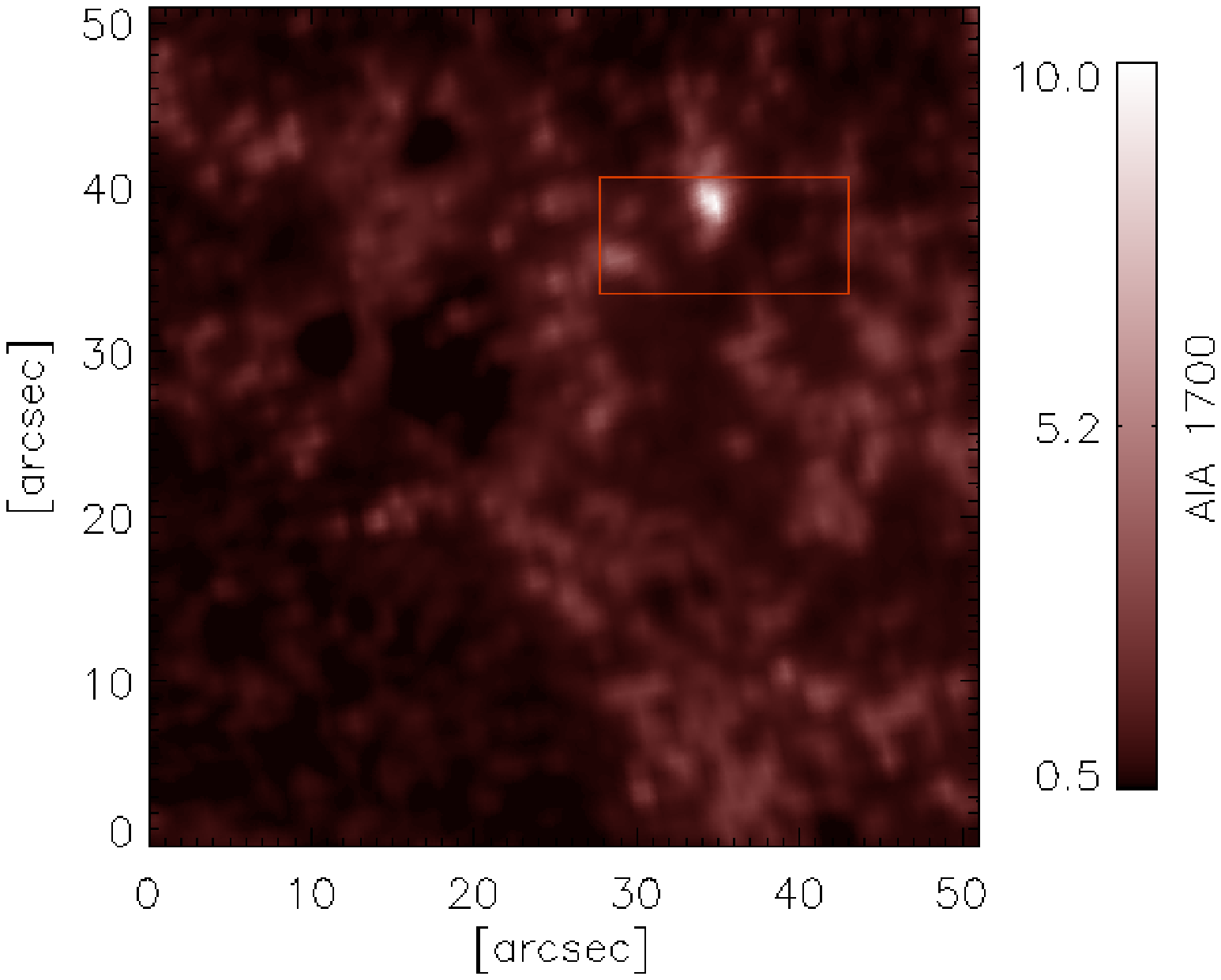}
  \caption{Line-of-sight magnetic field (top) over the whole FOV of \sunrise{}/IMaX and the corresponding AIA 1700 channel image (bottom) on June 12th 2013 at 23:47~UT when the event reached its highest brightness. Red rectangle outlines the subfield shown in Fig.~\ref{imax_inv_rest} and Fig.~\ref{imax_inv}. The FOV is inverted relative to that in \cite{Solanki:etal:this}, where the true orientation on the Sun is shown.}
\label{full_fov}
\end{figure*}

\begin{figure*}
  \centering
  \includegraphics[angle=90,width=0.327\linewidth ,trim=3.5cm 0cm 0.2cm 0cm,clip=true]{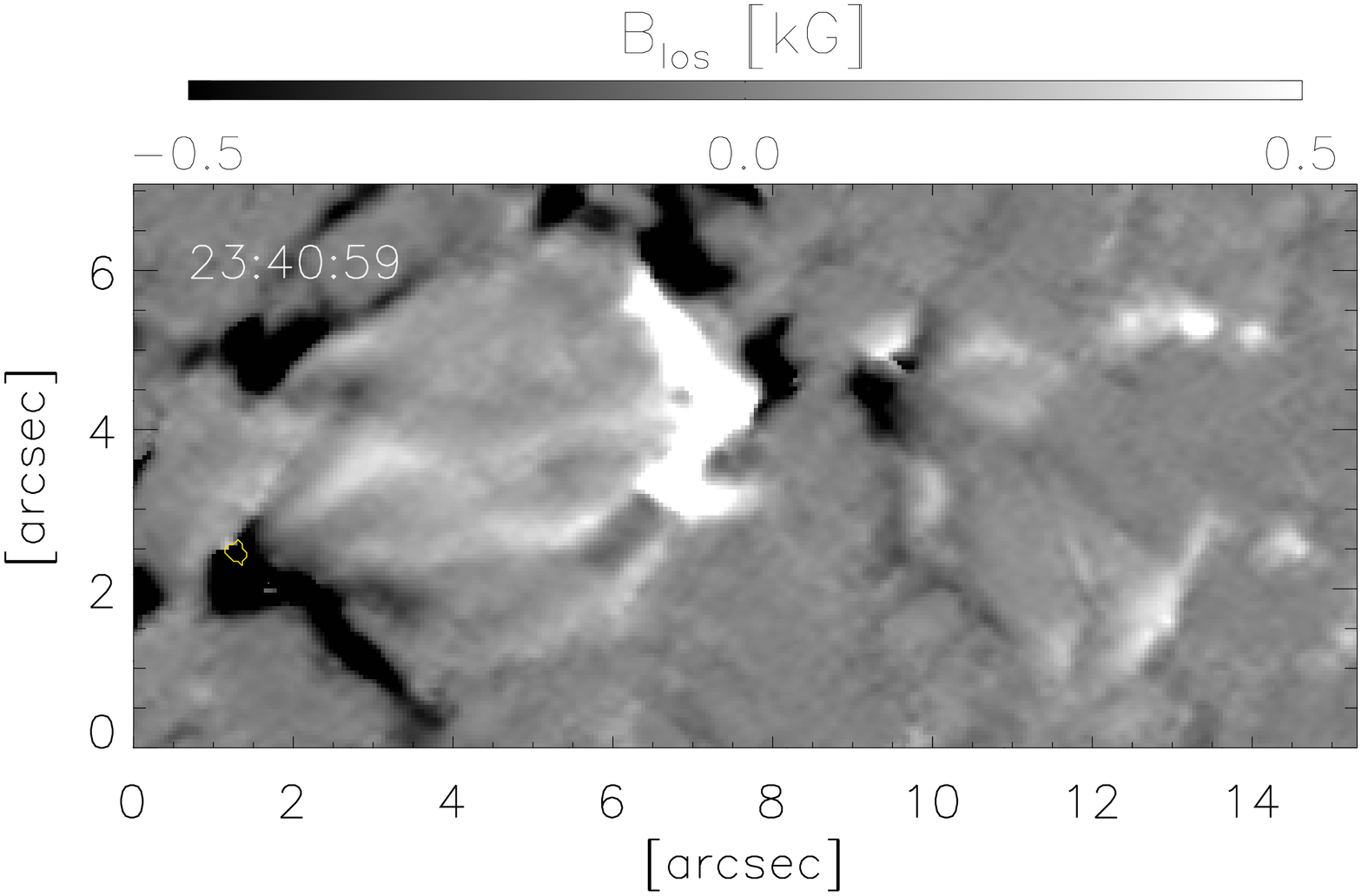} 
   \includegraphics[angle=90,width=0.30\linewidth ,trim=3.5cm 0cm 0.2cm 2.5cm,clip=true]{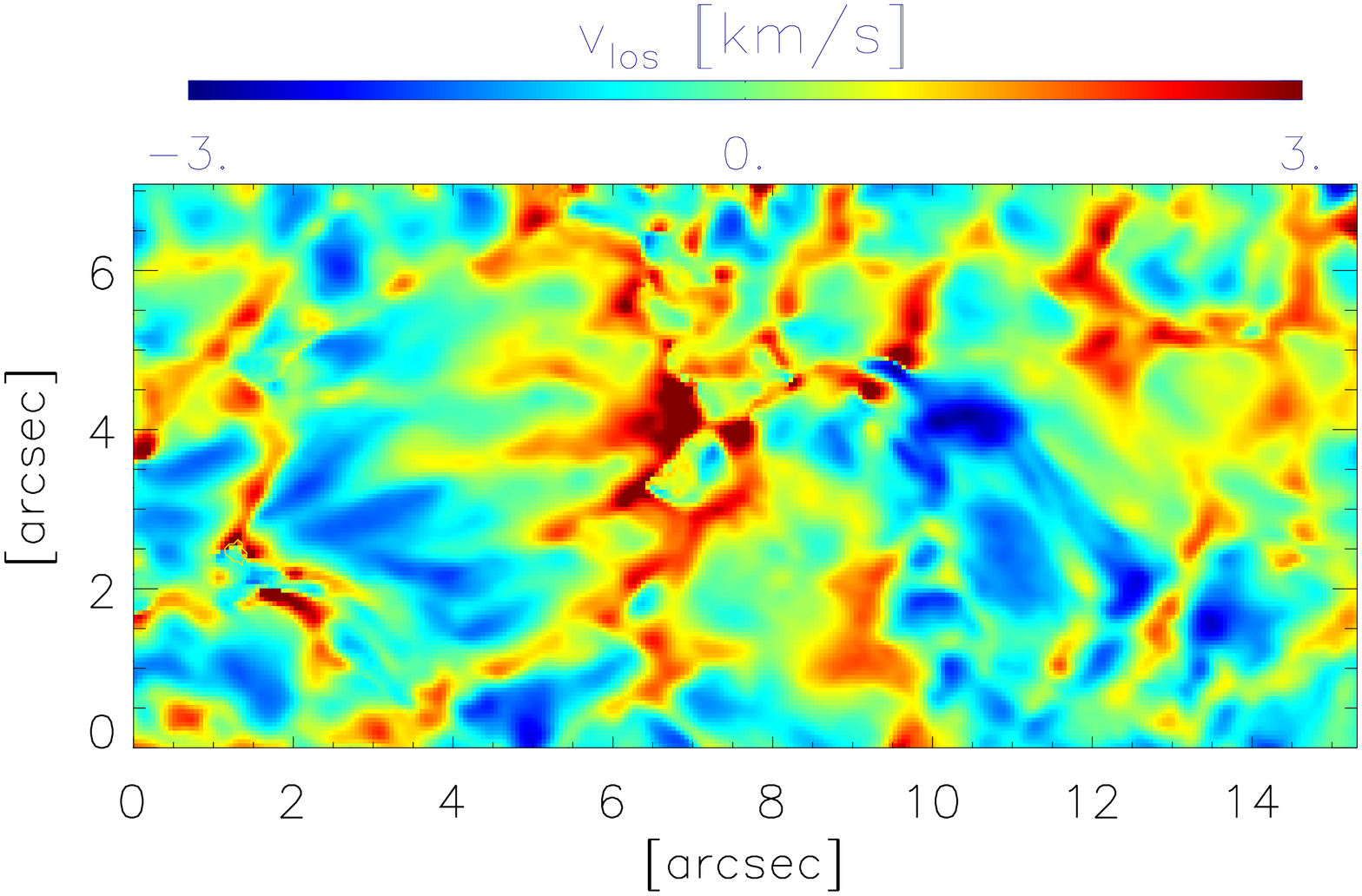}
    \includegraphics[angle=90,width=0.30\linewidth ,trim=3.5cm 0cm 0.2cm 2.5cm,clip=true]{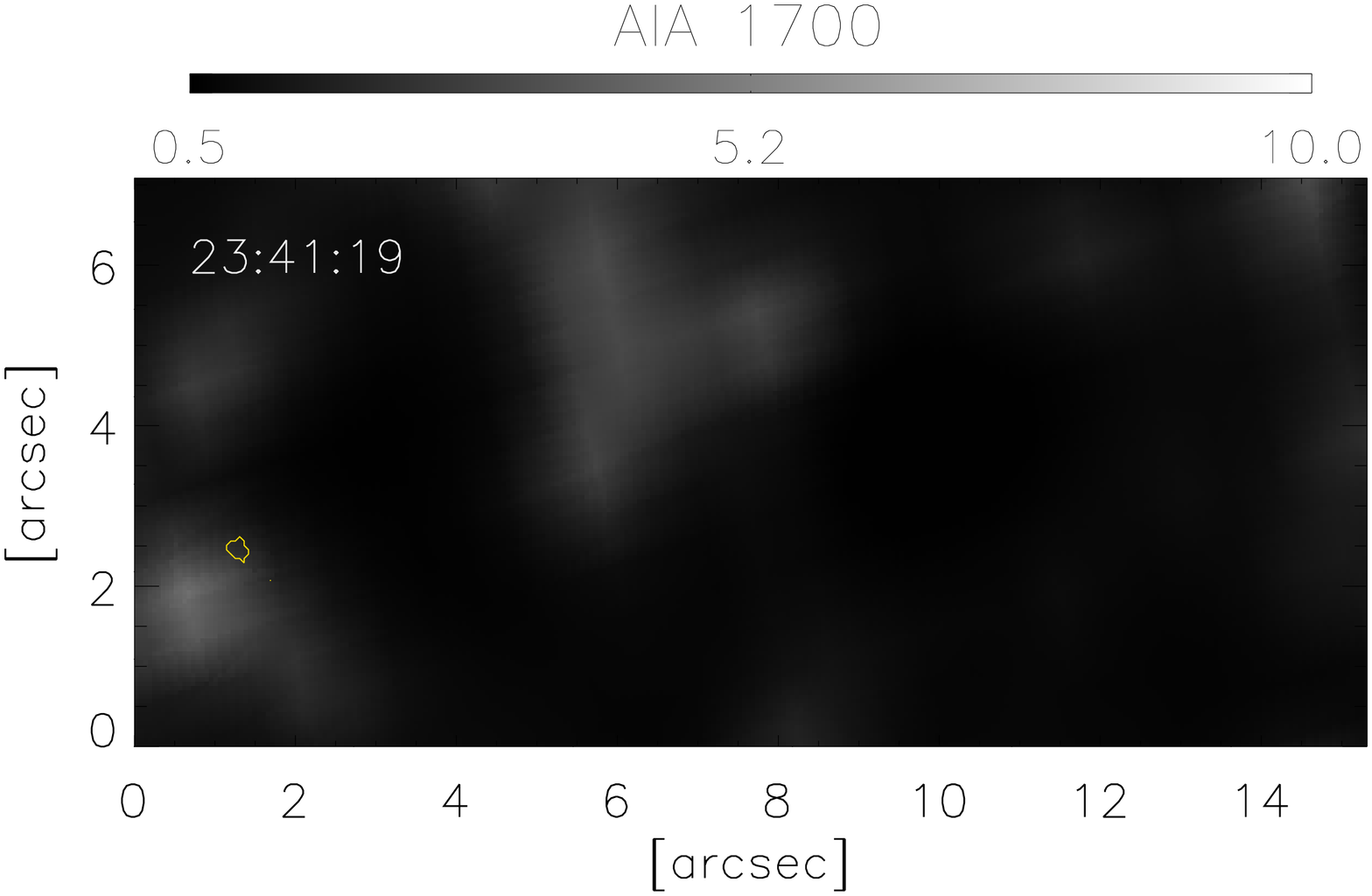}  
  \includegraphics[angle=90,width=0.327\linewidth ,trim=3.5cm 0cm 3.5cm 0cm,clip=true]{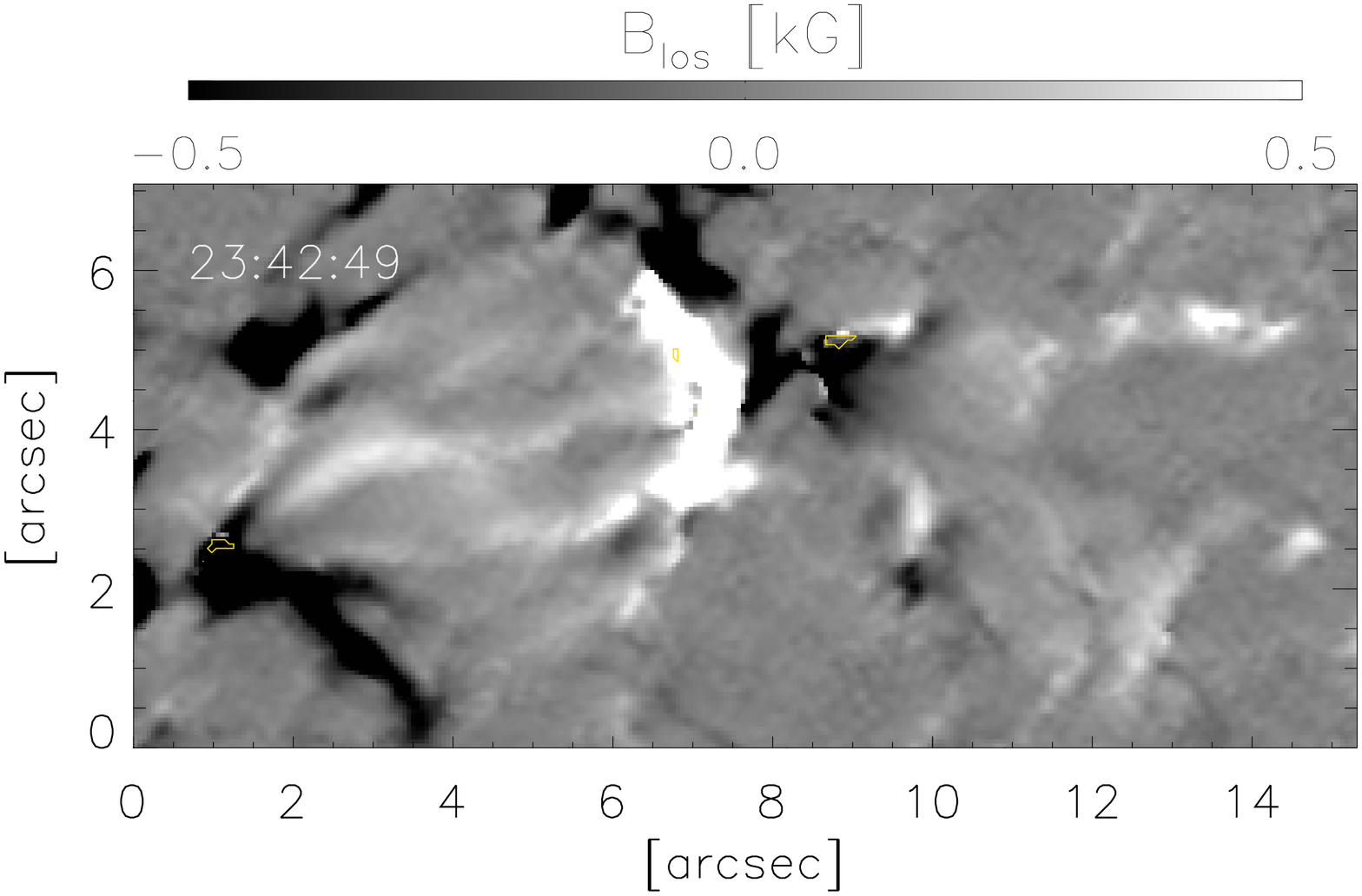}
     \includegraphics[angle=90,width=0.3\linewidth ,trim=3.5cm 0cm 3.5cm 2.5cm,clip=true]{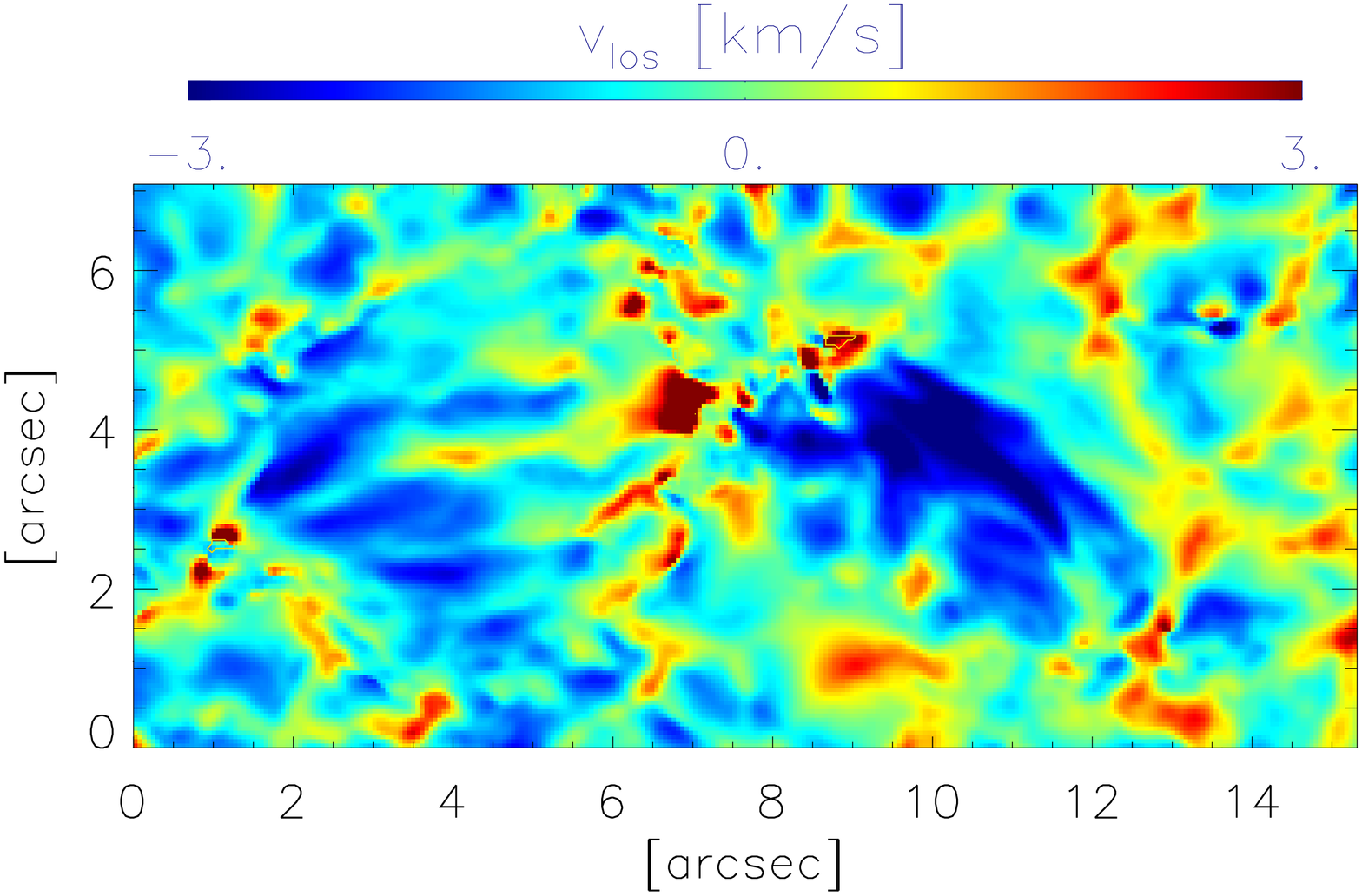}
    \includegraphics[angle=90,width=0.3\linewidth ,trim=3.5cm 0cm 3.5cm 2.5cm,clip=true]{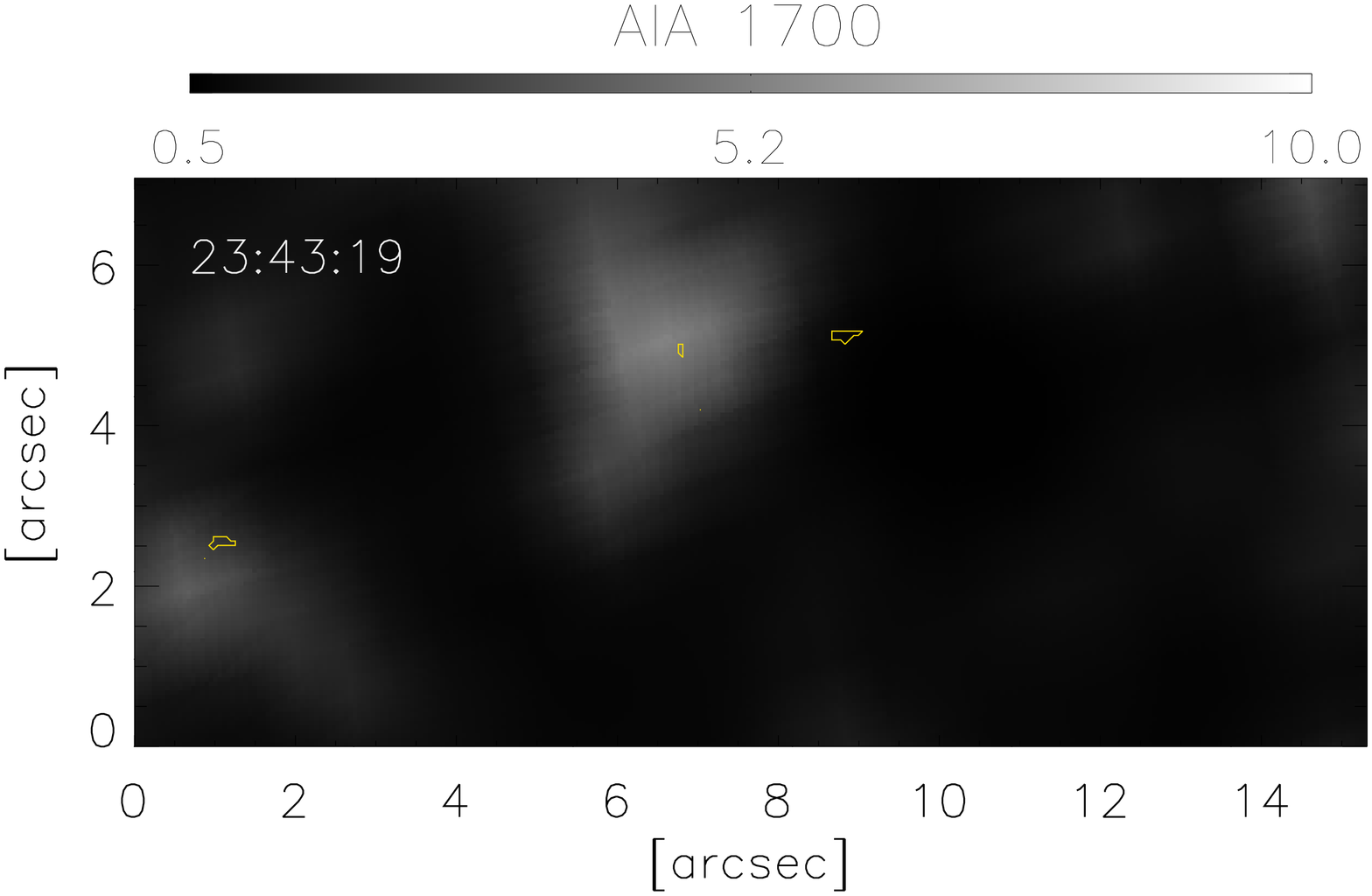} 
   \includegraphics[angle=90,width=0.327\linewidth ,trim=3.5cm 0cm 3.5cm 0cm,clip=true]{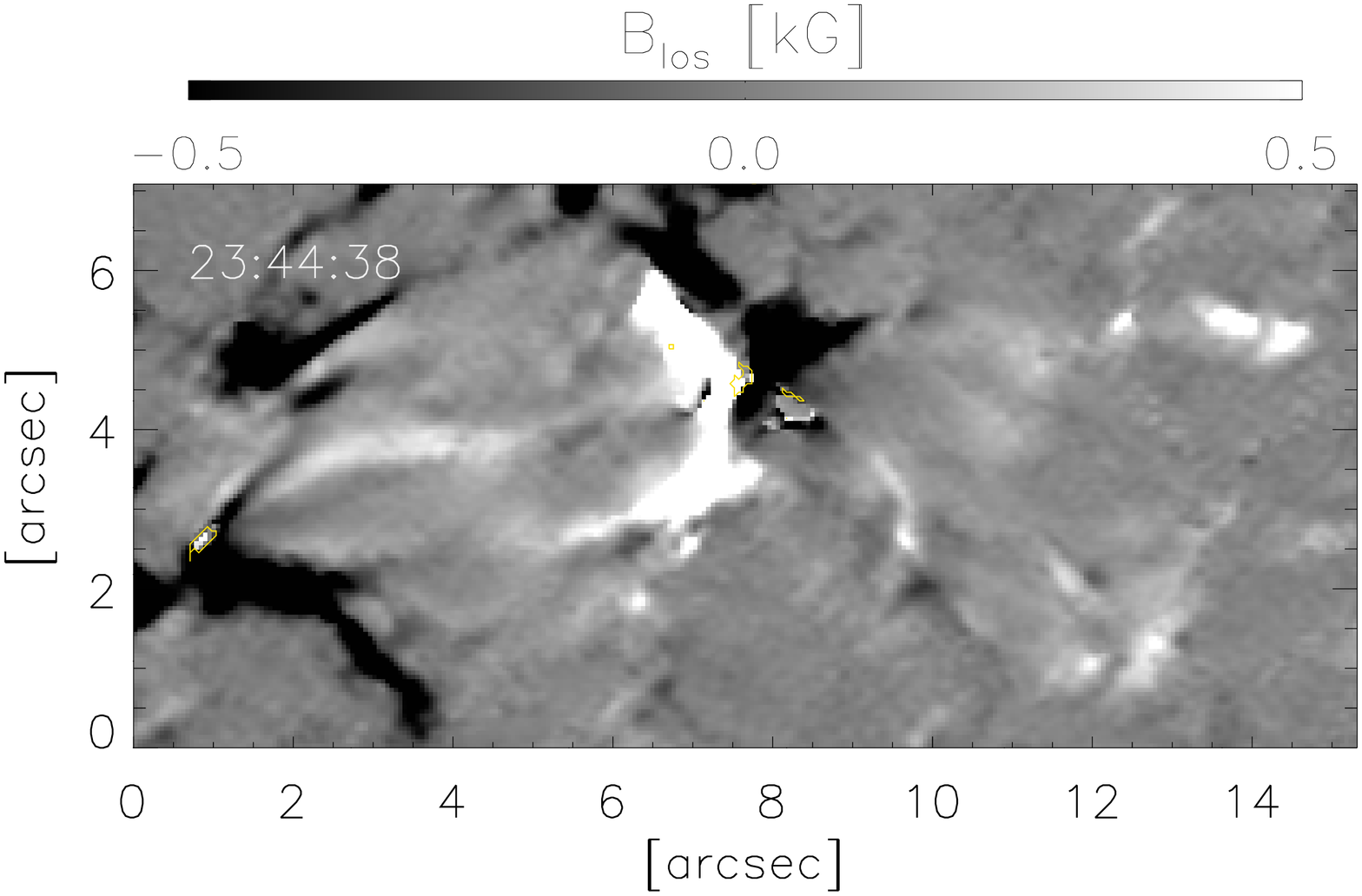}
     \includegraphics[angle=90,width=0.3\linewidth ,trim=3.5cm 0cm 3.5cm 2.5cm,clip=true]{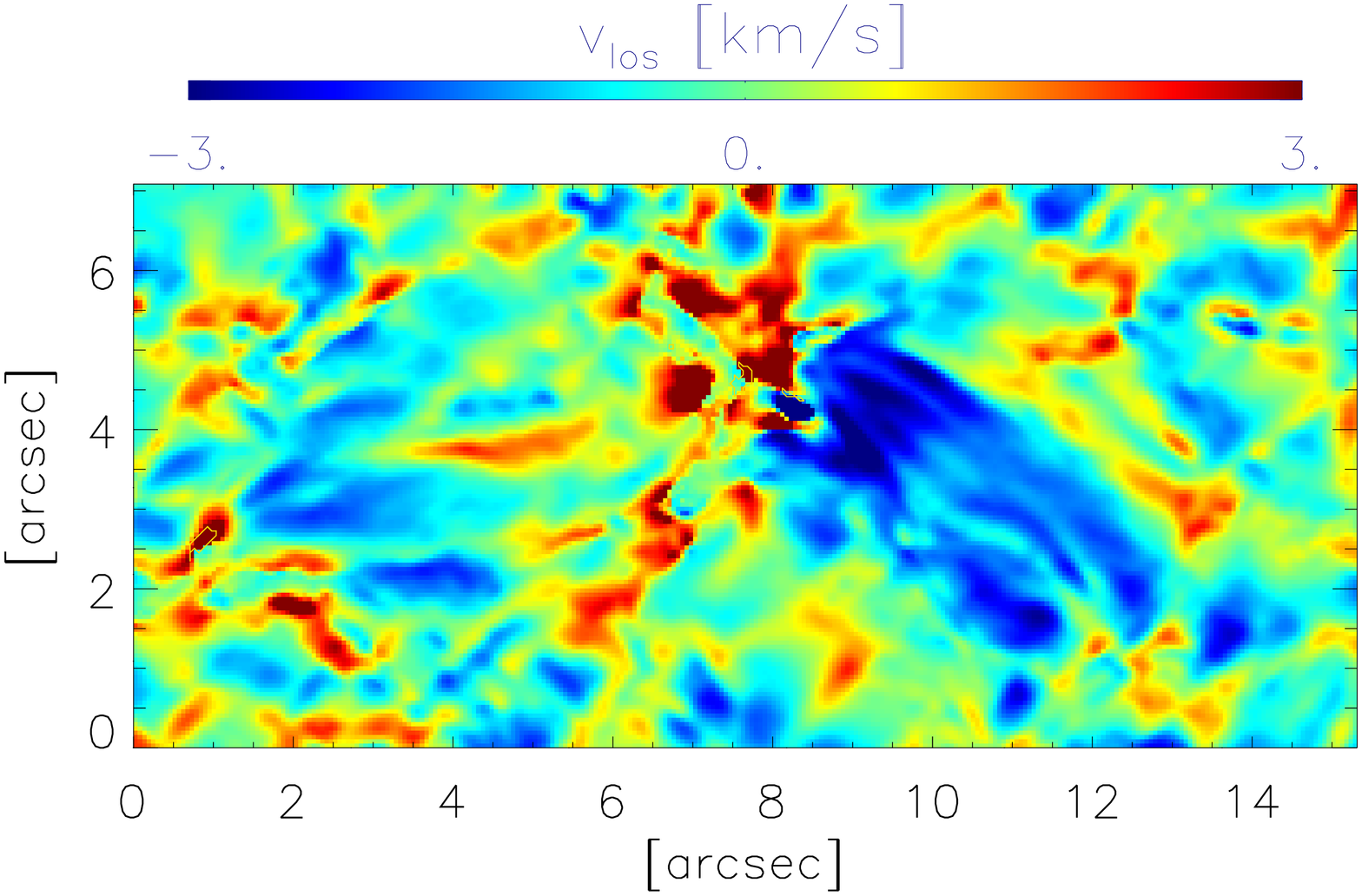}
    \includegraphics[angle=90,width=0.3\linewidth ,trim=3.5cm 0cm 3.5cm 2.5cm,clip=true]{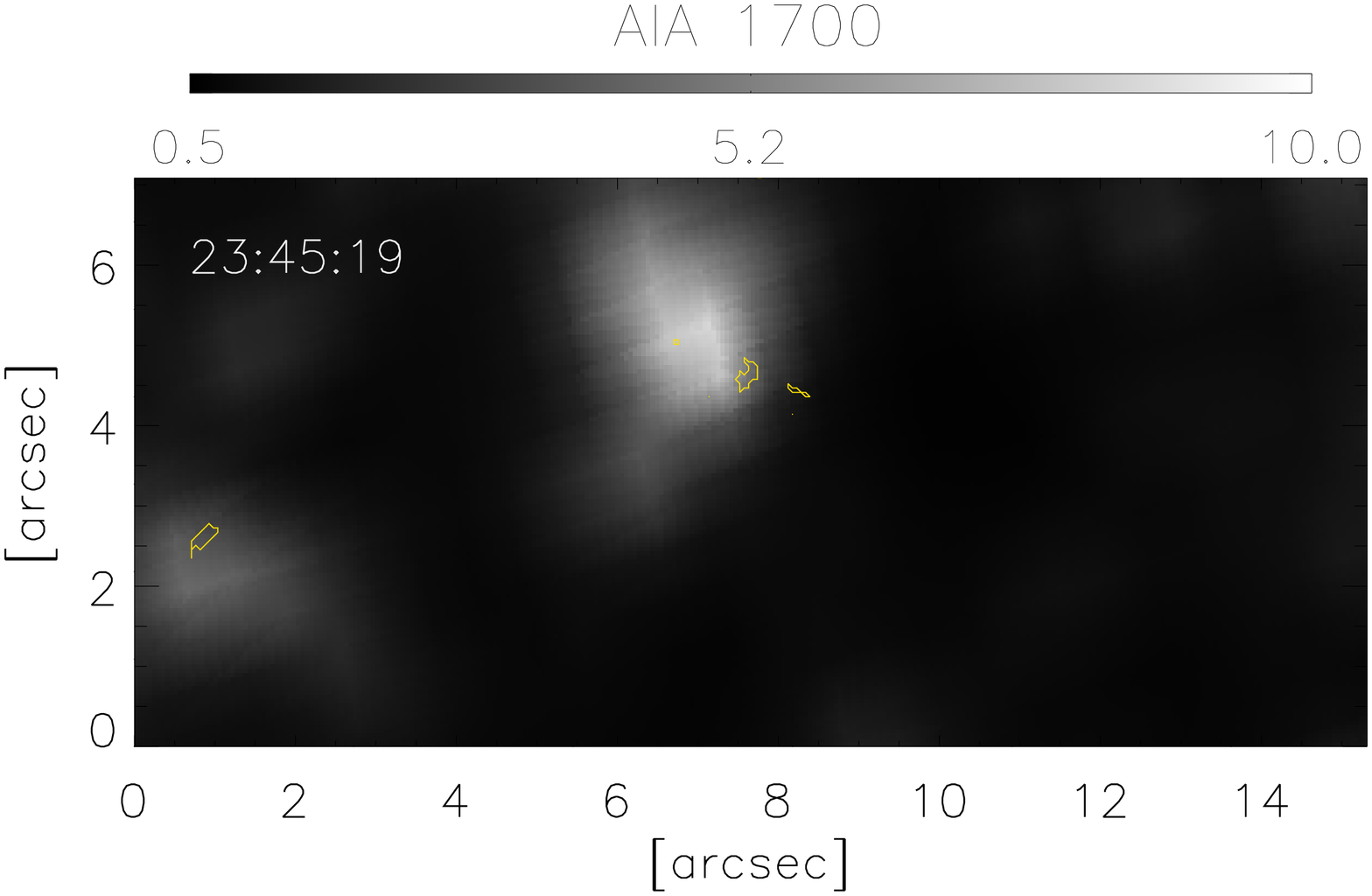}   
     \includegraphics[angle=90,width=0.327\linewidth ,trim=3.5cm 0cm 3.5cm 0cm,clip=true]{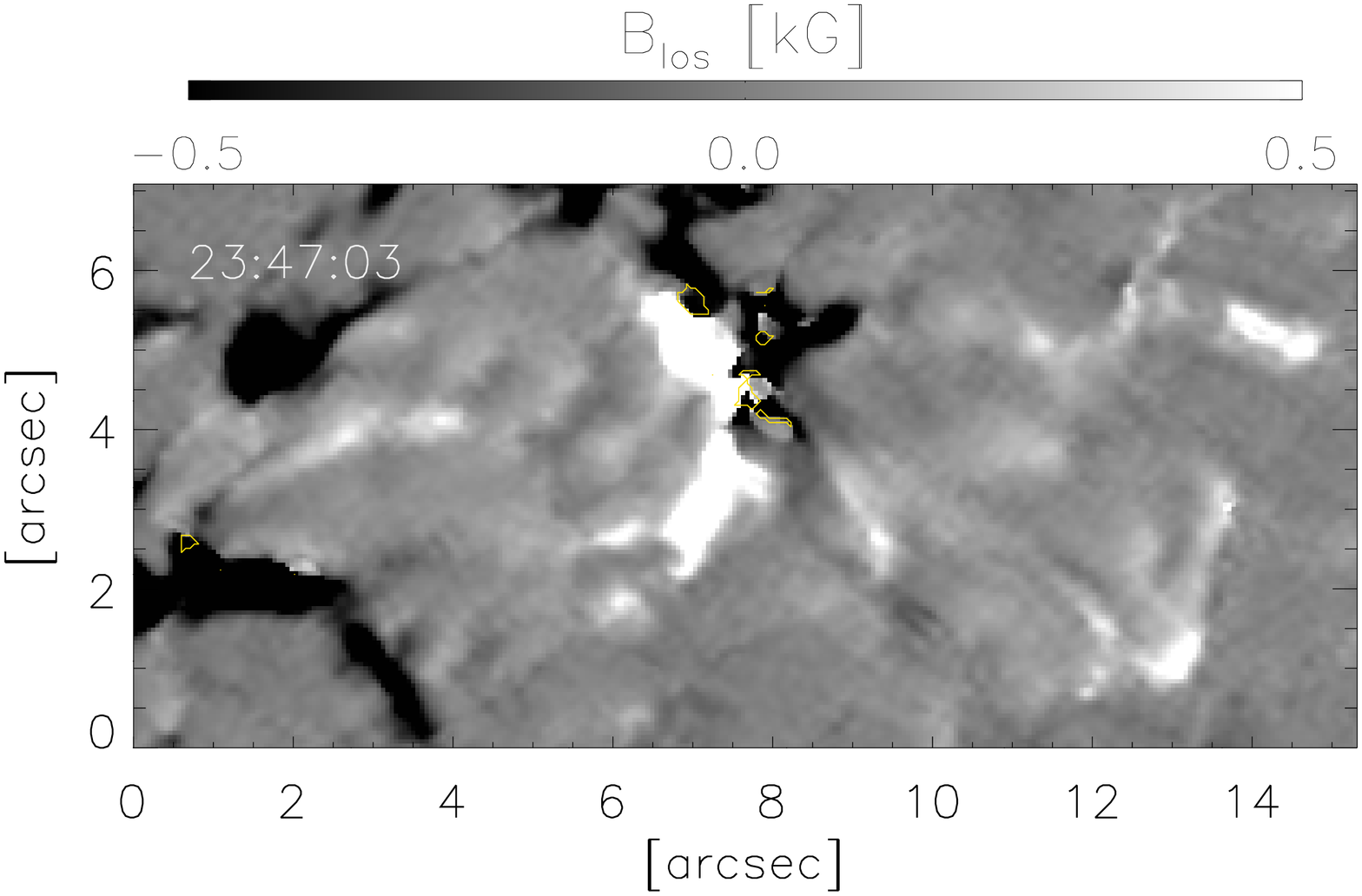}
     \includegraphics[angle=90,width=0.3\linewidth ,trim=3.5cm 0cm 3.5cm 2.5cm,clip=true]{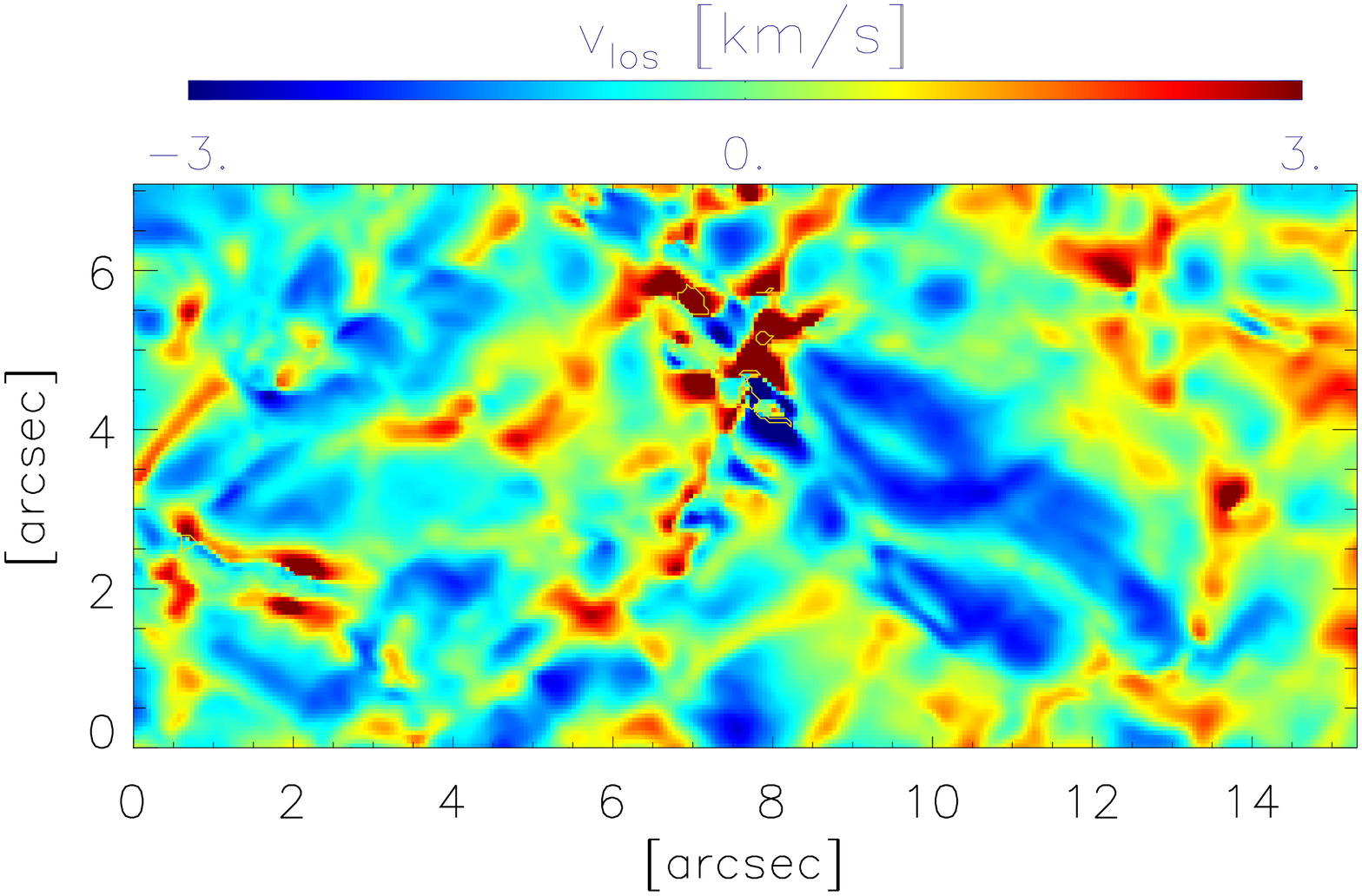}
    \includegraphics[angle=90,width=0.3\linewidth ,trim=3.5cm 0cm 3.5cm 2.5cm,clip=true]{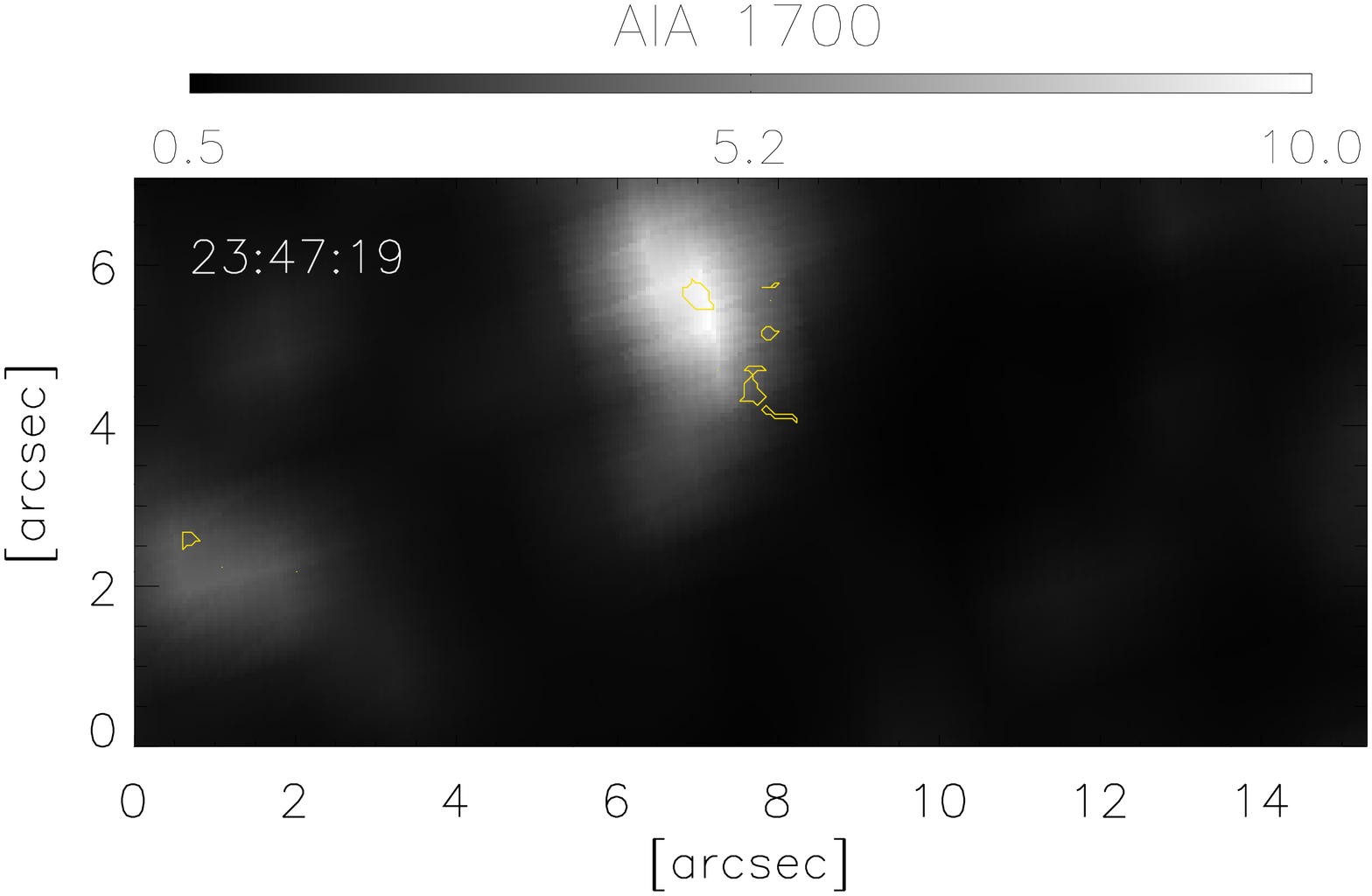} 
   \includegraphics[angle=90,width=0.327\linewidth ,trim=0cm 0cm 3.5cm 0cm,clip=true]{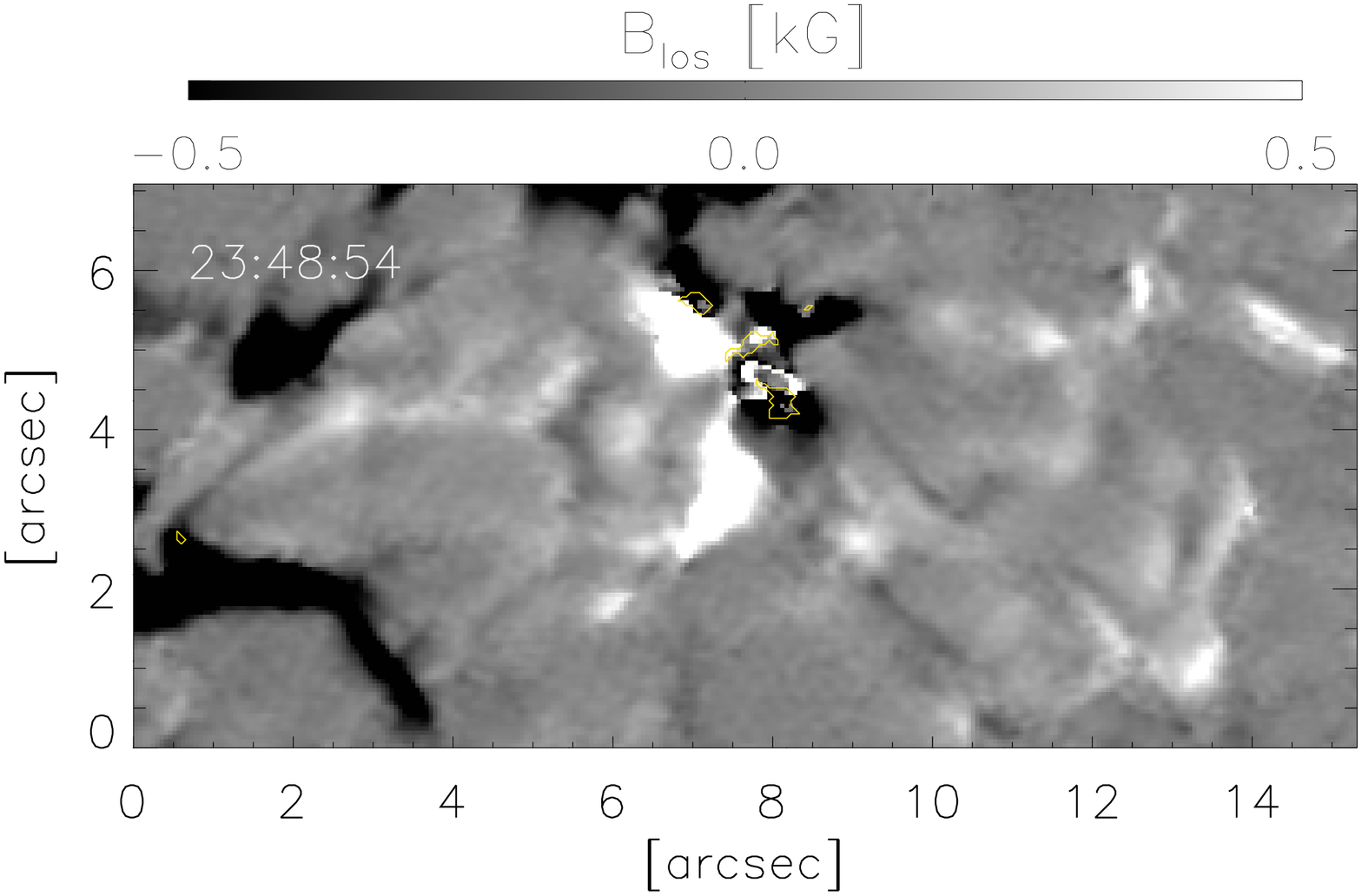}
   \includegraphics[angle=90,width=0.3\linewidth ,trim=0cm 0cm 3.5cm 2.5cm,clip=true]{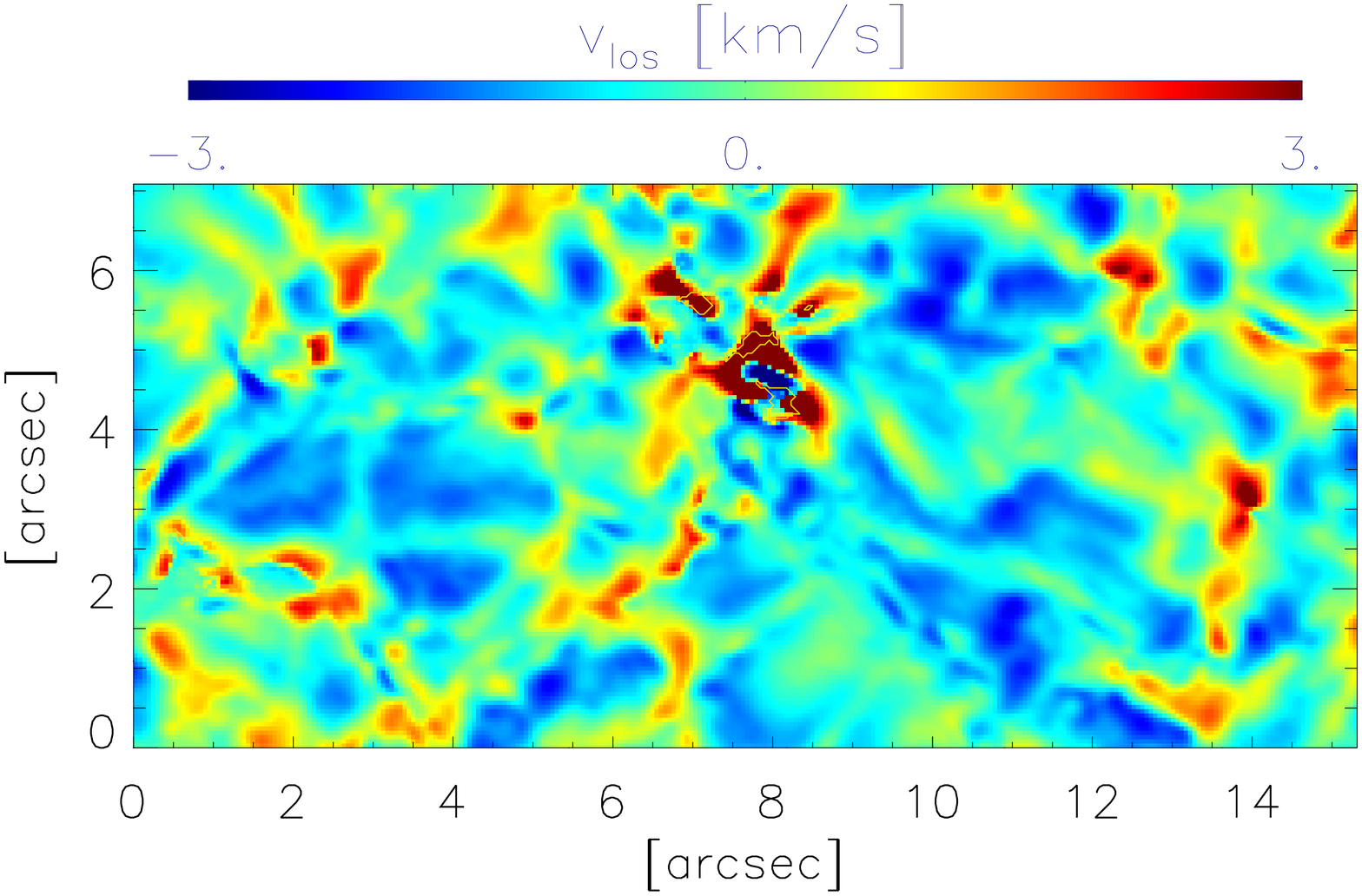}
   \includegraphics[angle=90,width=0.3\linewidth ,trim=0cm 0cm 3.5cm 2.5cm,clip=true]{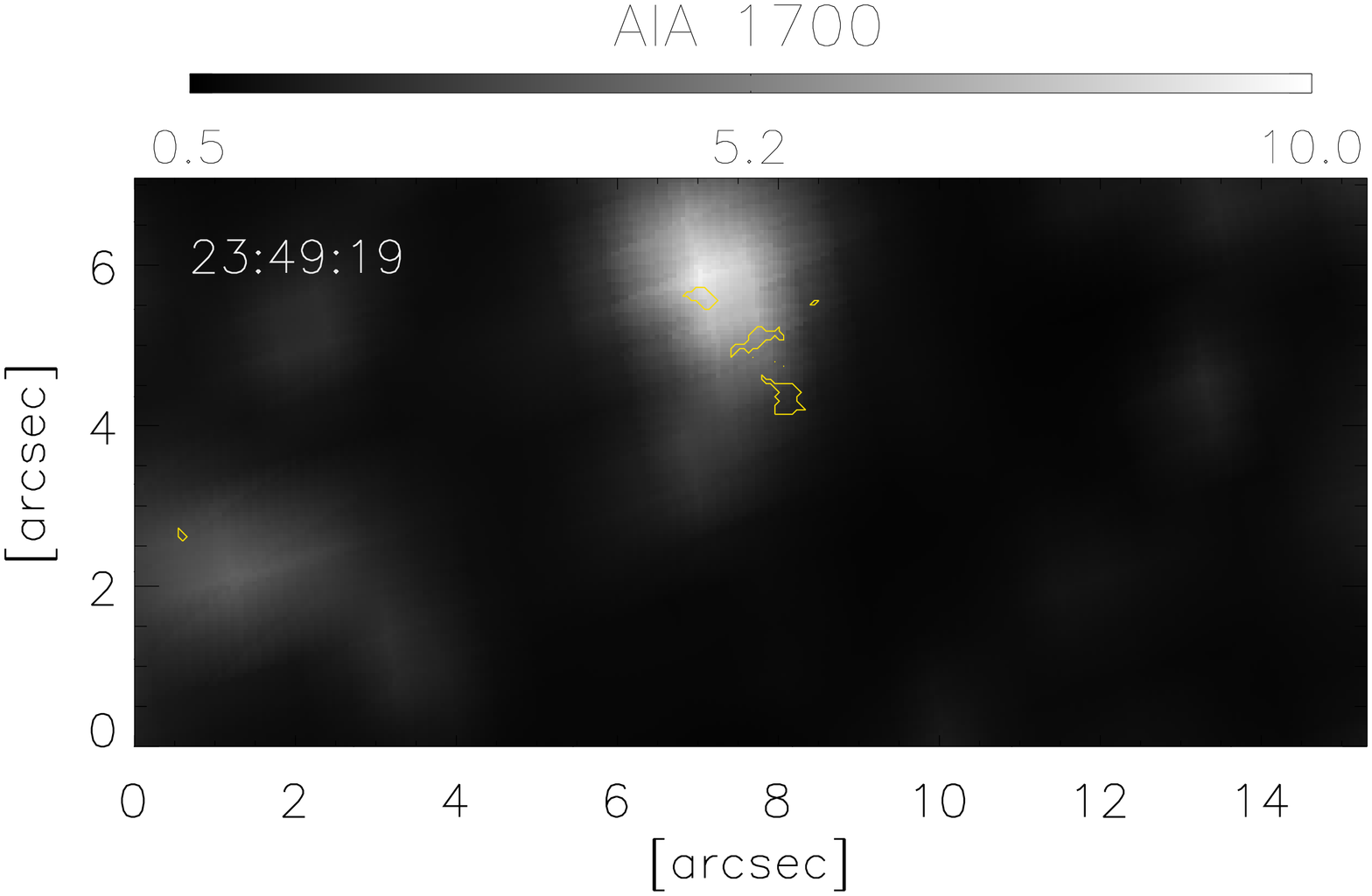}
   \caption{A small FOV around the observed event - Left and middle columns show the evolution of the line-of-sight magnetic field and line-of-sight velocity obtained from the inversions at five instances of time. Corresponding AIA 1700 channel images are given in the right column. Yellow contours mark the position where the temperature returned by the inversion at $\log\tau=-2.5$ exceeds $5400$~K. }
\label{imax_inv_rest}
\end{figure*}

\begin{figure*}
  \centering
  \includegraphics[angle=90,width=0.327\linewidth ,trim=3.5cm 0cm 0.2cm 0cm,clip=true]{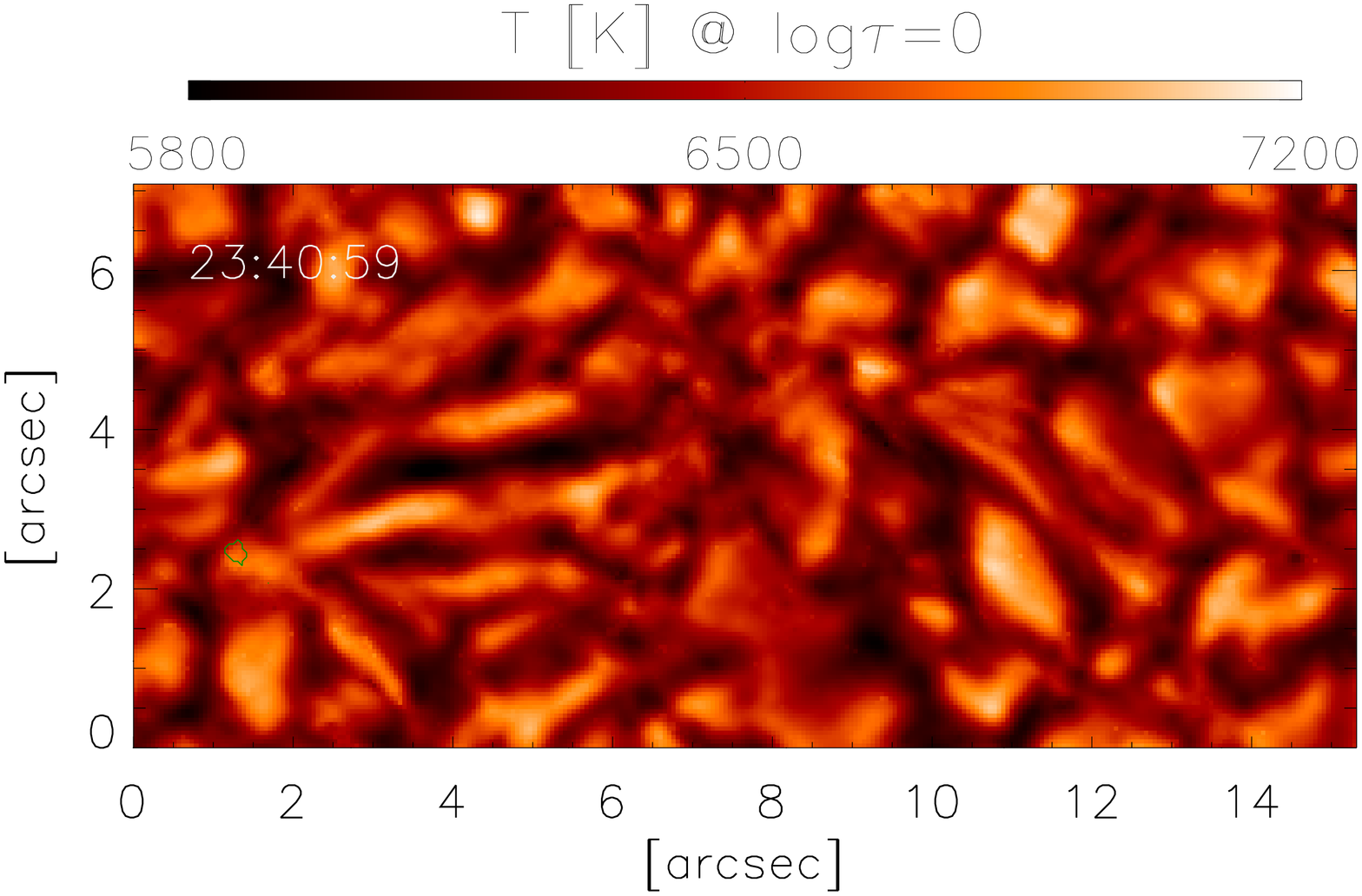} 
   \includegraphics[angle=90,width=0.30\linewidth ,trim=3.5cm 0cm 0.2cm 2.5cm,clip=true]{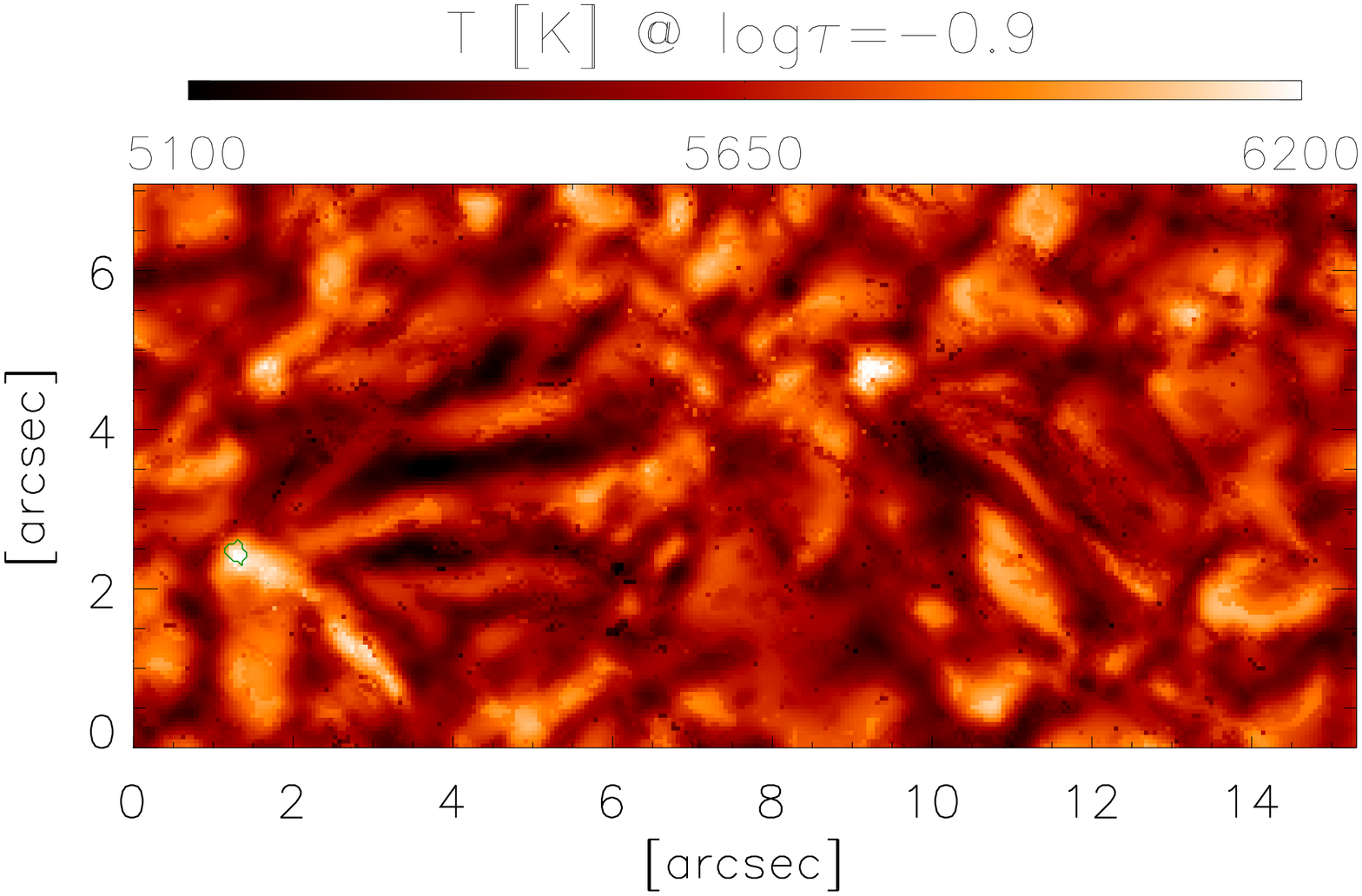}
    \includegraphics[angle=90,width=0.30\linewidth ,trim=3.5cm 0cm 0.2cm 2.5cm,clip=true]{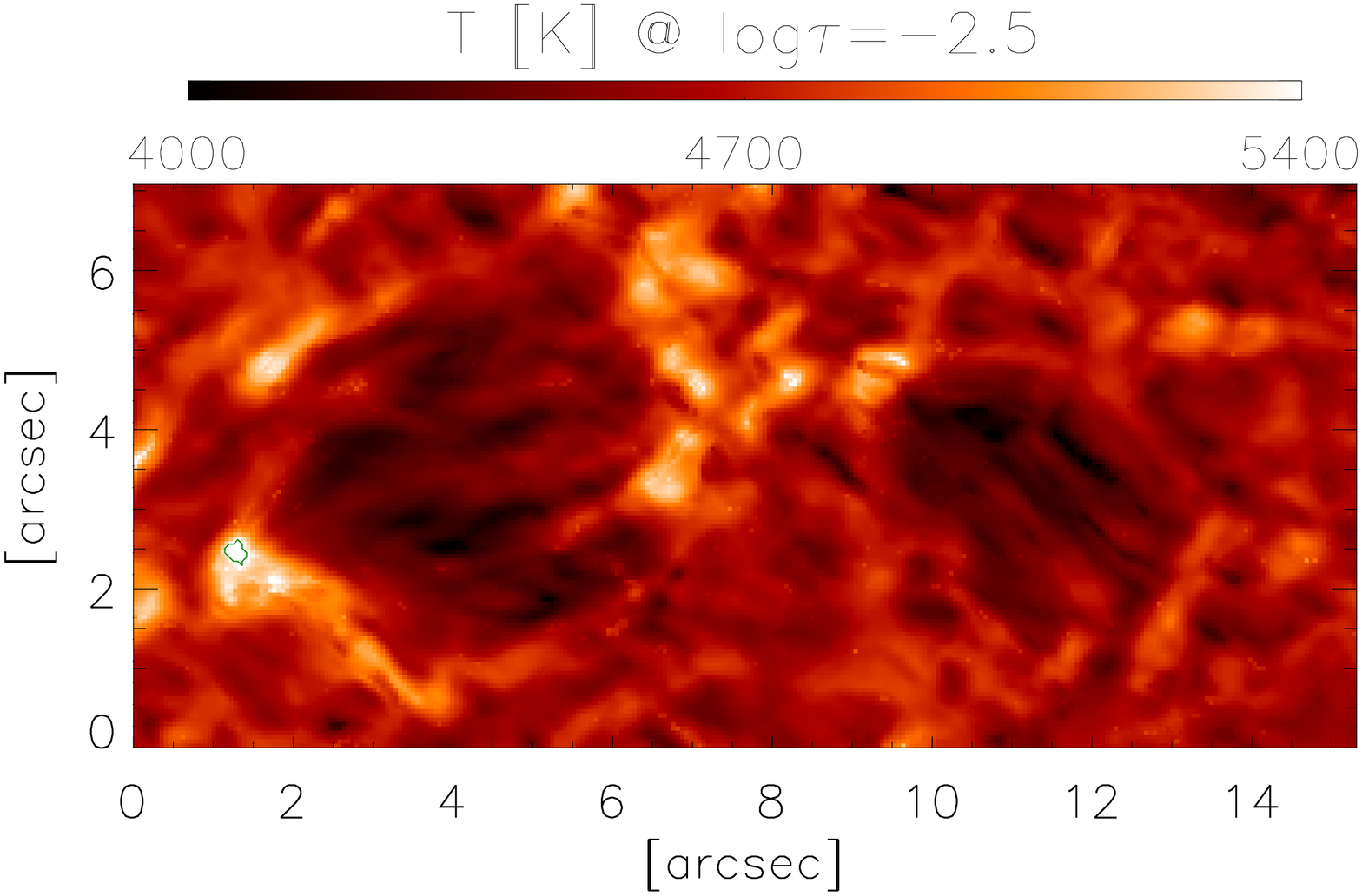}  
  \includegraphics[angle=90,width=0.327\linewidth ,trim=3.5cm 0cm 3.5cm 0cm,clip=true]{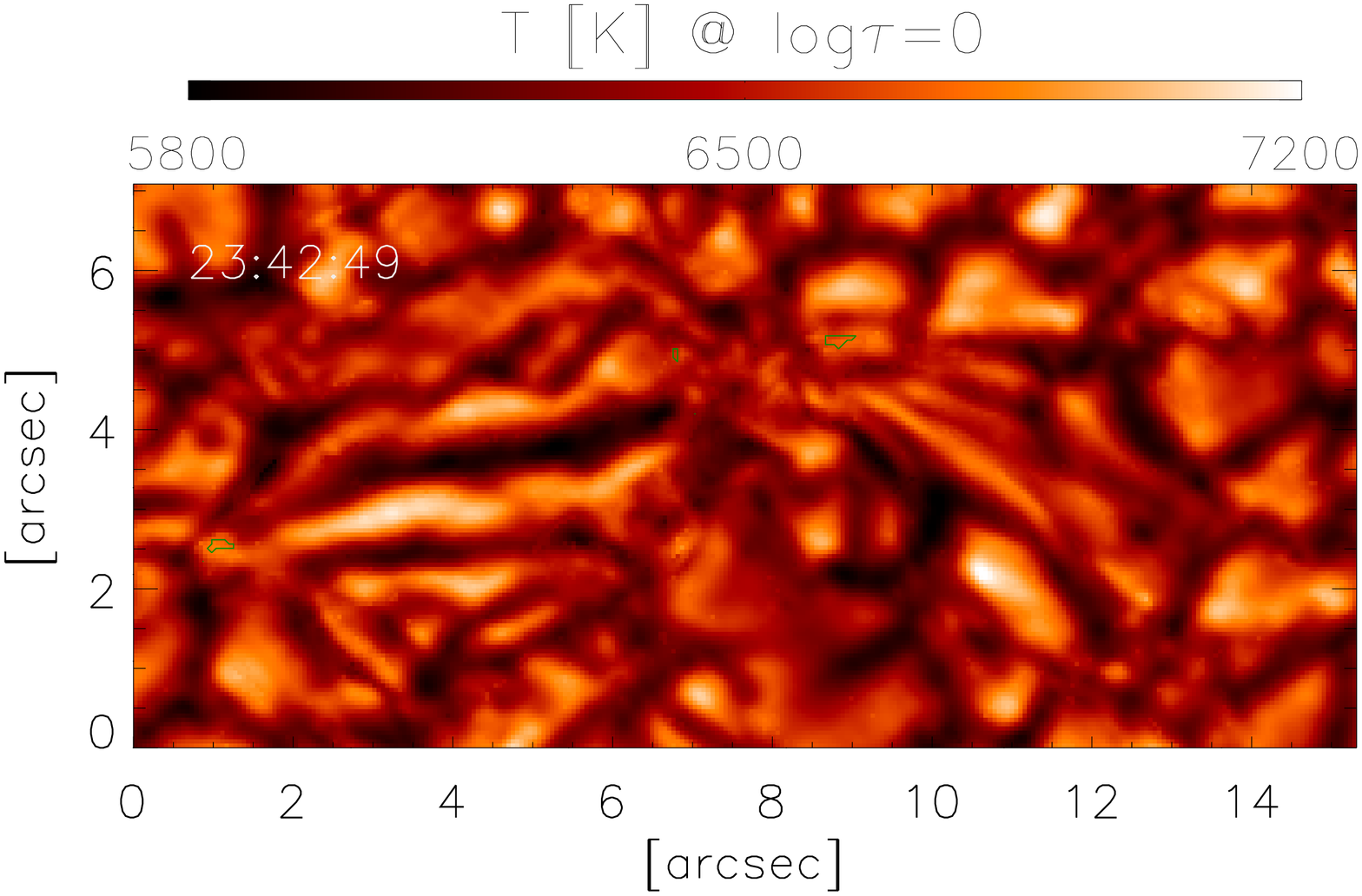}
     \includegraphics[angle=90,width=0.3\linewidth ,trim=3.5cm 0cm 3.5cm 2.5cm,clip=true]{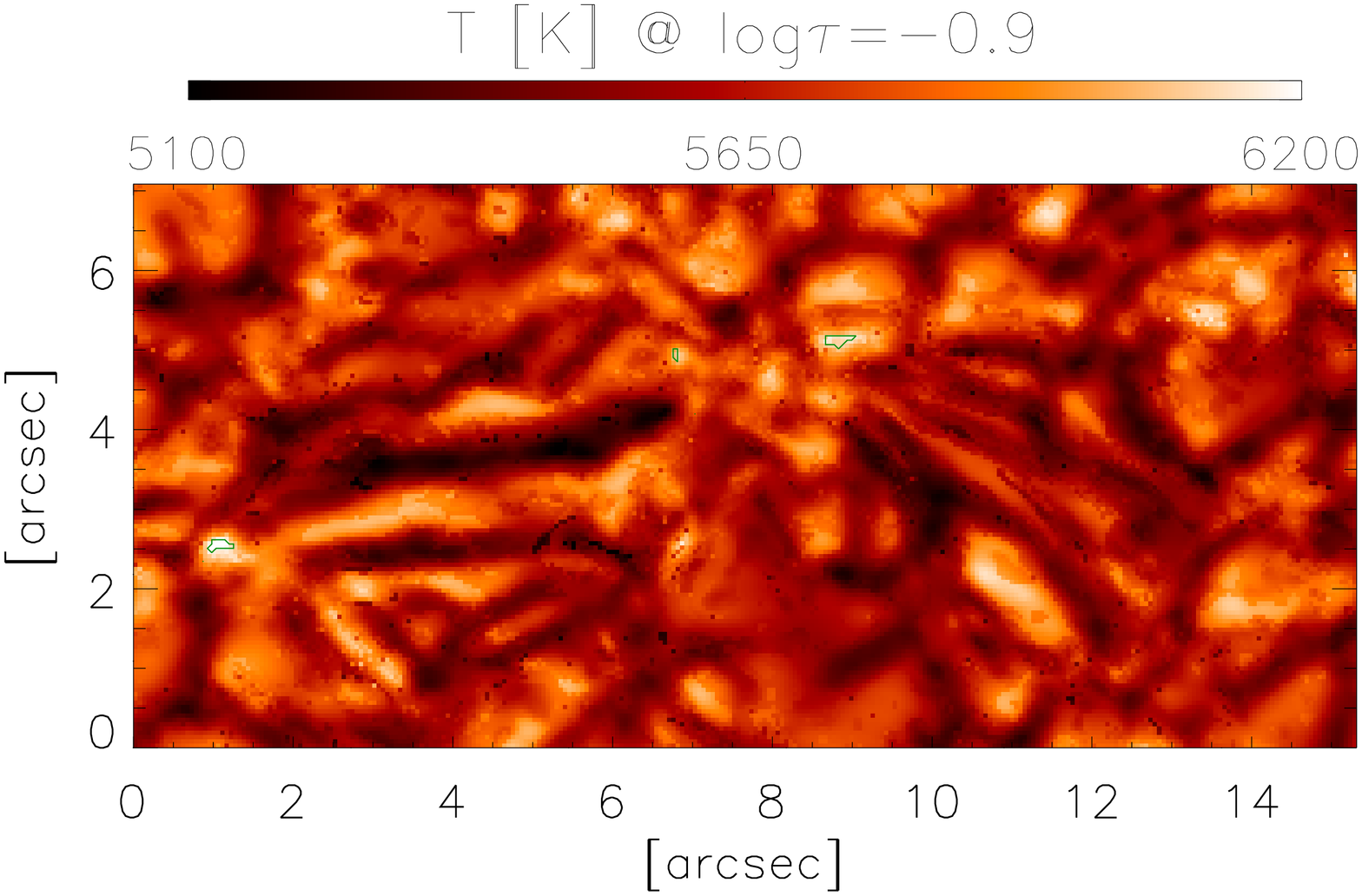}
    \includegraphics[angle=90,width=0.3\linewidth ,trim=3.5cm 0cm 3.5cm 2.5cm,clip=true]{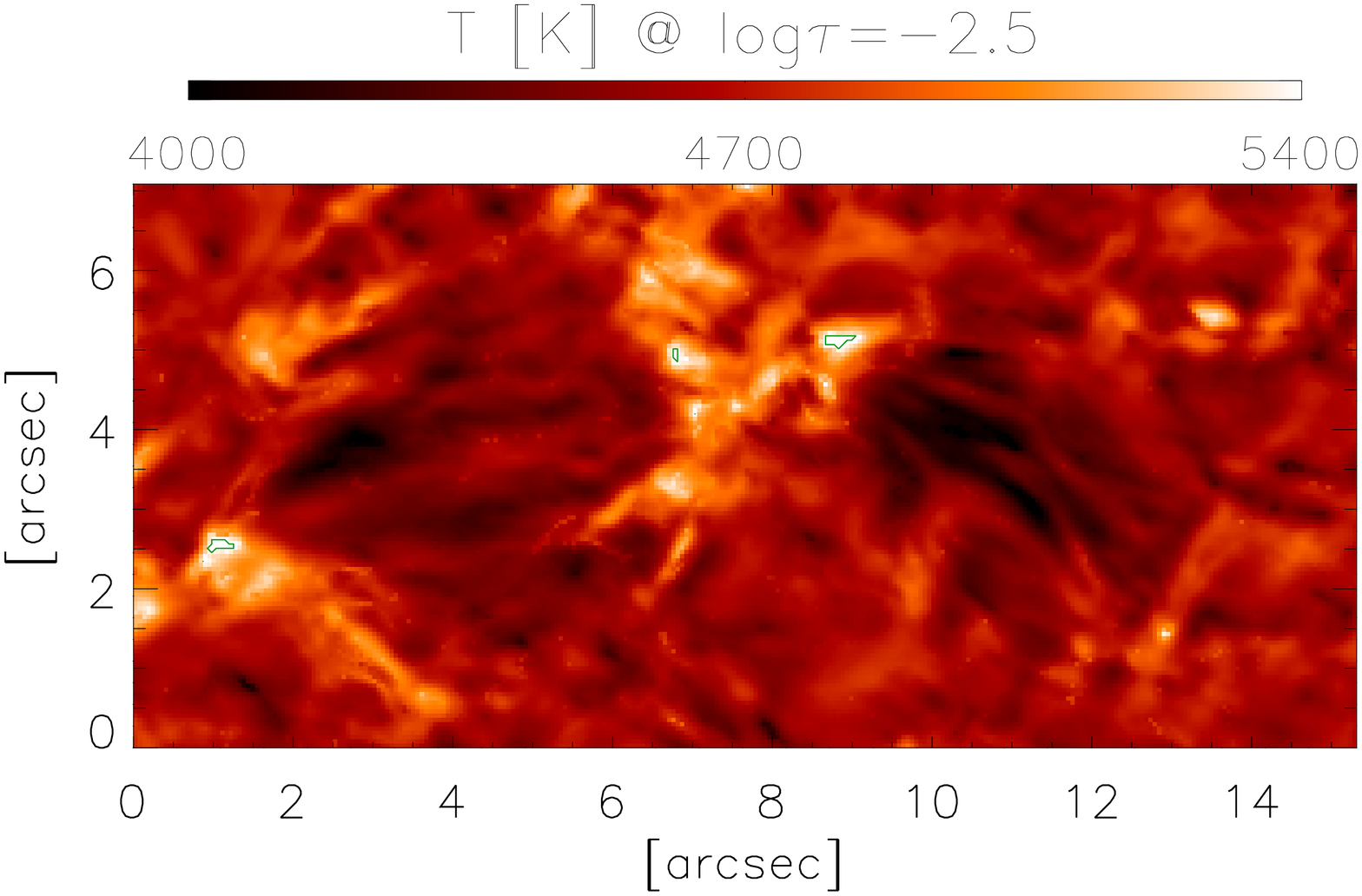} 
   \includegraphics[angle=90,width=0.327\linewidth ,trim=3.5cm 0cm 3.5cm 0cm,clip=true]{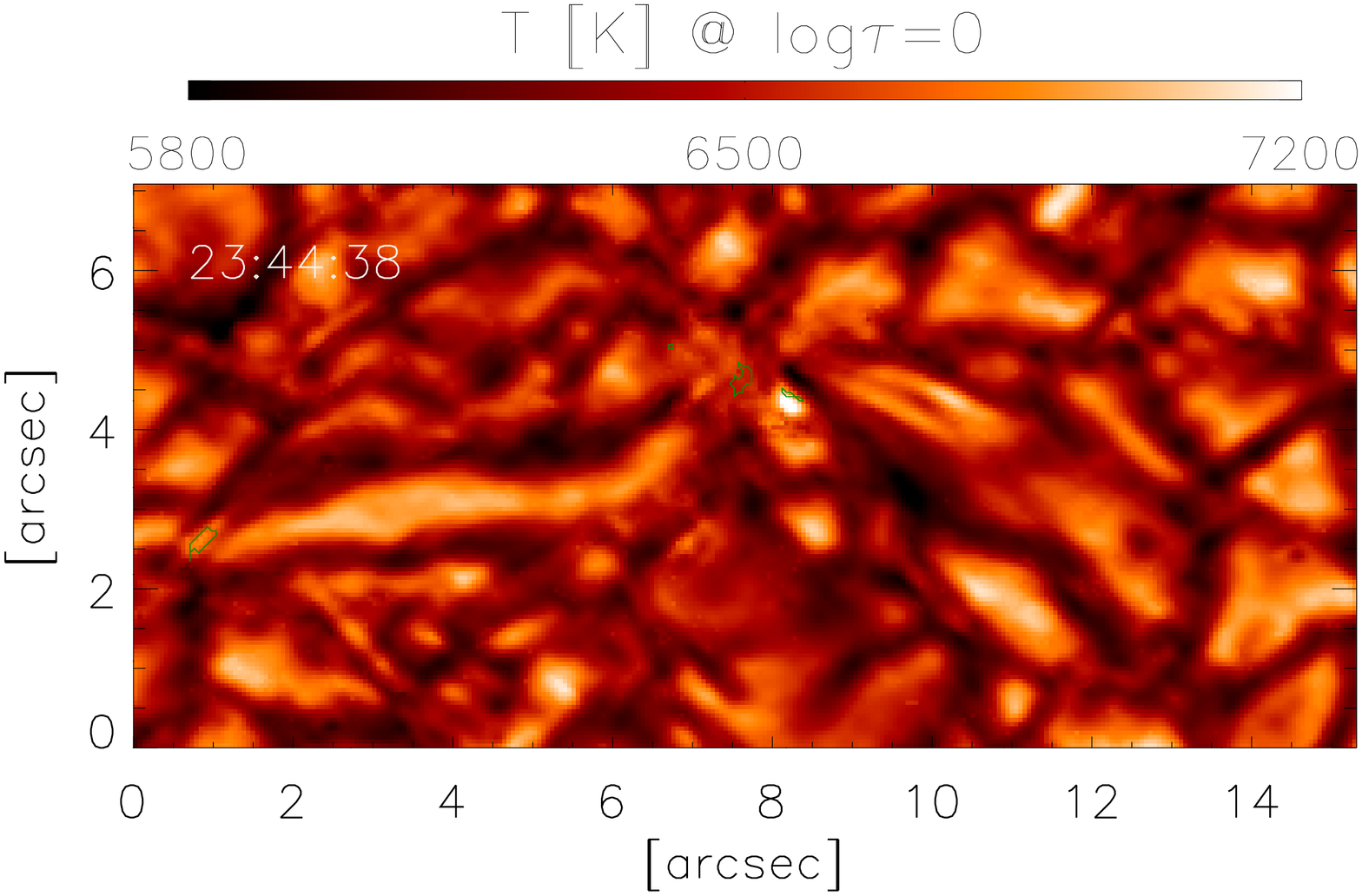}
     \includegraphics[angle=90,width=0.3\linewidth ,trim=3.5cm 0cm 3.5cm 2.5cm,clip=true]{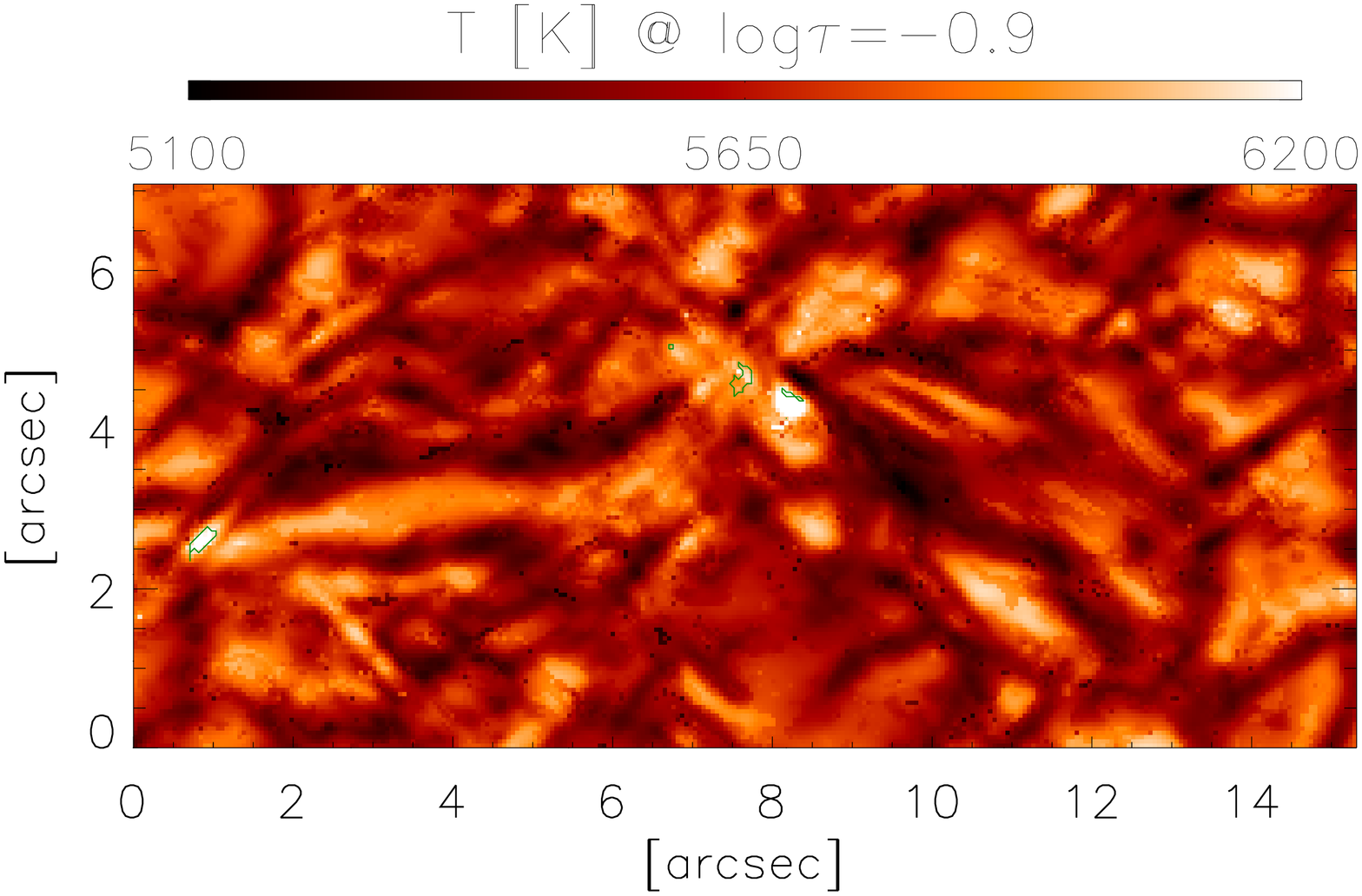}
    \includegraphics[angle=90,width=0.3\linewidth ,trim=3.5cm 0cm 3.5cm 2.5cm,clip=true]{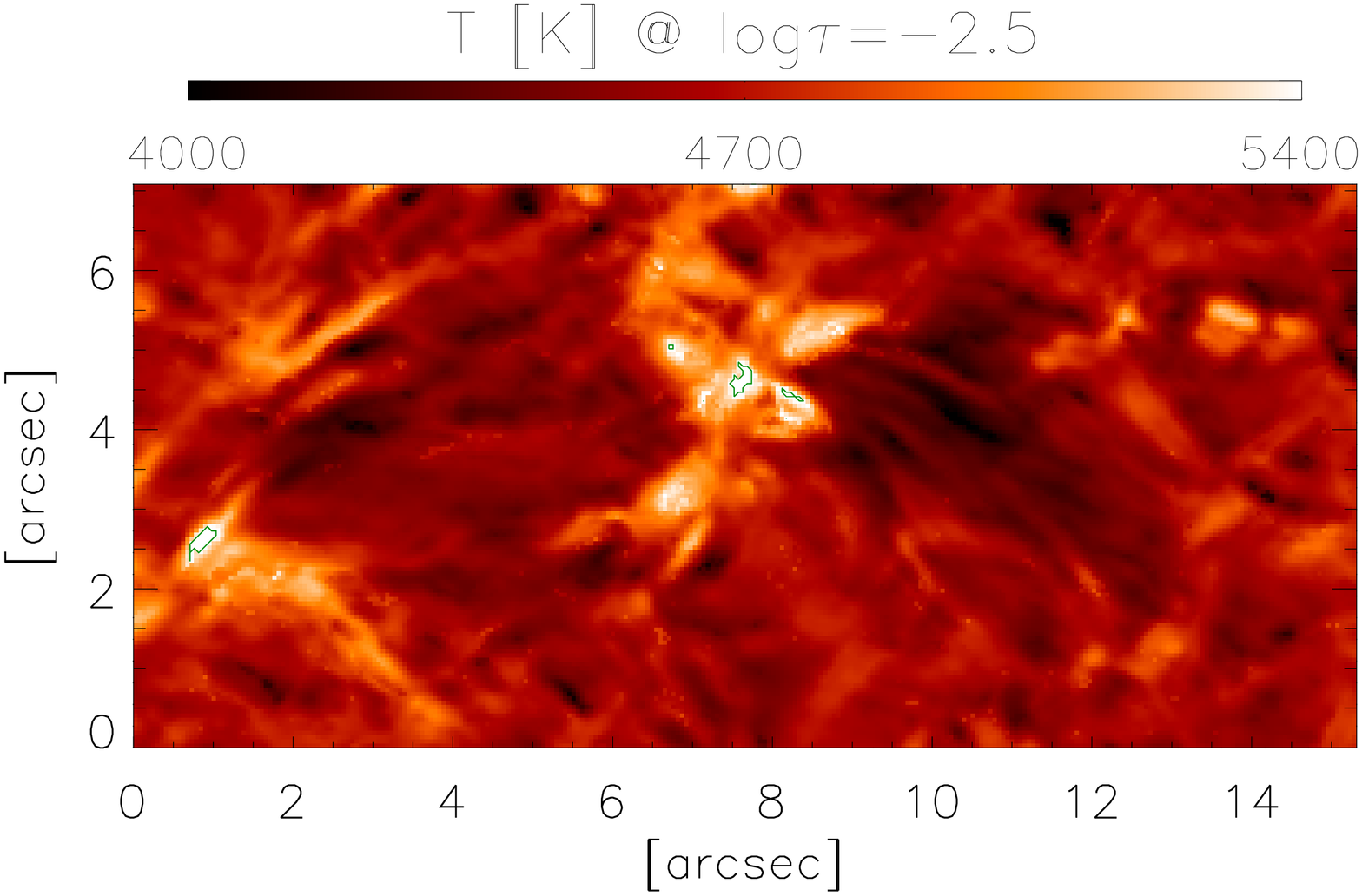}   
     \includegraphics[angle=90,width=0.327\linewidth ,trim=3.5cm 0cm 3.5cm 0cm,clip=true]{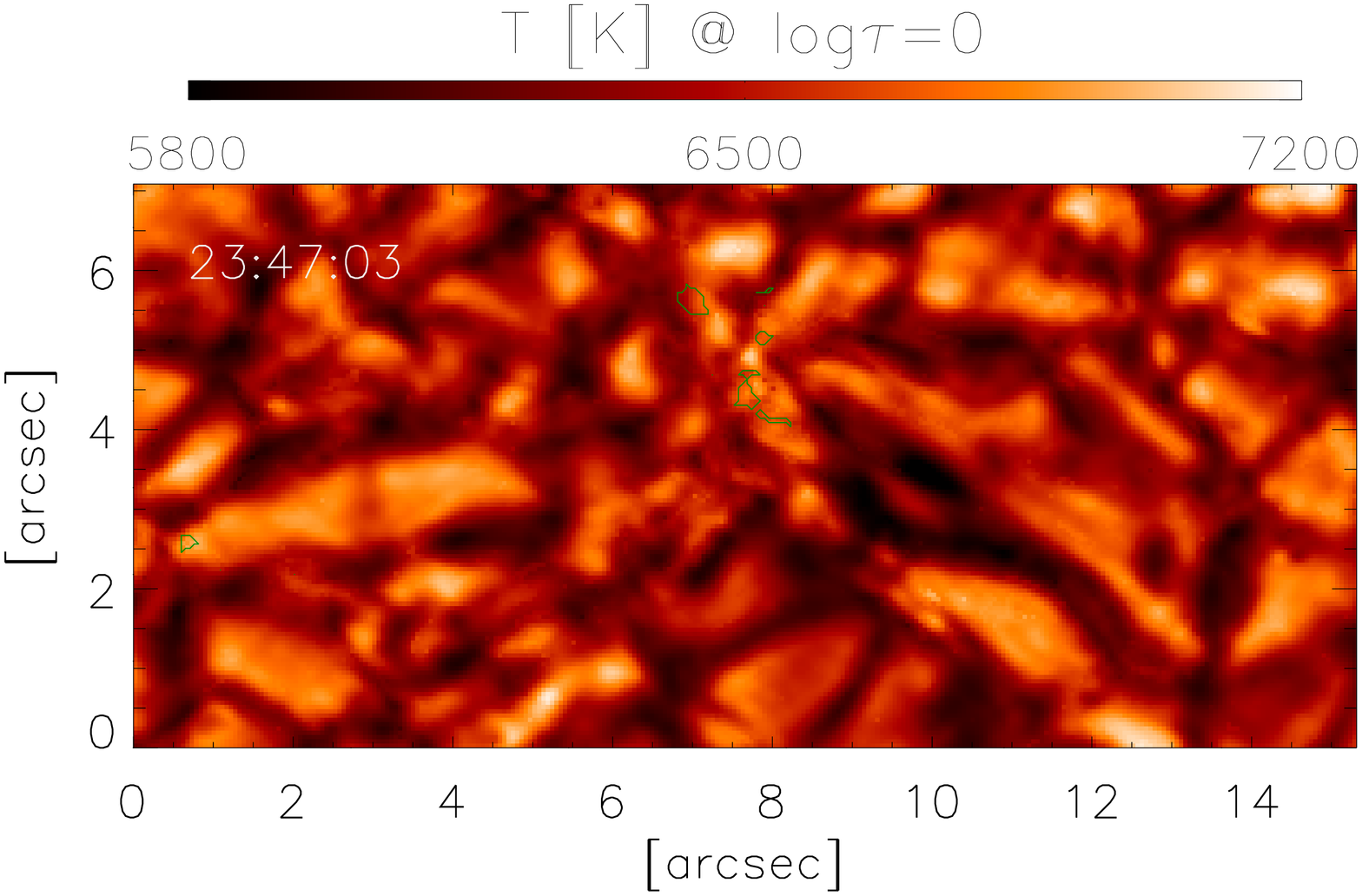}
     \includegraphics[angle=90,width=0.3\linewidth ,trim=3.5cm 0cm 3.5cm 2.5cm,clip=true]{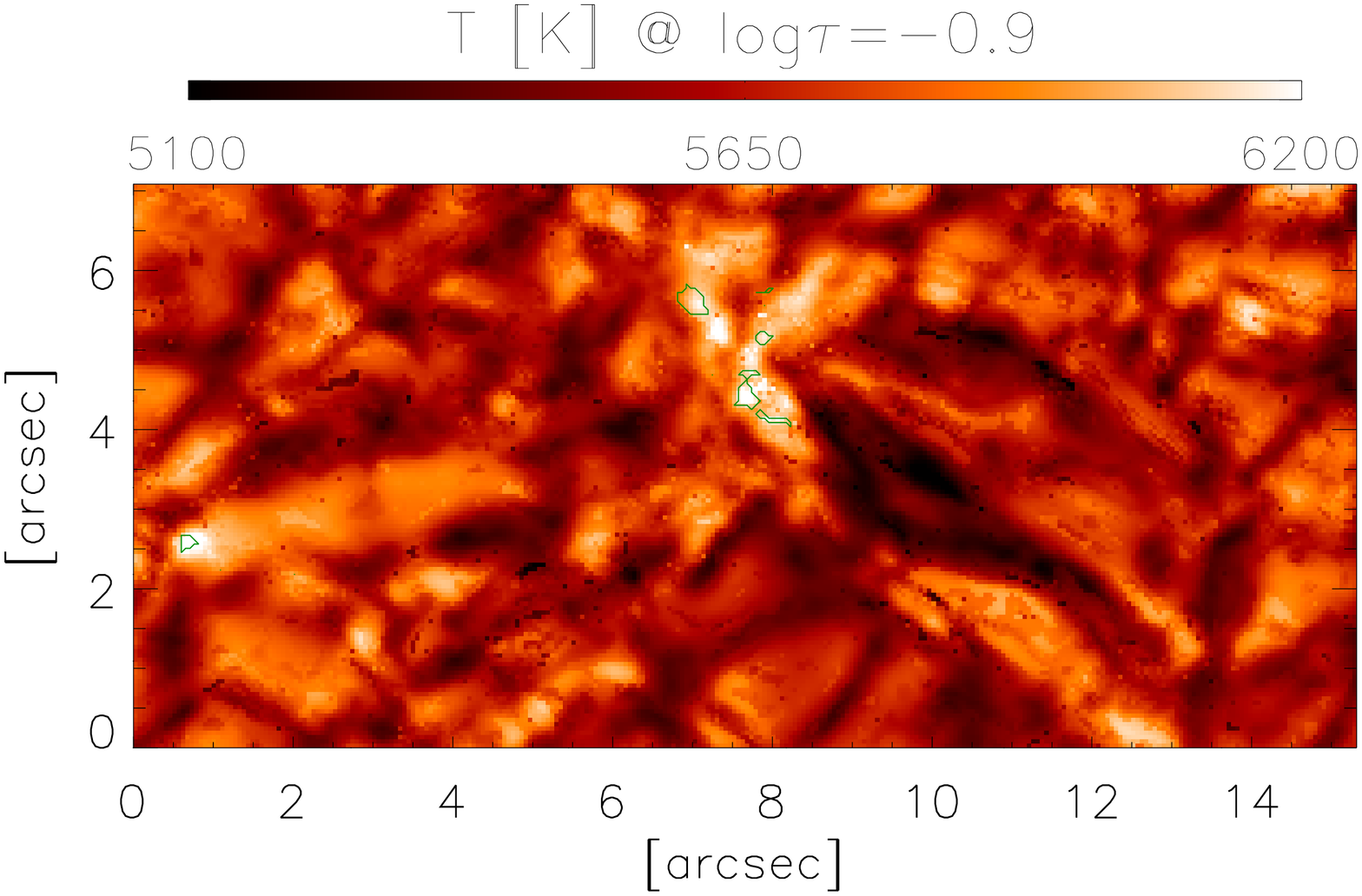}
    \includegraphics[angle=90,width=0.3\linewidth ,trim=3.5cm 0cm 3.5cm 2.5cm,clip=true]{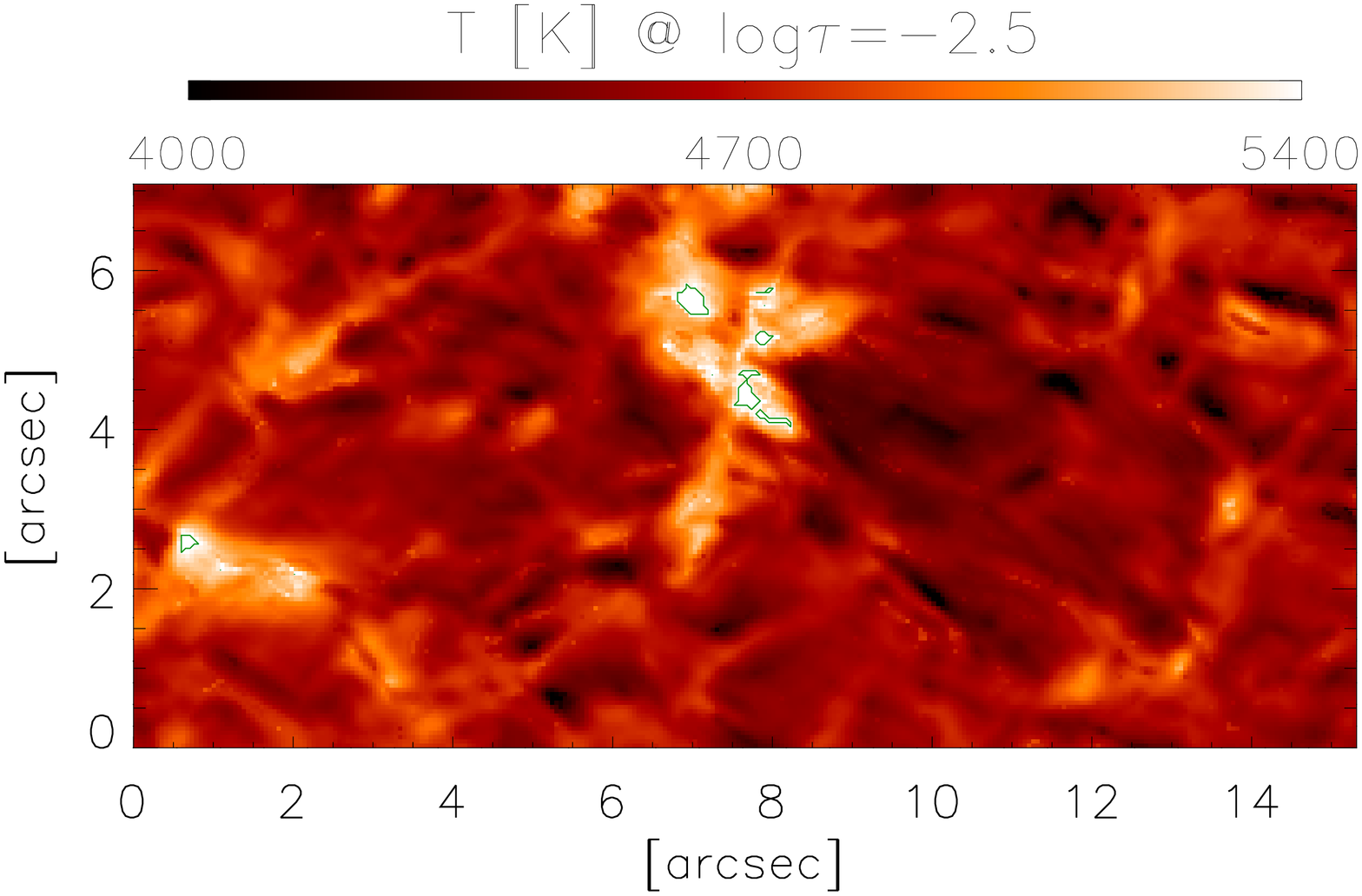} 
   \includegraphics[angle=90,width=0.327\linewidth ,trim=0cm 0cm 3.5cm 0cm,clip=true]{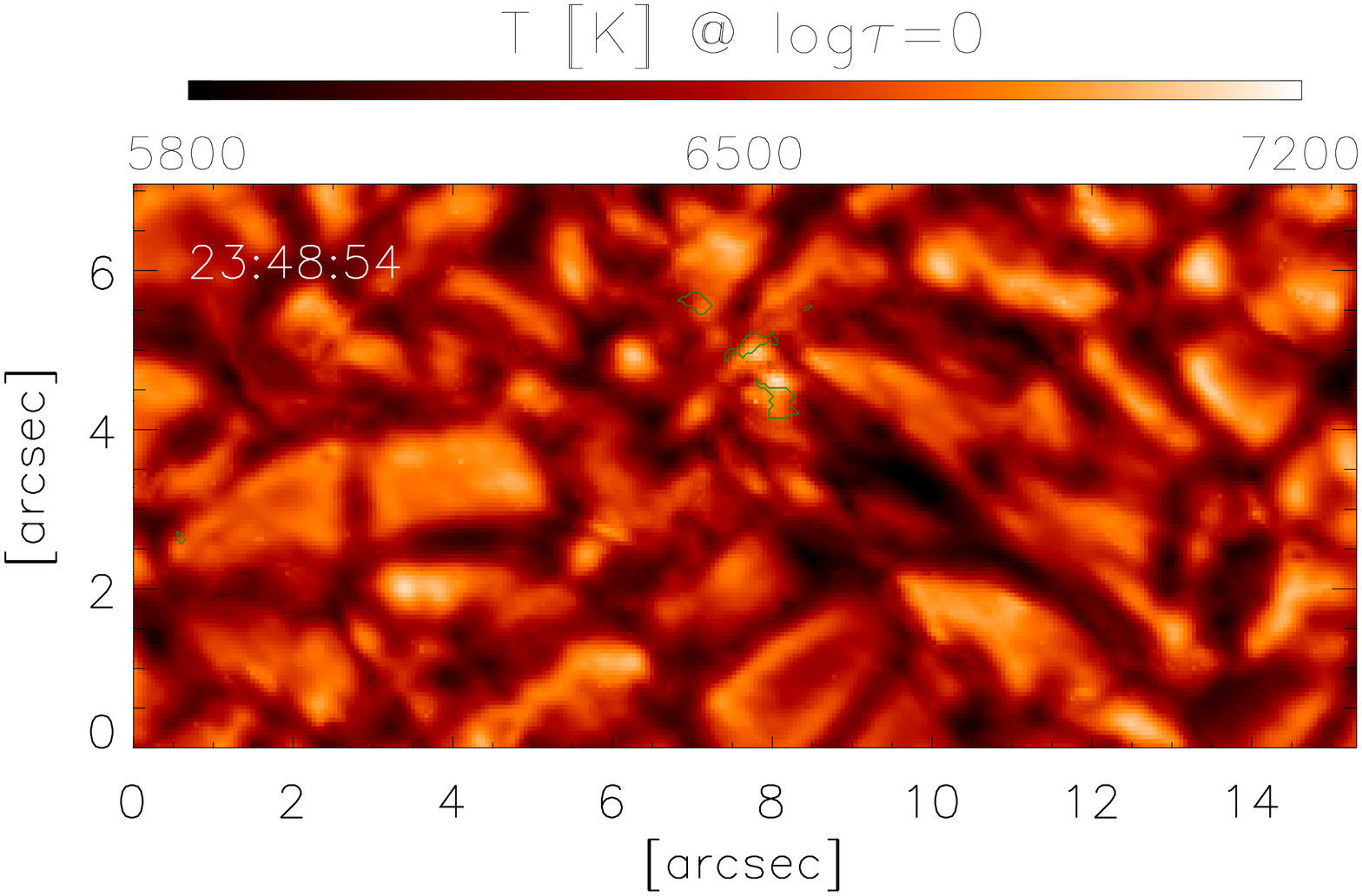}
   \includegraphics[angle=90,width=0.3\linewidth ,trim=0cm 0cm 3.5cm 2.5cm,clip=true]{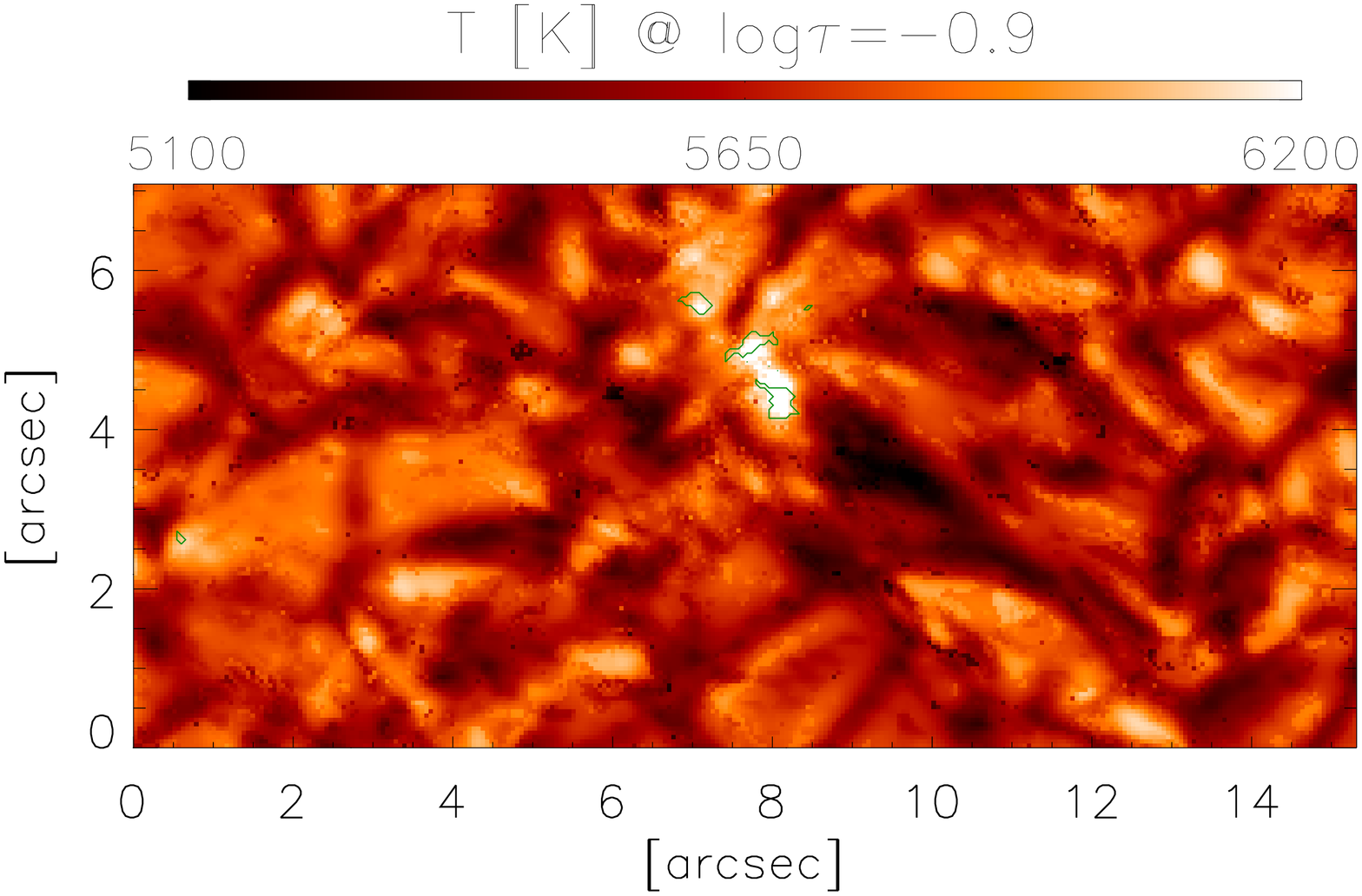}
   \includegraphics[angle=90,width=0.3\linewidth ,trim=0cm 0cm 3.5cm 2.5cm,clip=true]{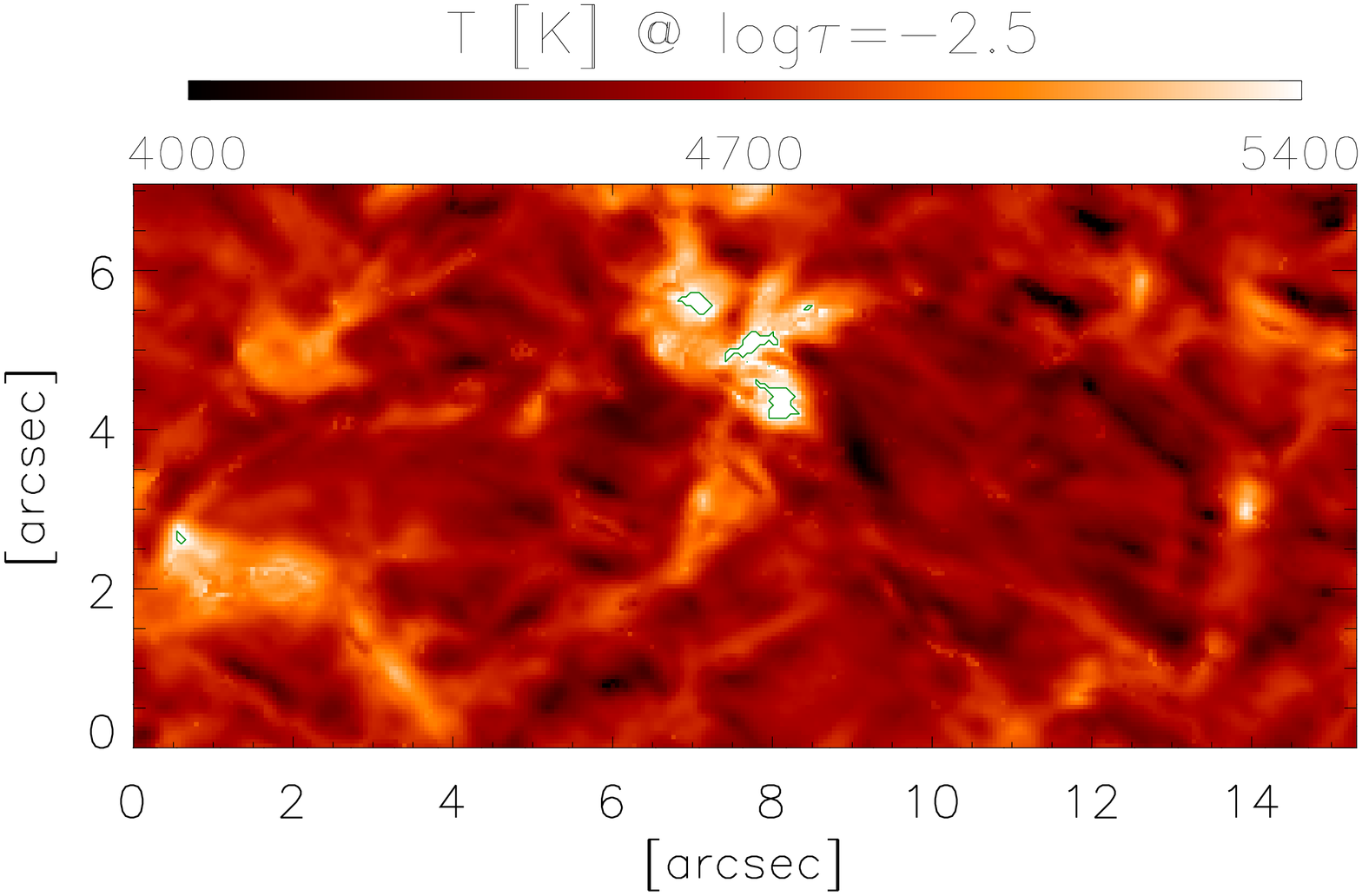}
   \caption{Maps of parameters obtained from inversions of the Sunrise/IMaX observations - \textit{Left to right:} temperatures at $\log\tau=0, -0.9$ and $-2.5$ at the same times as in Fig~\ref{imax_inv_rest}. Green contours mark the position where the temperature at $\log\tau=-2.5$ exceeds $5400$~K. }
\label{imax_inv}
\end{figure*}

The investigated event appeared during the second flight of \sunrise~\citep{Solanki:etal:this}, in a young and growing region NOAA AR 11768 on June 12th 2013 at 23:40~UT, when the AR was located at $\mu=0.93$. Fig.~\ref{full_fov} shows the full field of view (FOV) of the \sunrise{}/IMaX instrument and the corresponding image recorded in the SDO/AIA 1700 channel \citep{Pesnell2012,Lemen2012} when the event reached its highest brightness at 23:47~UT. \cite{Centeno:etal:this} described, in detail, the flux emergence that led to its onset and its chromospheric signatures seen by the Sunrise Filter Imager \citep[SuFI;][]{Gandorfer2011}, the other instrument on board of \sunrise. We concentrate only on the photosphere, in particular the \sunrise{}/IMaX data that continuously recorded the evolution of emerging magnetic field from approximately 2 minutes before the appearance of the event till its extended decay some 15 minutes after.  Figures~\ref{imax_inv_rest} and ~\ref{imax_inv} show the results of the 1D version of the SPINOR inversion code \citep{Frutiger:etal:2000} applied to the data \citep{Solanki:etal:this} at five instances of time.\footnote{The whole evolution can be found as a part of on-line material.} The inversion strategy allows temperature to be modified at three nodes along the line-of-sight (LOS) while the magnetic field vector and LOS velocity are assumed to be constant with height.

\sunrise{}/IMaX time series start shortly after the footpoints of the second emerging loop appear \citep{Centeno:etal:this}. Inverted maps of LOS velocity show fast upflows already at the beginning of the time series and they seem to continue for over 7 min. During that period, the spatial extent of the upflowing region increases from $2$\arcsec ~to approximately $6$\arcsec ~and the maximum projected LOS velocity exceeds 4~km/s. The emergence is, however, not symmetric. Faster upflows appear near the negative footpoint, which is also more concentrated and moves much faster than the positive one. It approaches the previously formed opposite polarity, i.e. the positive footpoint of the other emerging event, with a projected horizontal speed of approximately 3.5 km/s. Once it comes into contact with pre-existing magnetic features, the brightness in the AIA 1700 channel starts to increase. The brightness in the AIA 1700 channel reaches its maximum almost at the same time when upflows cease. It then gradually fades, but it does not dim completely as the two opposite polarities are still visible at the end of the \sunrise{}/IMaX time series. 

During the emergence and cancellation, a temperature increase is detected at several locations. Particularly interesting are two for which we find counterparts in simulations: at the negative polarity footpoint and the neutral line between the two polarities. The region that coincides with the negative polarity footpoint of the emerging system shows a temperature increase very early on, at the beginning of the emergence  ([11\arcsec, 5\arcsec] at t=23:40:59 UT, top panels in Figs.~\ref{imax_inv_rest} and ~\ref{imax_inv}). There, at all instances, the inversion code returns increased temperature at all nodes. At first, the temperature rises by less than 50 K at all three levels, but it increases by more than 1000 K at the lowest and more than 1500 K at the highest node when the footpoint reaches the opposite polarity (t=23:44:38 UT). At the second location of interest, the neutral line between the opposite polarities, the code returns the temperature increase at the highest node first (t=23:44:38 UT) and only in the next frame (i.e. 36.5 sec later), is the temperature jump retrieved also at the lower nodes. At this region, the highest temperature increase again does not go over $1500$~K and is largest at the highest node.

\begin{figure}
  \centering 
   \includegraphics[width=0.7\linewidth,trim=3.5cm 2cm 3.5cm 1cm,clip=true ]{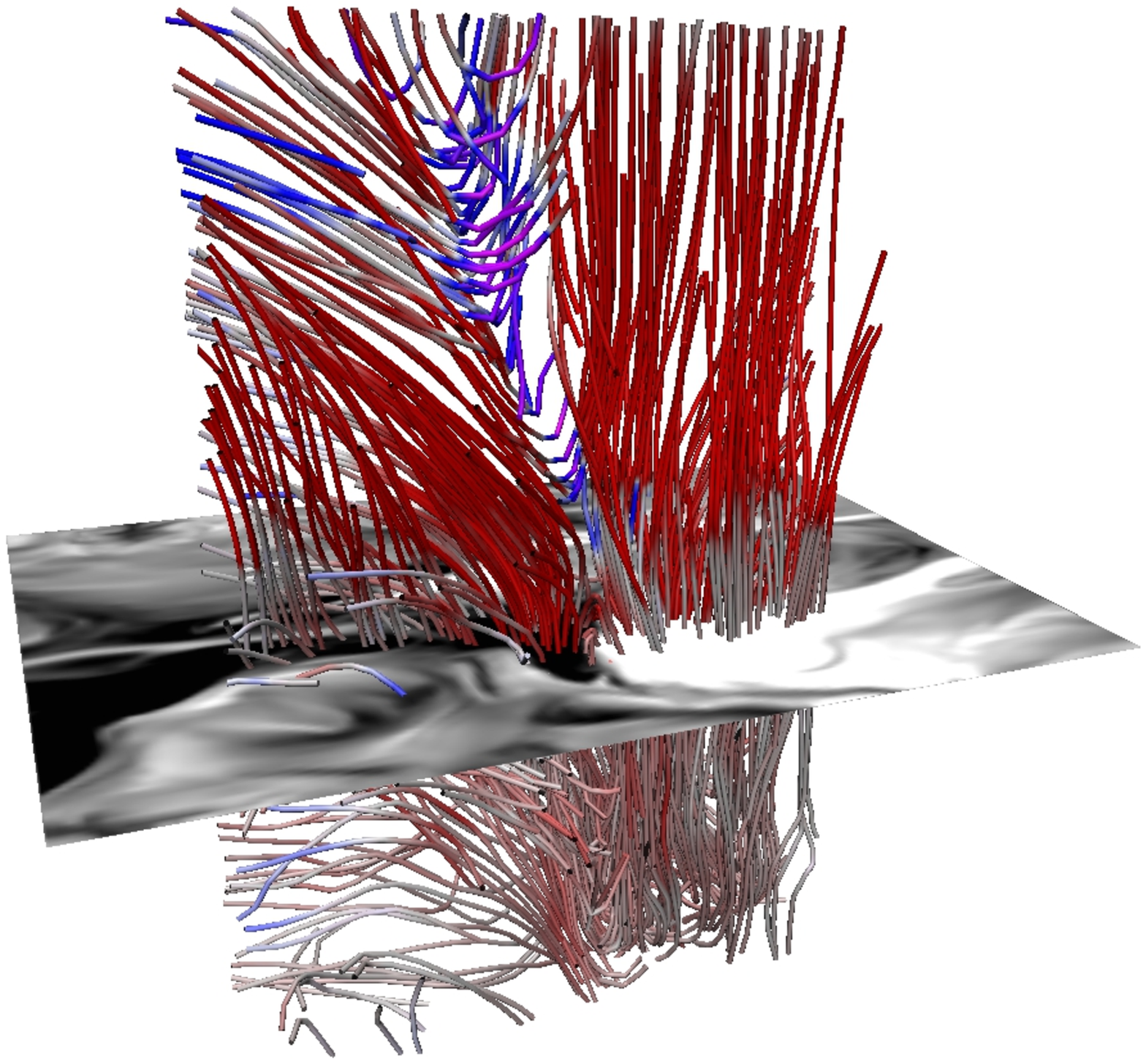}  
    \includegraphics[width=0.6\linewidth,trim=3.5cm 11cm 17cm 1cm,clip=true]{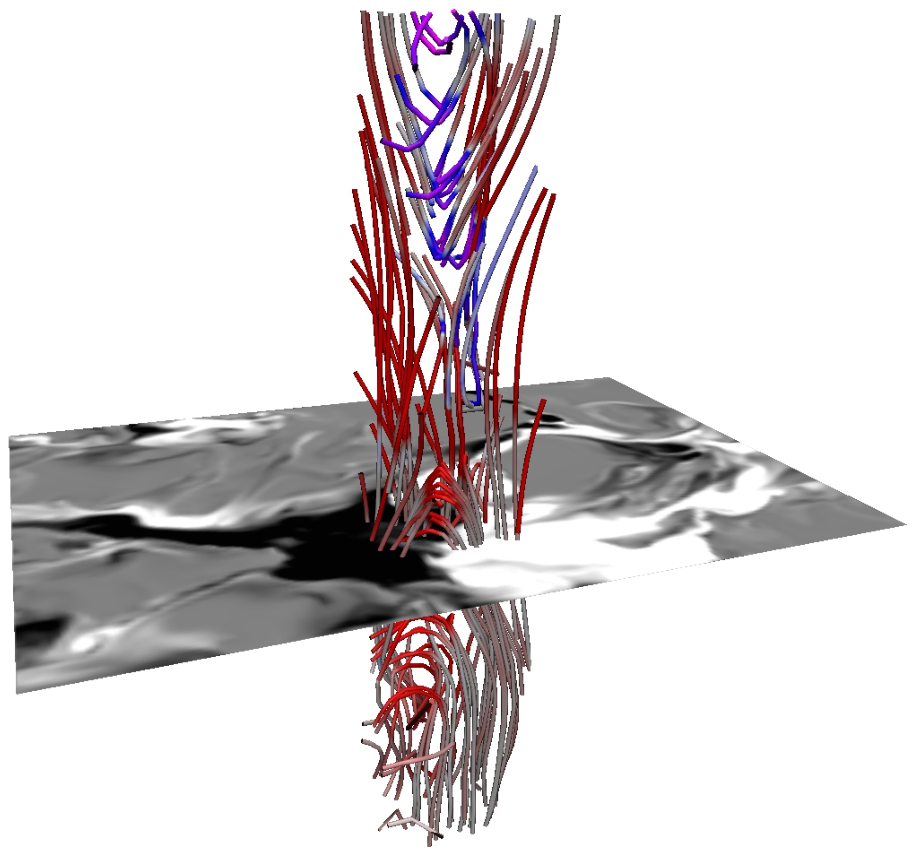} 
    \caption{Magnetic field lines at two time instances: t=17.5~min (top panel) and t=25.8~min (bottom panel). The color coding corresponds to the vertical velocity with upflows being blue.}
\label{topology}
\end{figure}

\begin{figure*}
  \centering
  \includegraphics[angle=90,width=0.327\linewidth ,trim=3.5cm 0cm 0.2cm 0cm,clip=true]{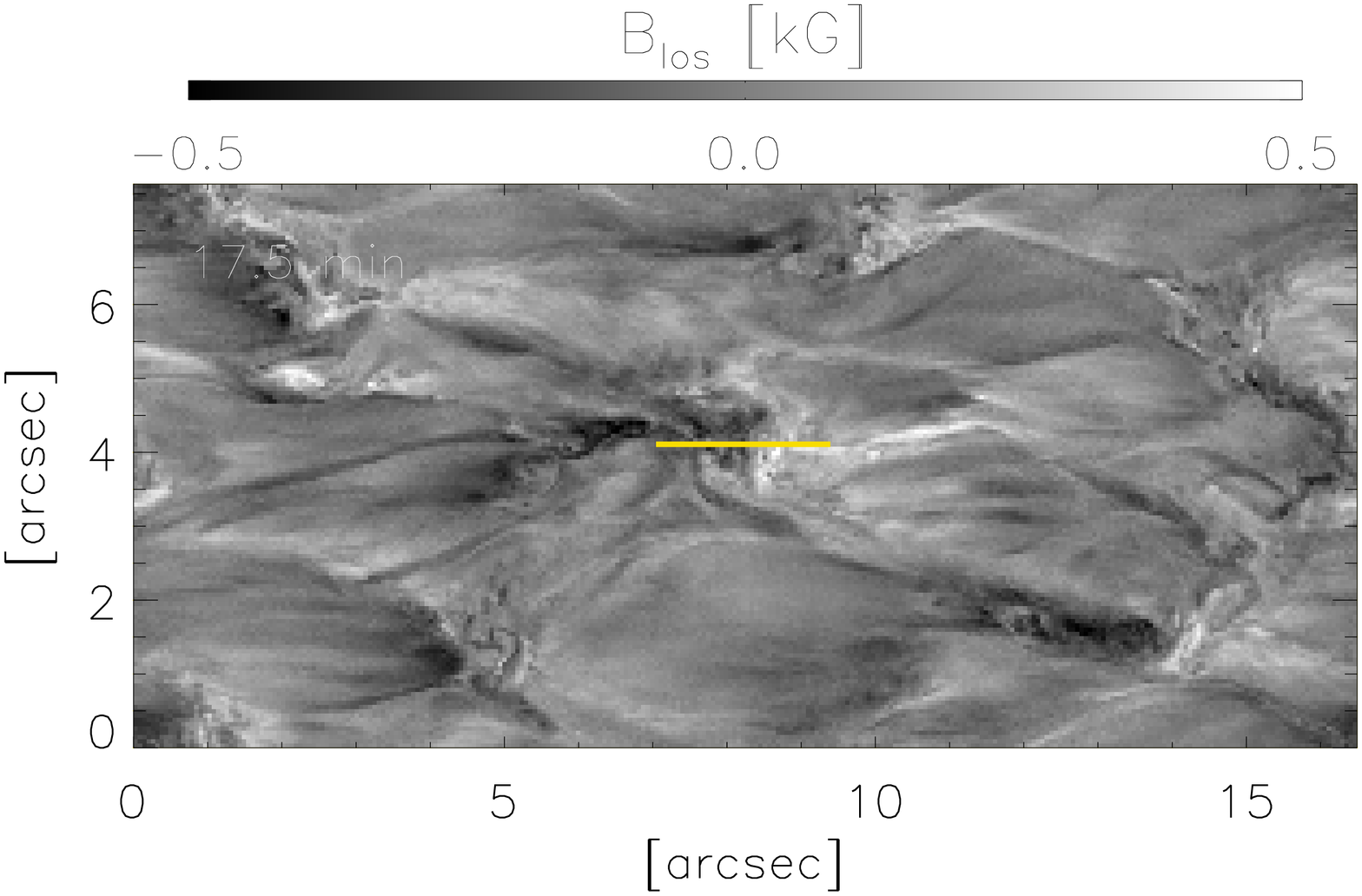} 
   \includegraphics[angle=90,width=0.30\linewidth ,trim=3.5cm 0cm 0.2cm 2.5cm,clip=true]{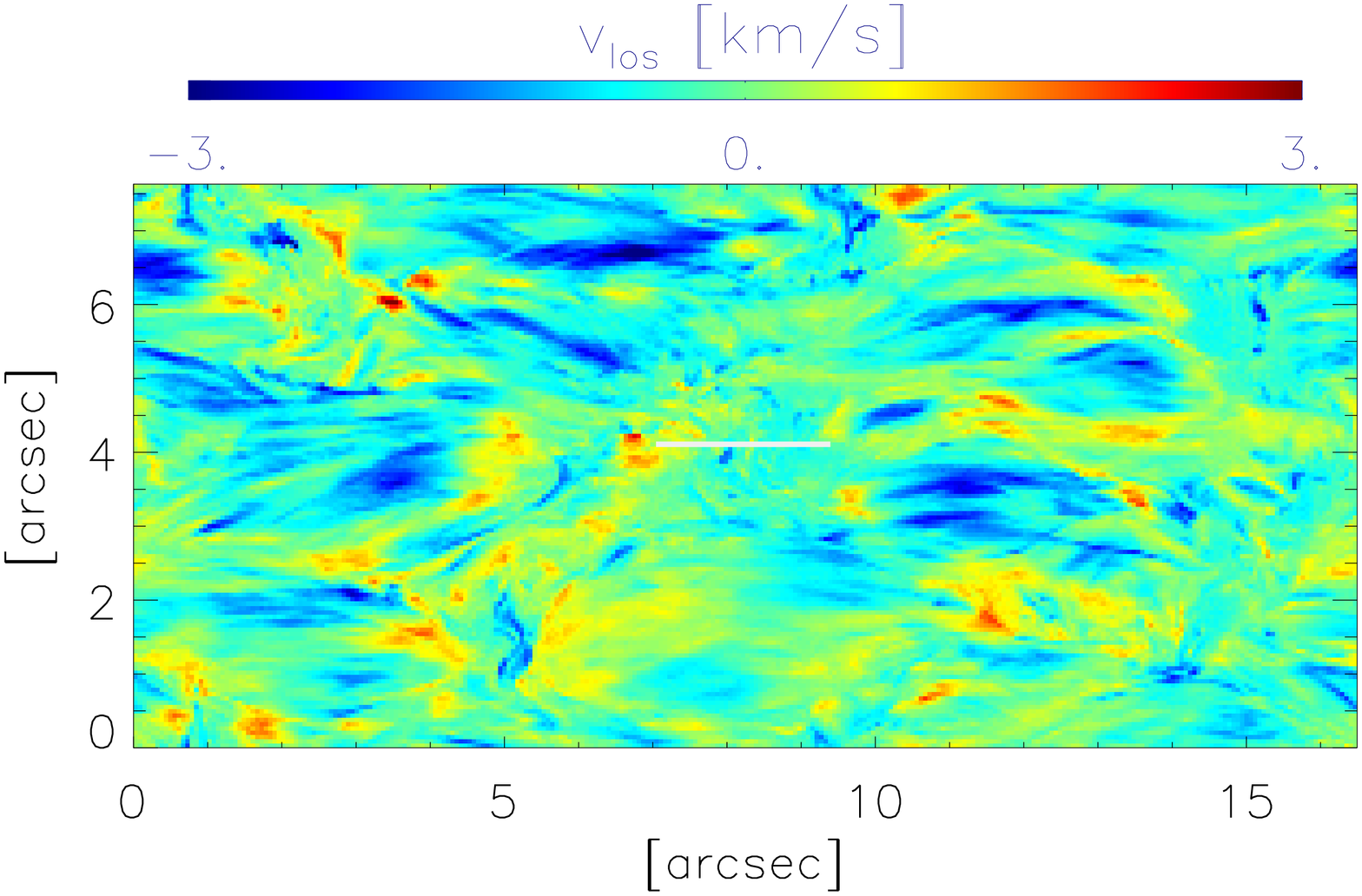}
    \includegraphics[angle=90,width=0.30\linewidth ,trim=3.5cm 0cm 0.2cm 2.5cm,clip=true]{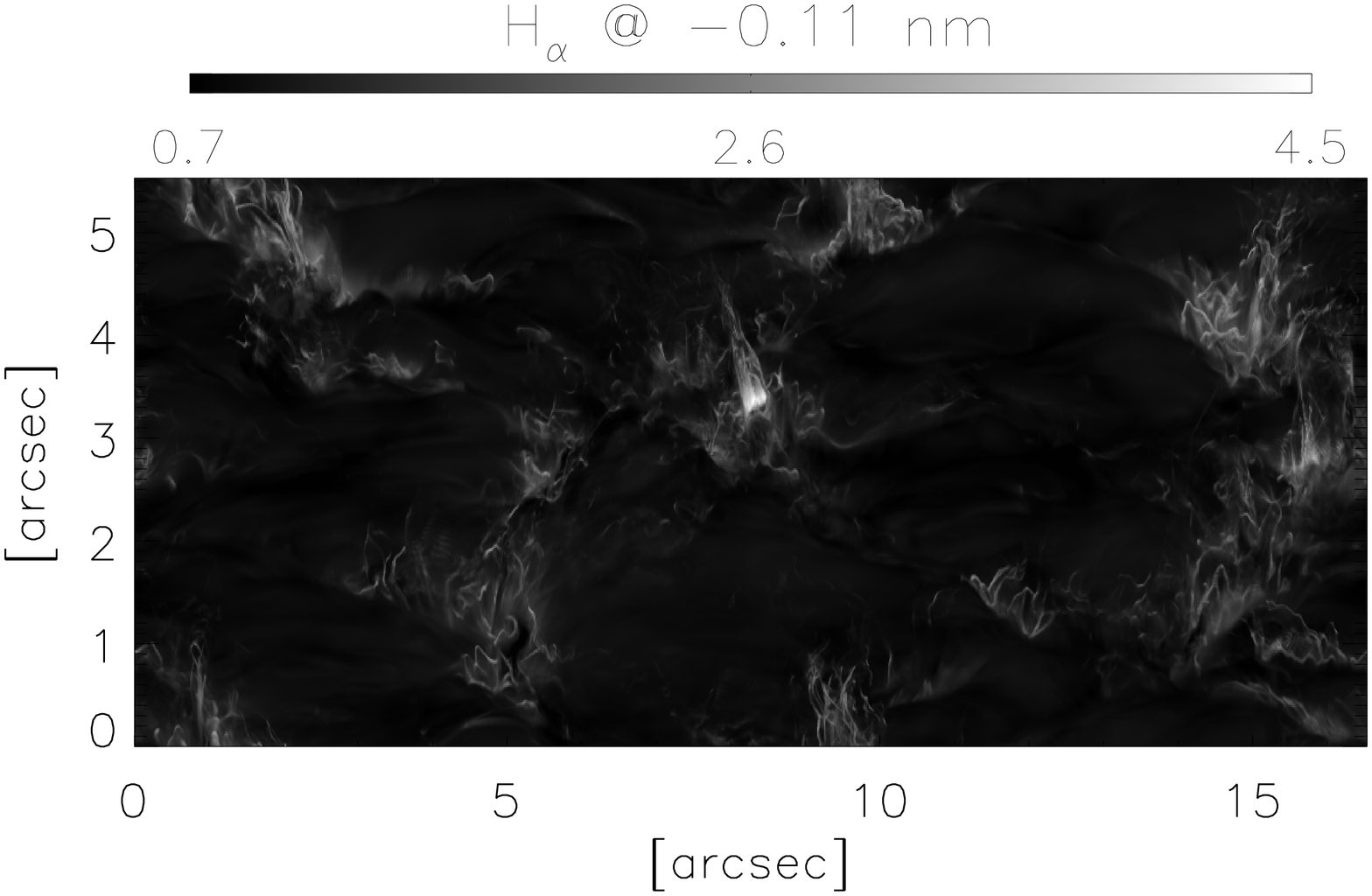}   \includegraphics[angle=90,width=0.327\linewidth ,trim=3.5cm 0cm 3.5cm 0cm,clip=true]{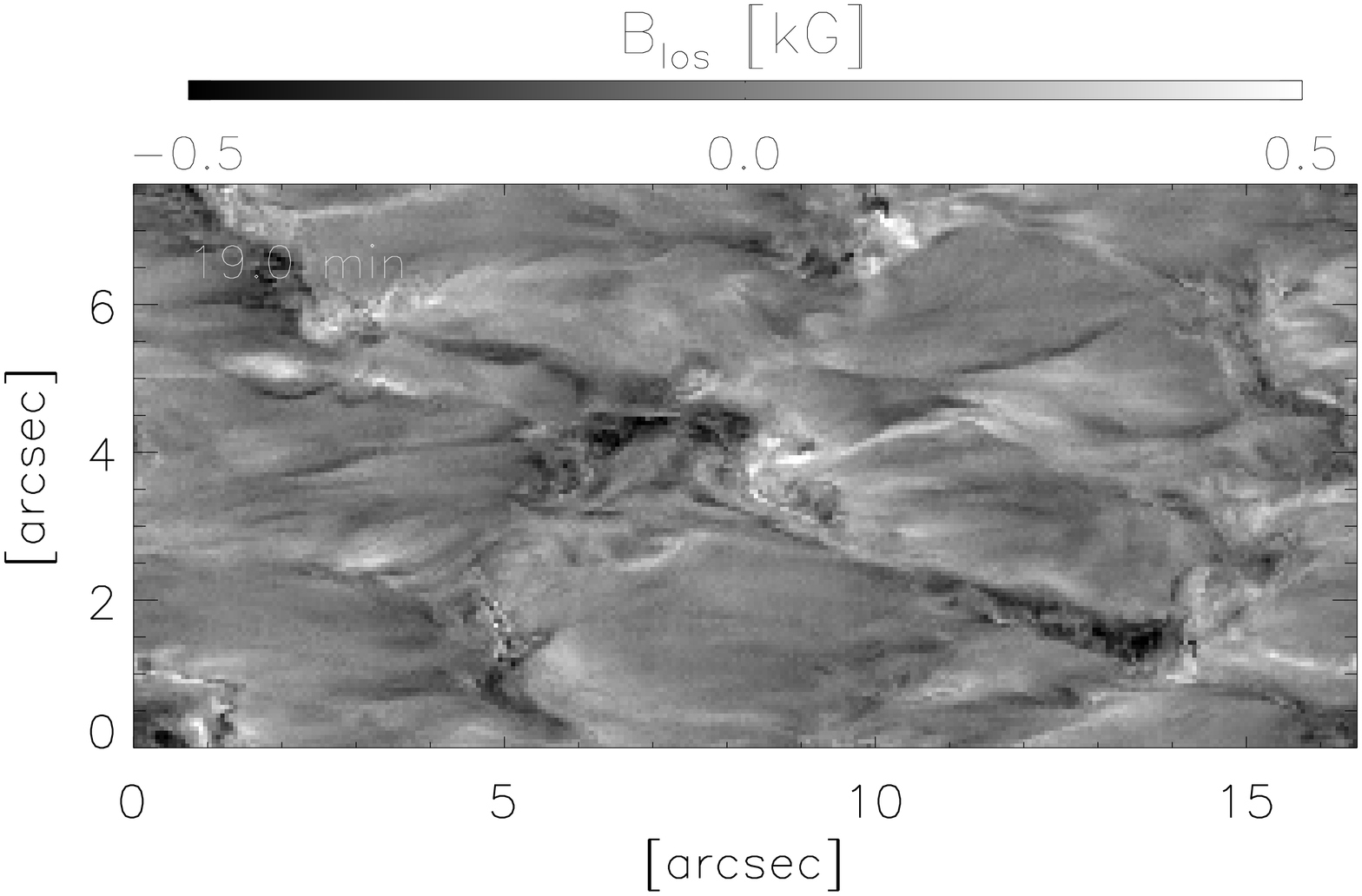}
     \includegraphics[angle=90,width=0.3\linewidth ,trim=3.5cm 0cm 3.5cm 2.5cm,clip=true]{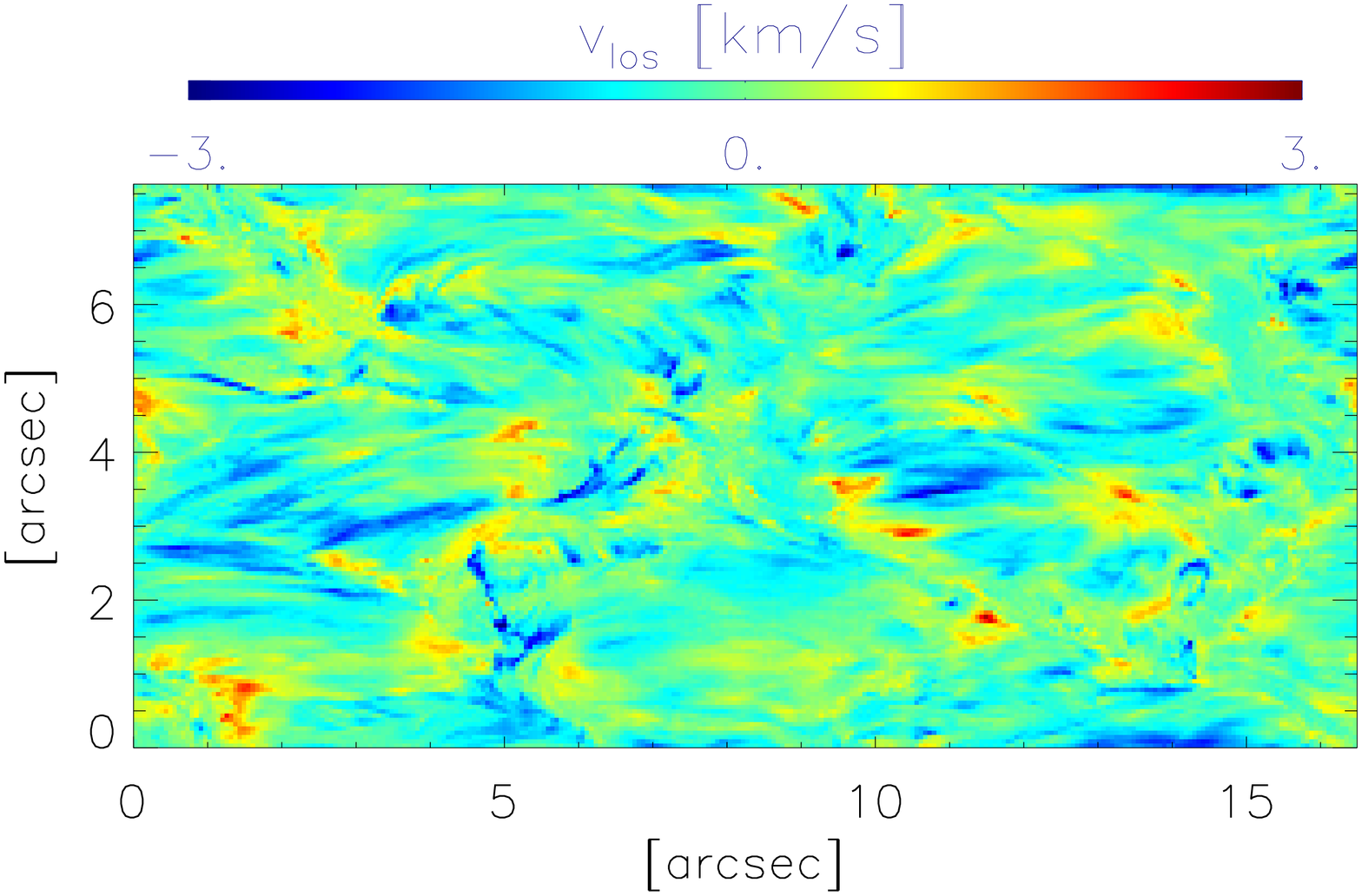}
    \includegraphics[angle=90,width=0.3\linewidth ,trim=3.5cm 0cm 3.5cm 2.5cm,clip=true]{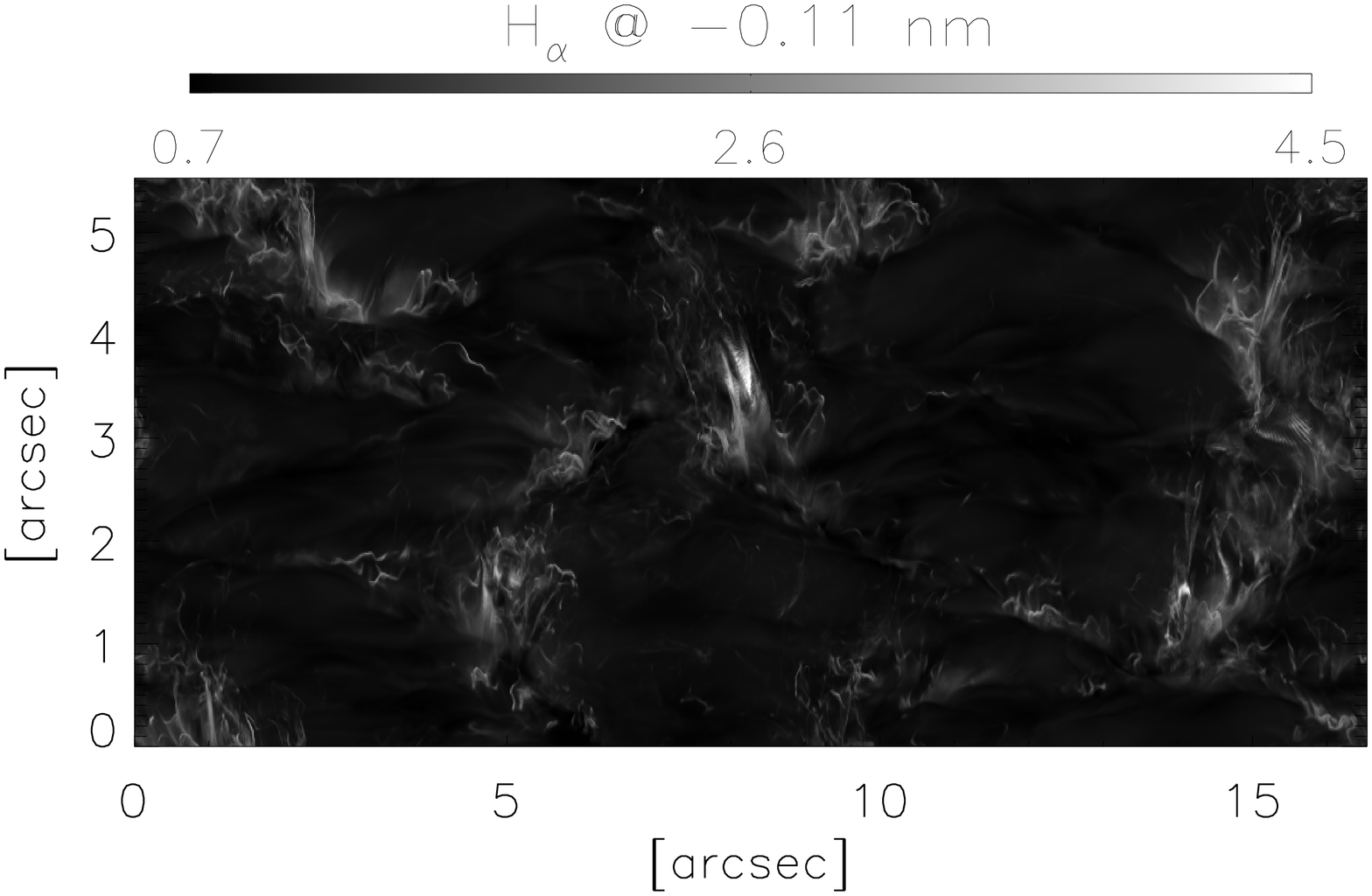}  
  \includegraphics[angle=90,width=0.327\linewidth ,trim=3.5cm 0cm 3.5cm 0cm,clip=true]{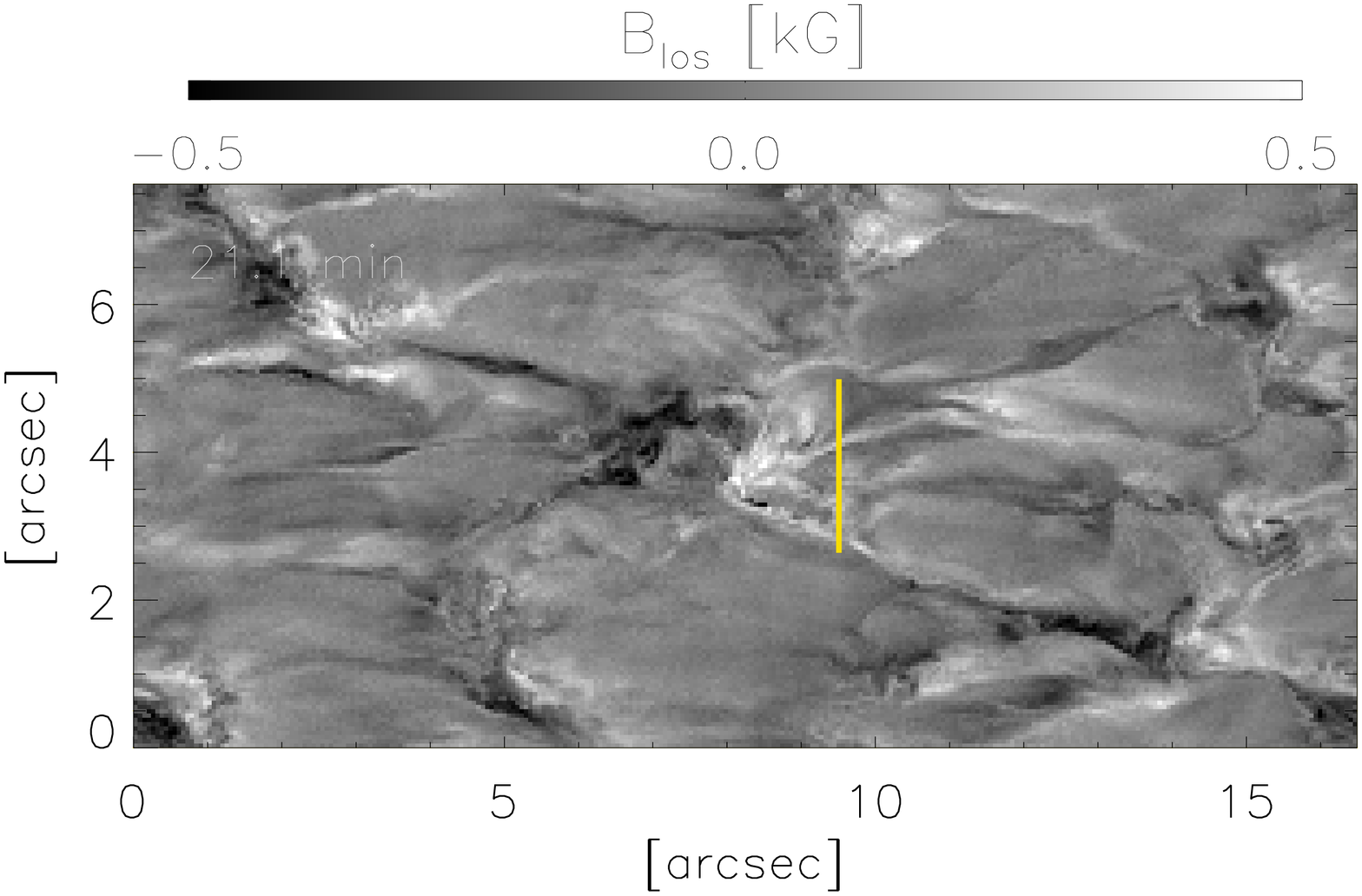}
     \includegraphics[angle=90,width=0.3\linewidth ,trim=3.5cm 0cm 3.5cm 2.5cm,clip=true]{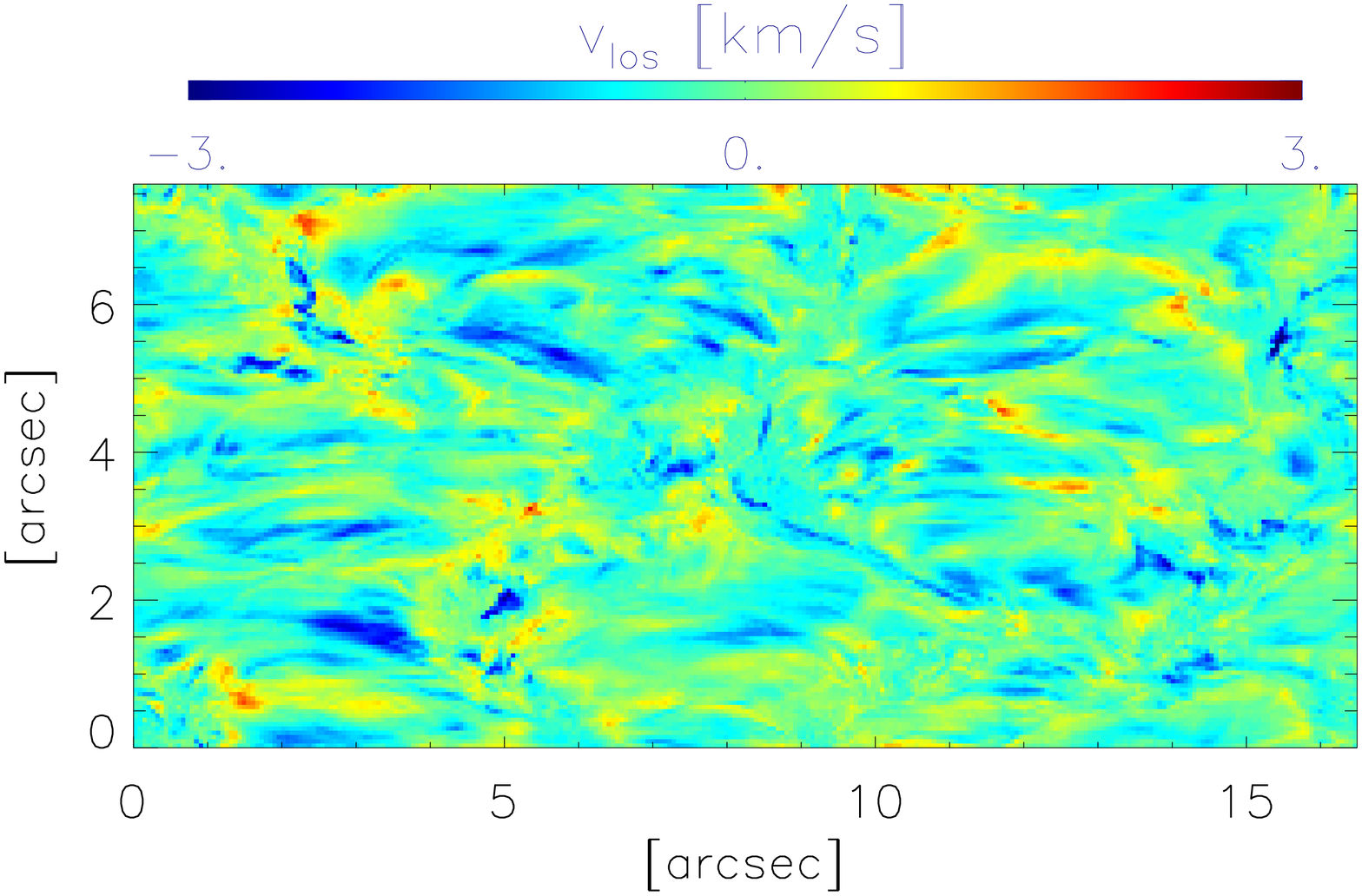}
    \includegraphics[angle=90,width=0.3\linewidth ,trim=3.5cm 0cm 3.5cm 2.5cm,clip=true]{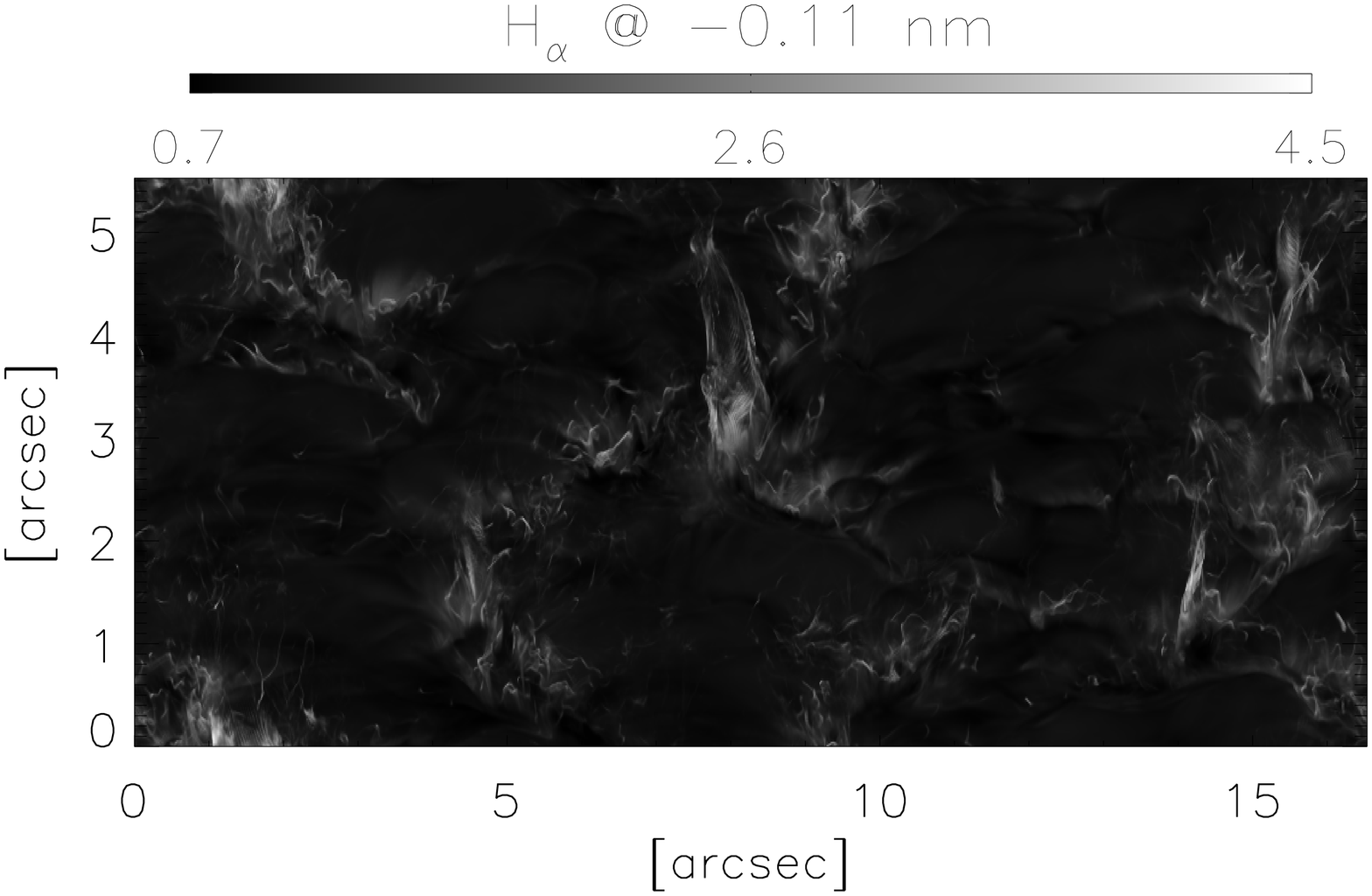} 
   \includegraphics[angle=90,width=0.327\linewidth ,trim=3.5cm 0cm 3.5cm 0cm,clip=true]{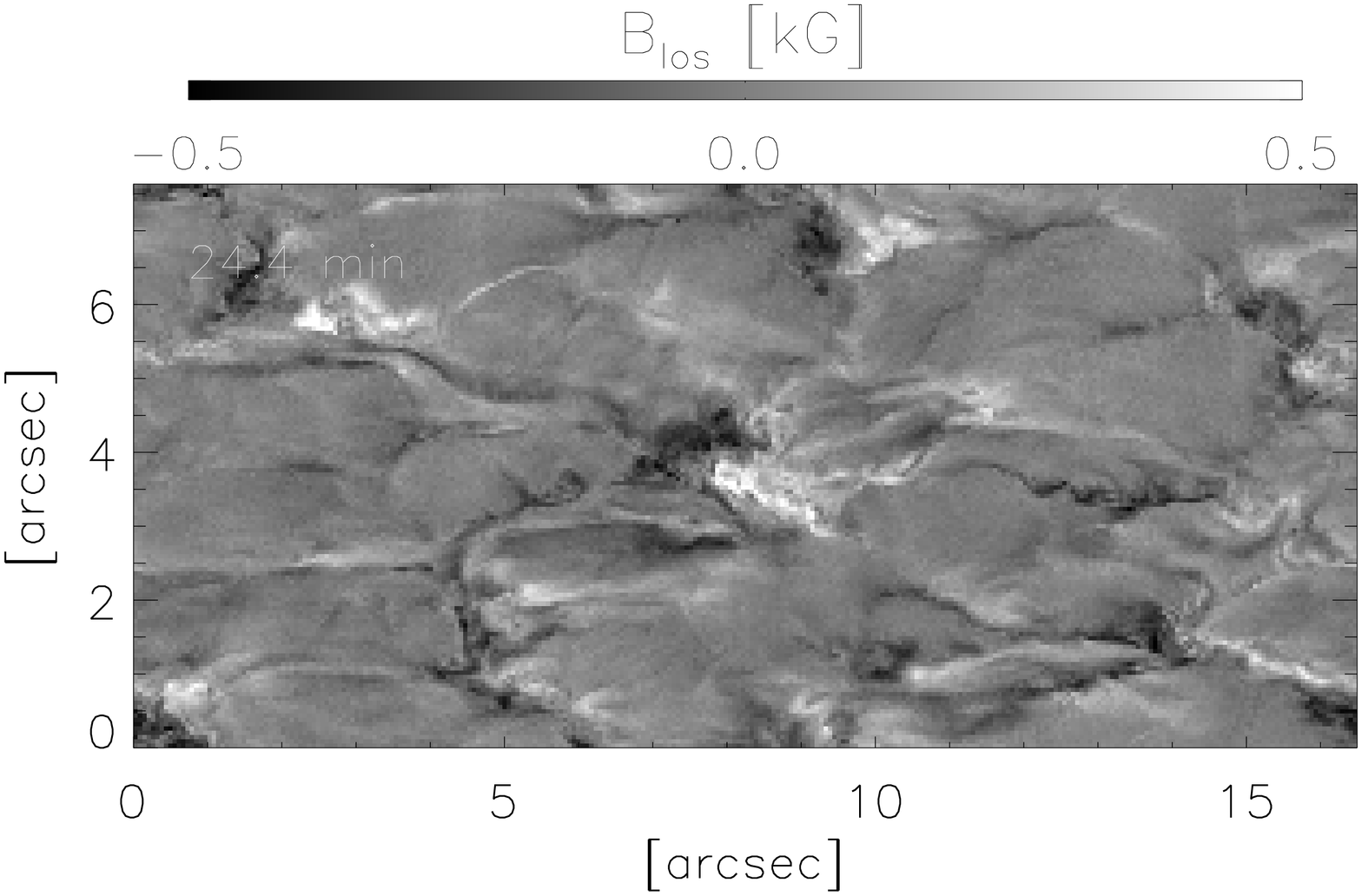}
     \includegraphics[angle=90,width=0.3\linewidth ,trim=3.5cm 0cm 3.5cm 2.5cm,clip=true]{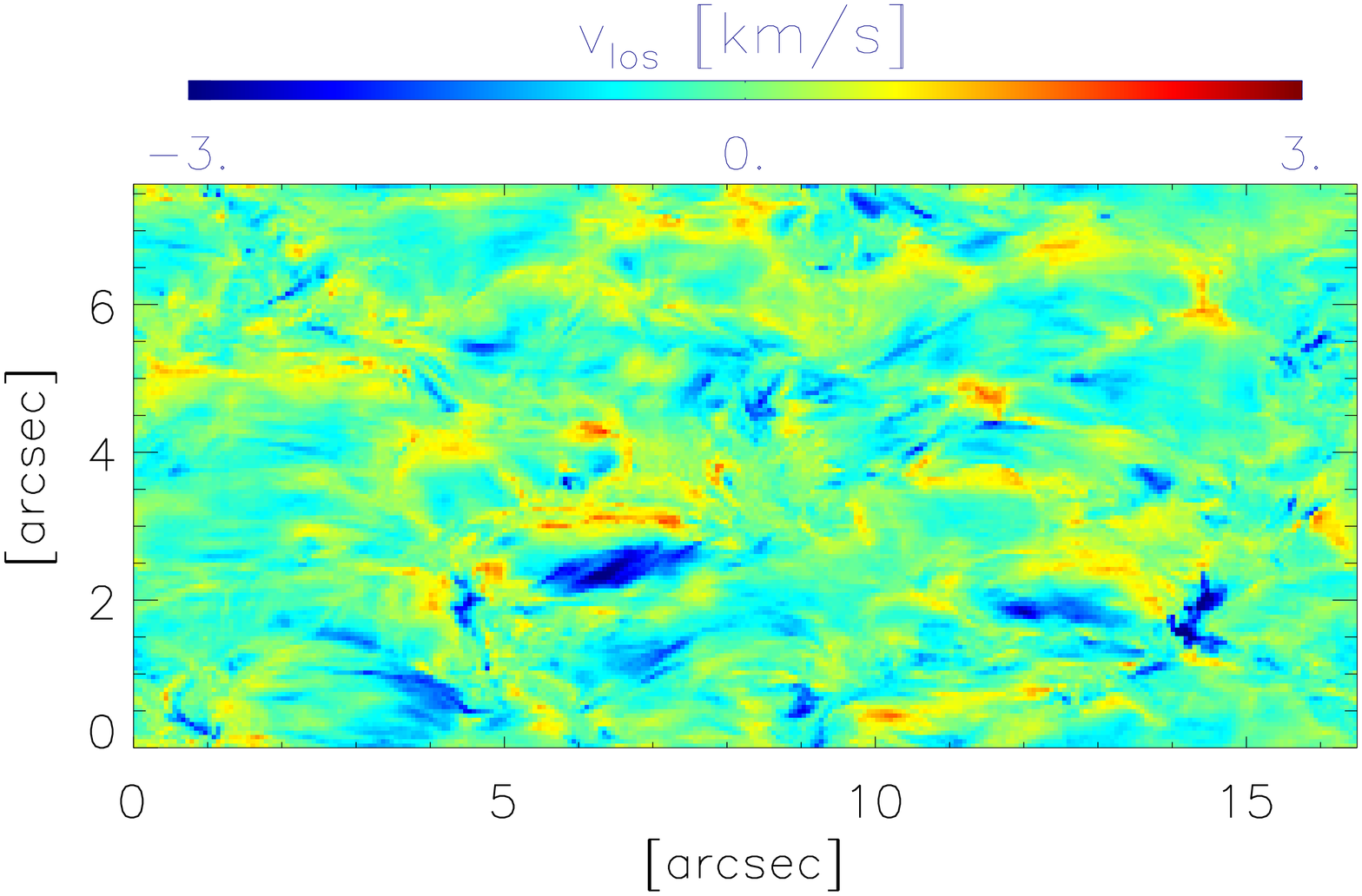}
    \includegraphics[angle=90,width=0.3\linewidth ,trim=3.5cm 0cm 3.5cm 2.5cm,clip=true]{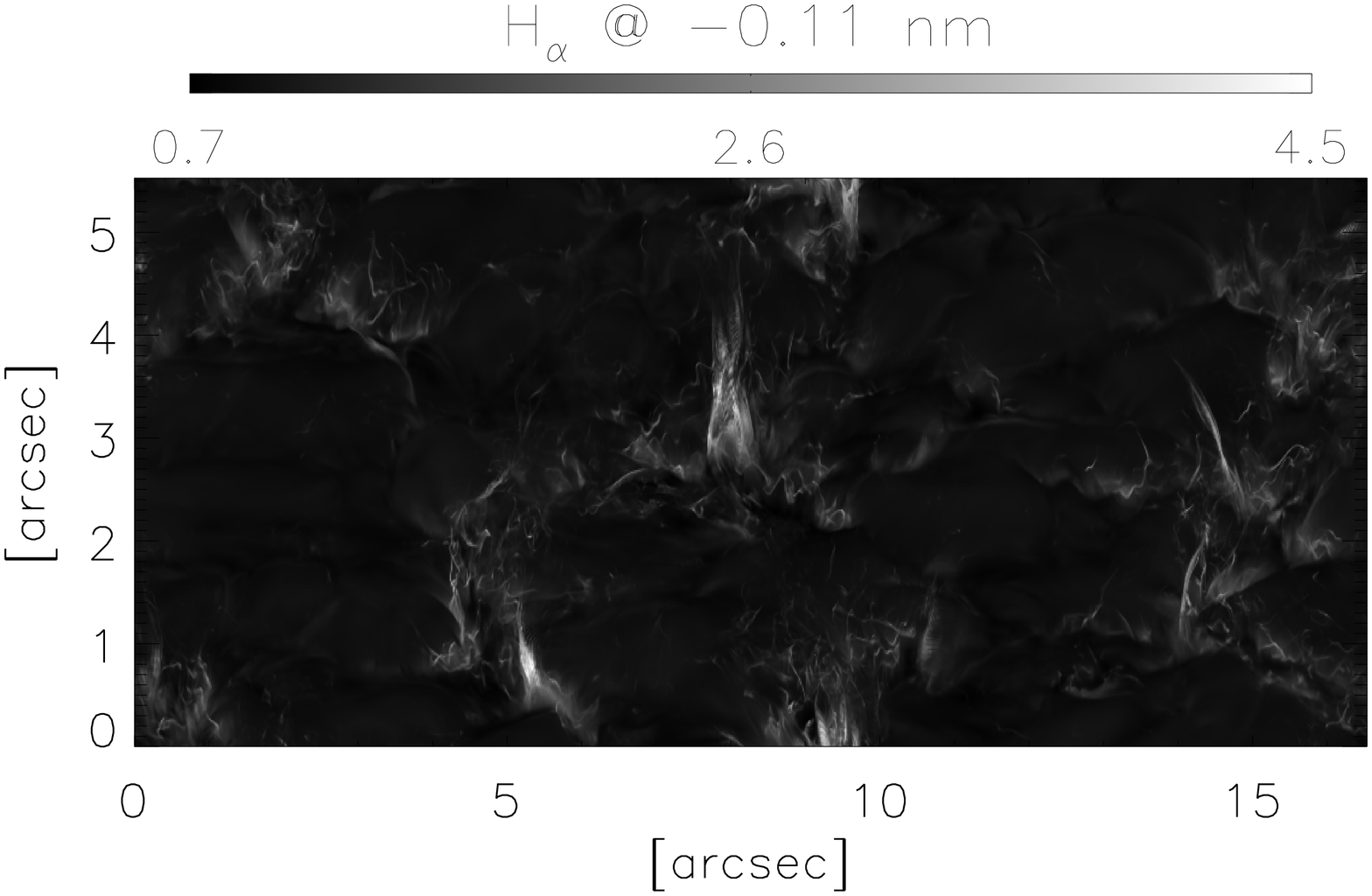} 
   \includegraphics[angle=90,width=0.327\linewidth ,trim=0cm 0cm 3.5cm 0cm,clip=true]{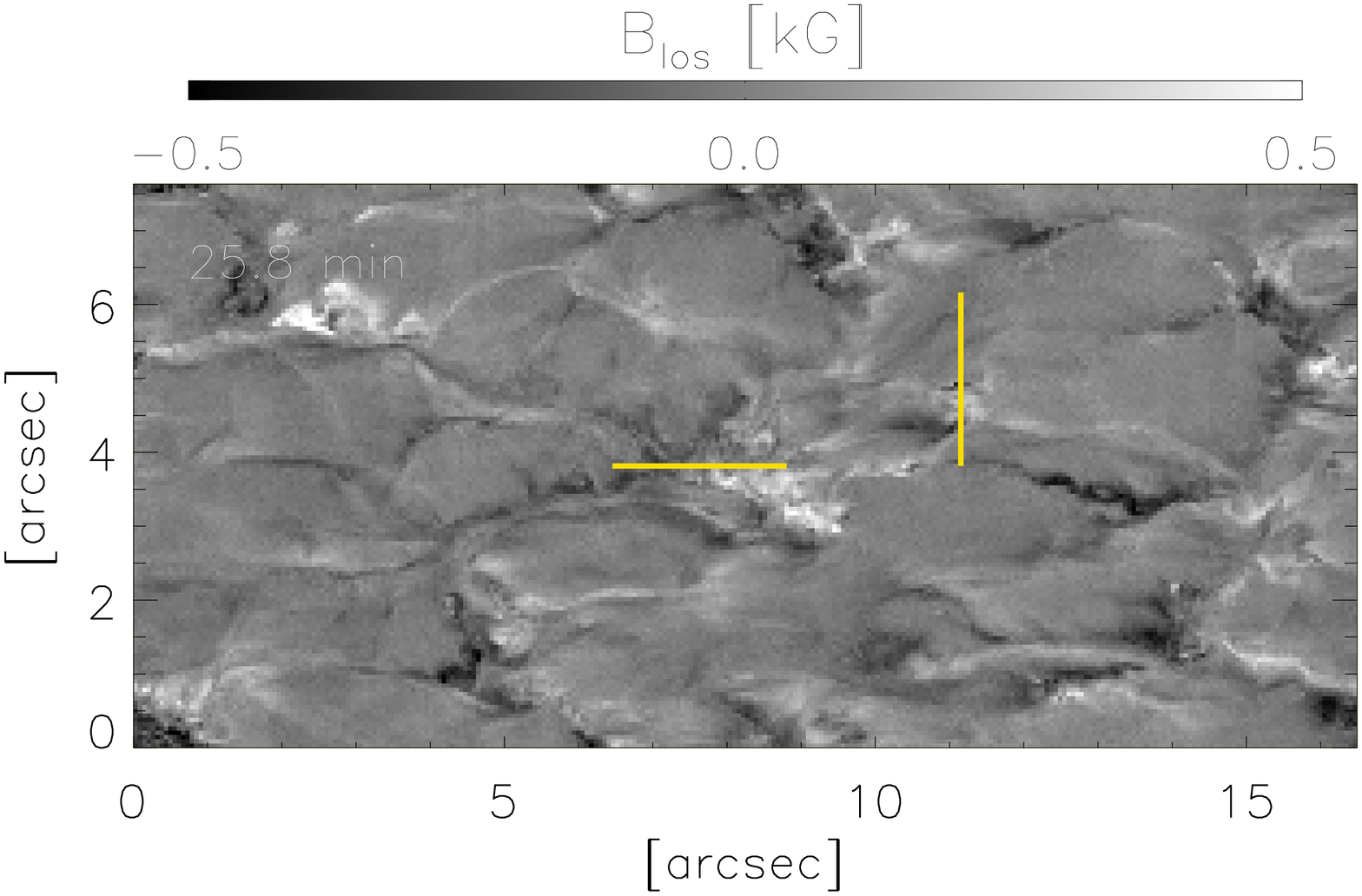}
   \includegraphics[angle=90,width=0.3\linewidth ,trim=0cm 0cm 3.5cm 2.5cm,clip=true]{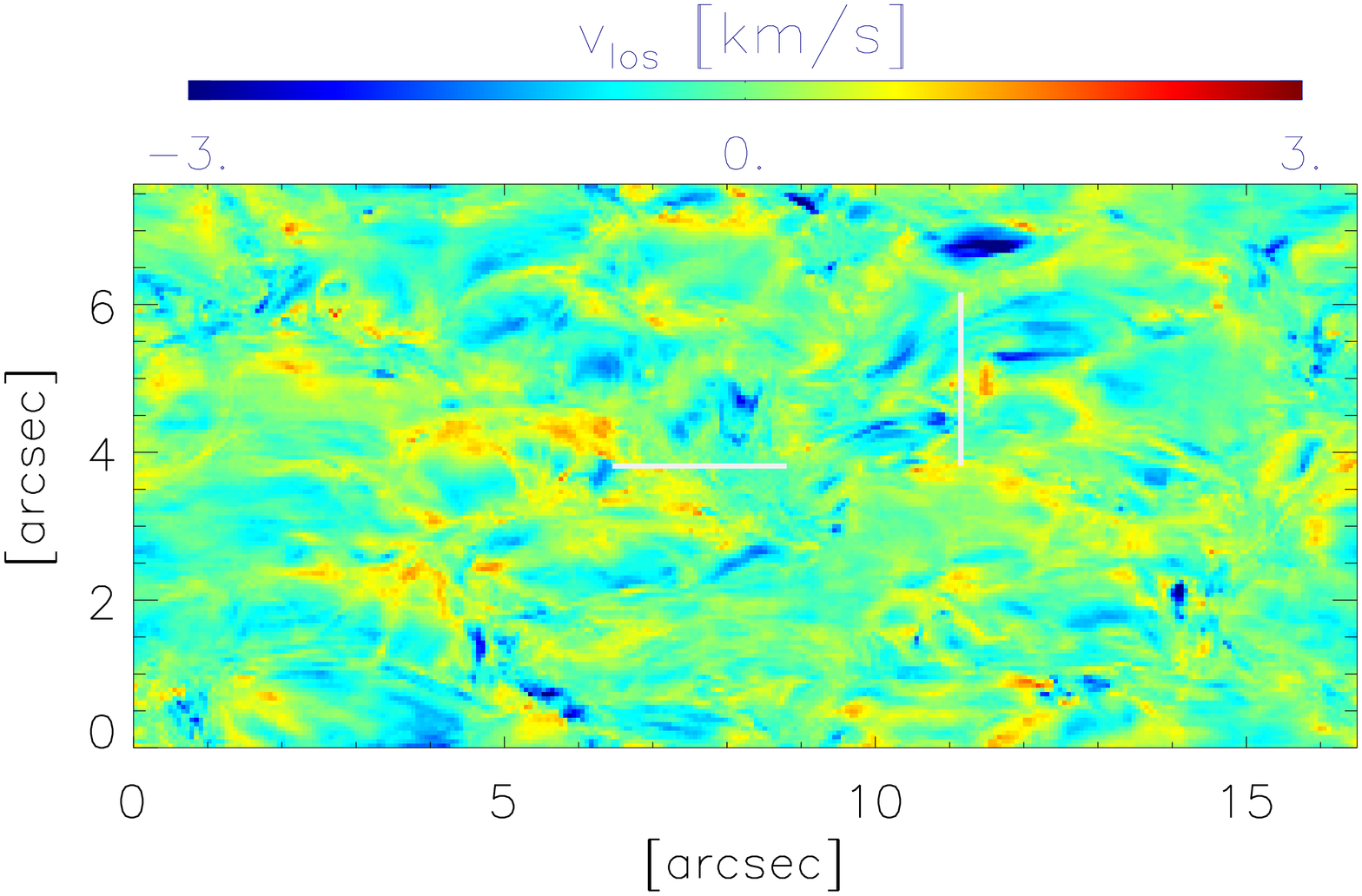}
   \includegraphics[angle=90,width=0.3\linewidth ,trim=0cm 0cm 3.5cm 2.5cm,clip=true]{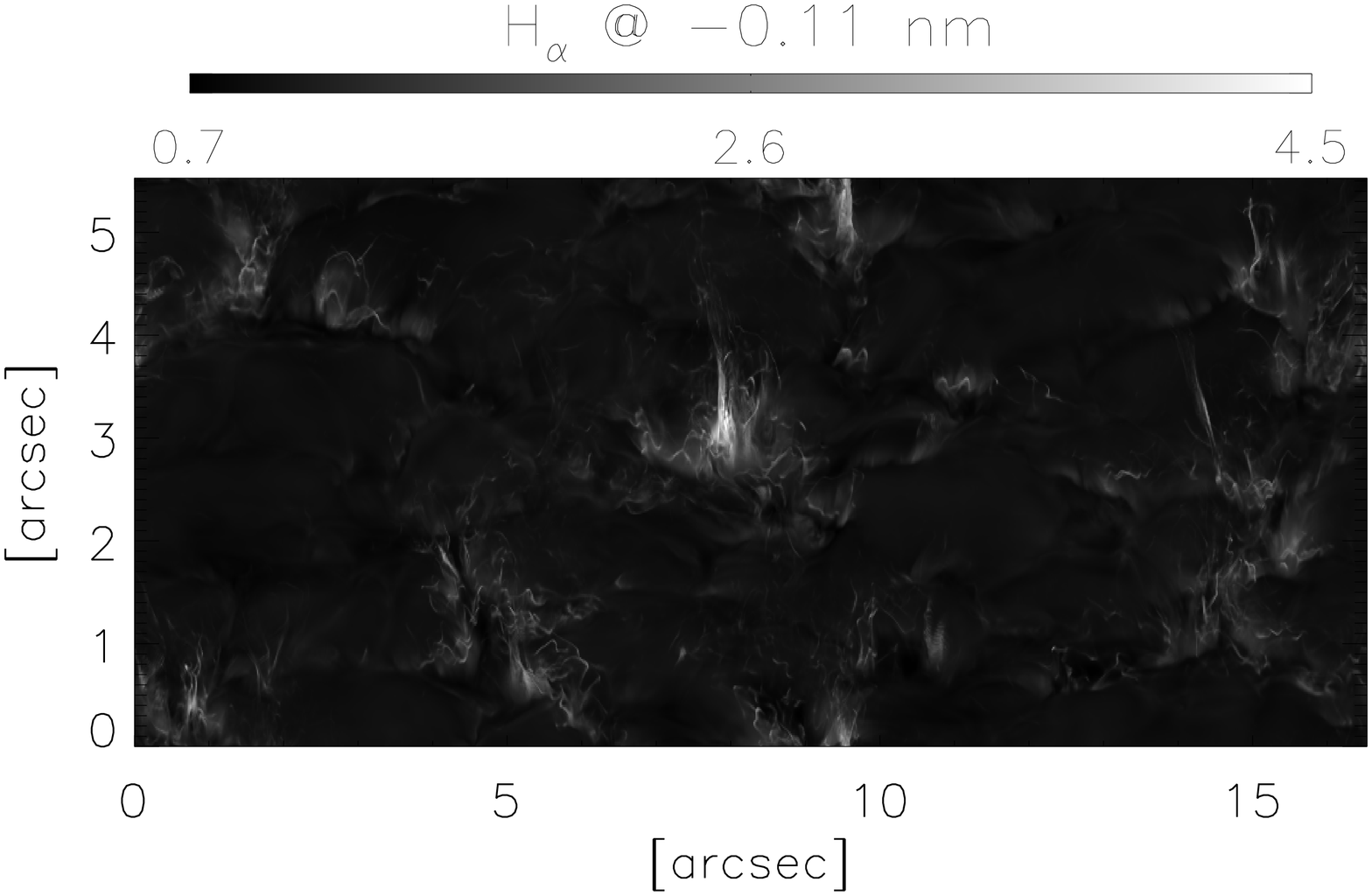}
   \caption{Simulated Sunrise/IMaX observations - Line-of-sight magnetic field strength and line-of-sight velocity (upflows are coloured blue) obtained from inversions of the simulations at five different times during the evolution of the simulated EB-like event. Vertical/horizontal lines mark the position of the cuts shown in Fig.~\ref{sim_cut_vert}/~\ref{sim_cut_hor}. The rightmost column shows the H$\alpha$-wing images synthesized from the same snapshots. Bright flame-like features mark positions of the simulated EBs.}
\label{sim_inv_rest}
\end{figure*}

\begin{figure*}
  \centering
  \includegraphics[angle=90,width=0.327\linewidth ,trim=3.5cm 0cm 0.2cm 0cm,clip=true]{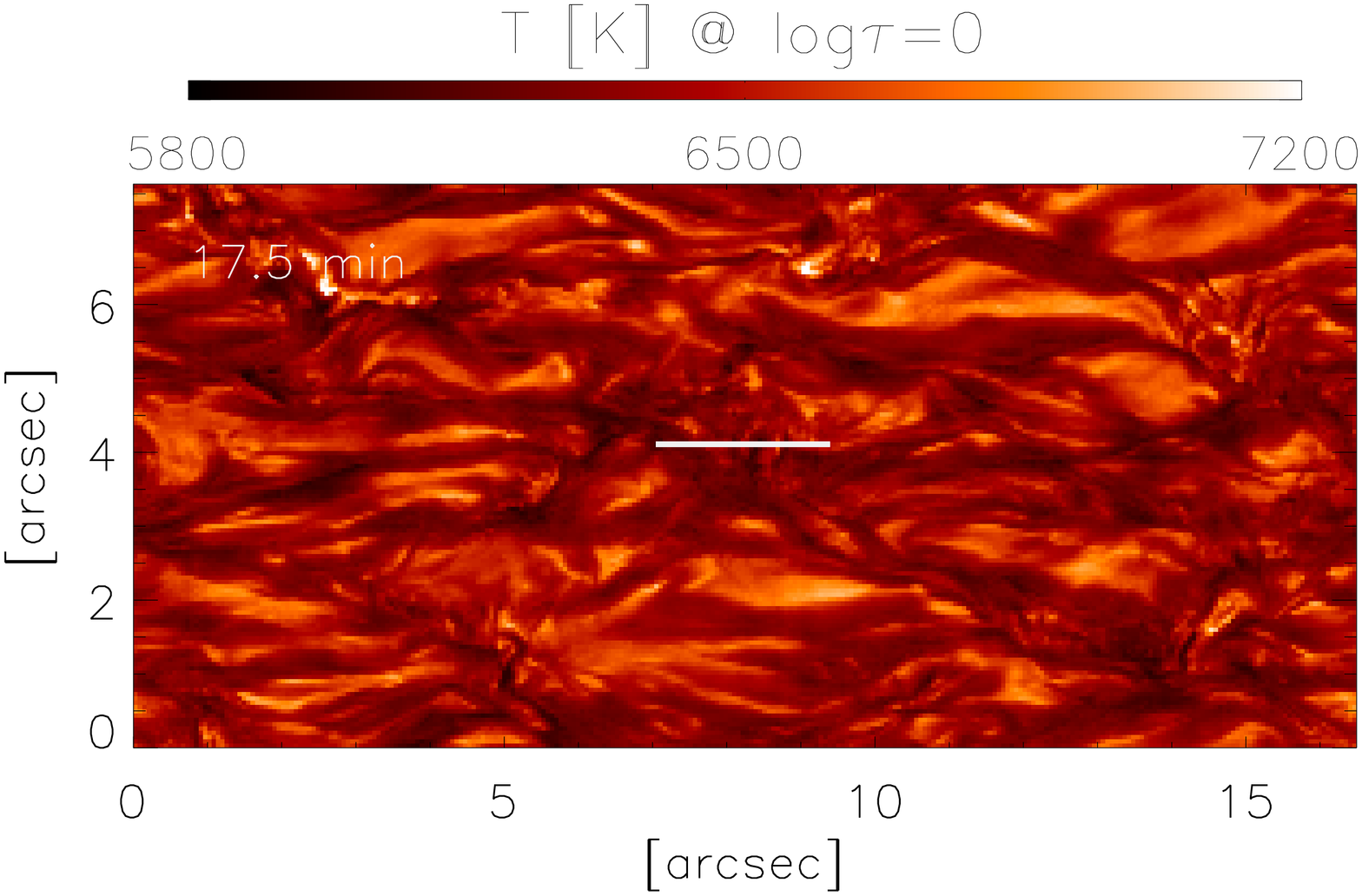} 
   \includegraphics[angle=90,width=0.30\linewidth ,trim=3.5cm 0cm 0.2cm 2.5cm,clip=true]{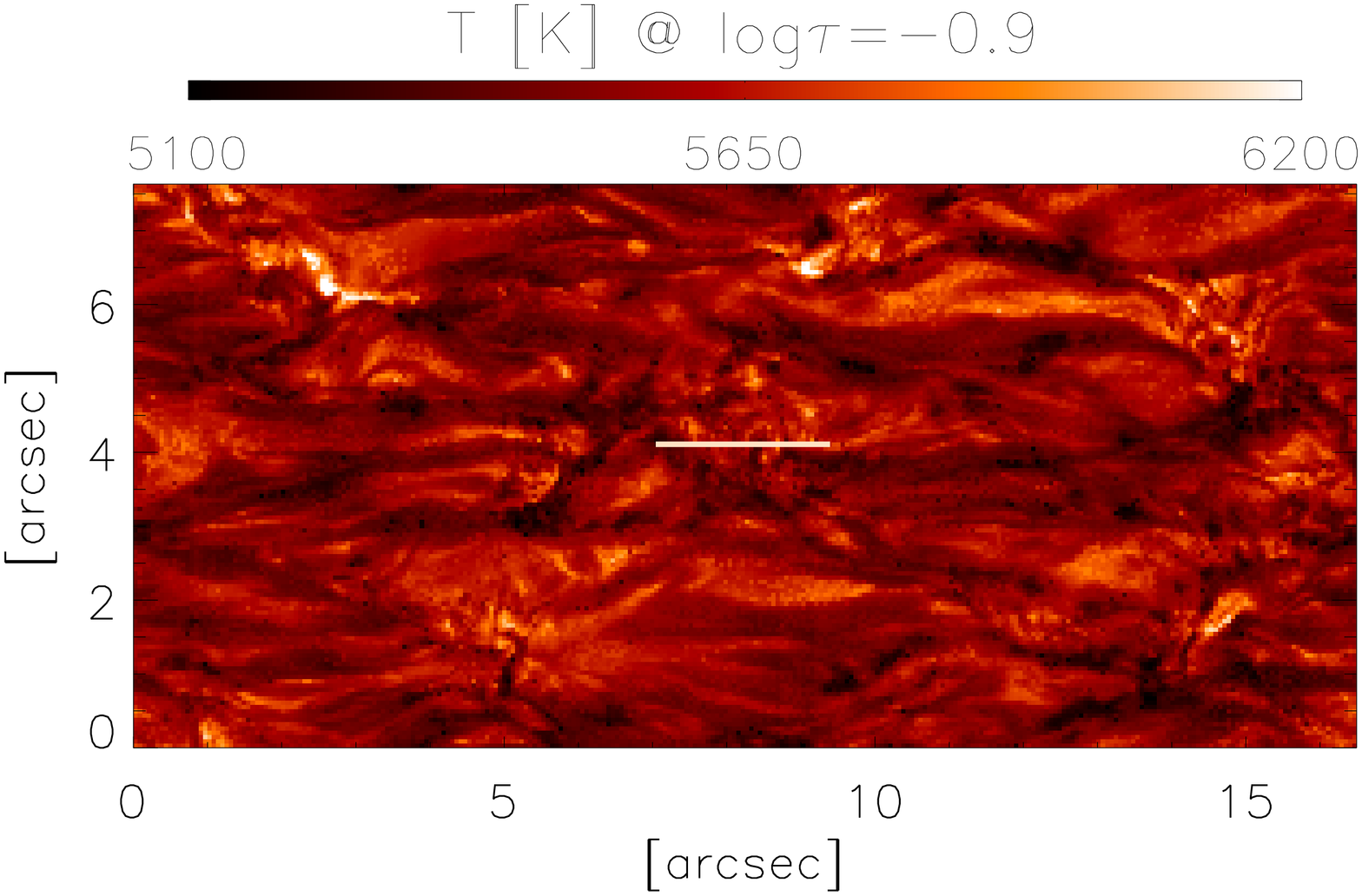}
    \includegraphics[angle=90,width=0.30\linewidth ,trim=3.5cm 0cm 0.2cm 2.5cm,clip=true]{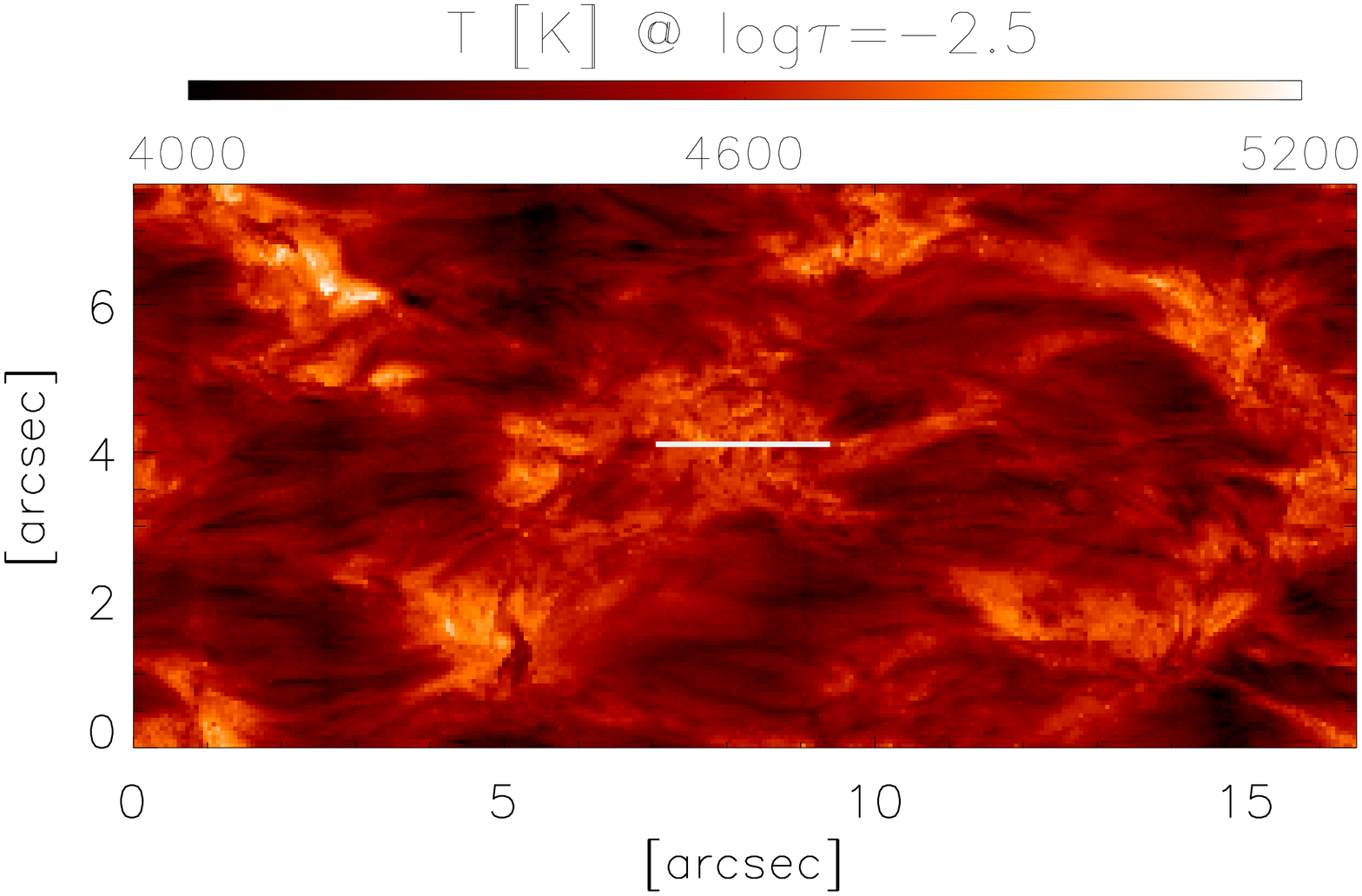}  
  \includegraphics[angle=90,width=0.327\linewidth ,trim=3.5cm 0cm 3.5cm 0cm,clip=true]{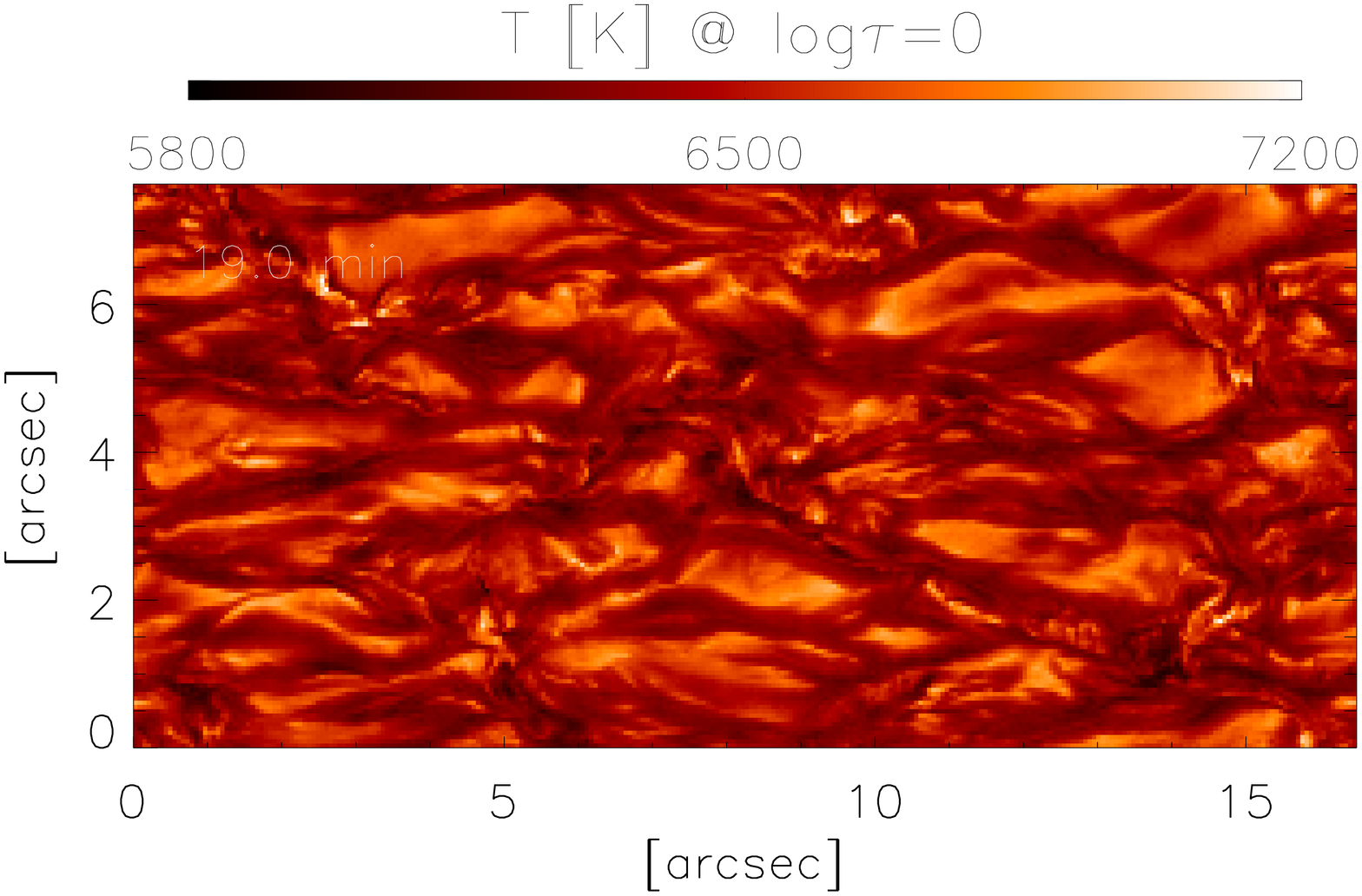}
     \includegraphics[angle=90,width=0.3\linewidth ,trim=3.5cm 0cm 3.5cm 2.5cm,clip=true]{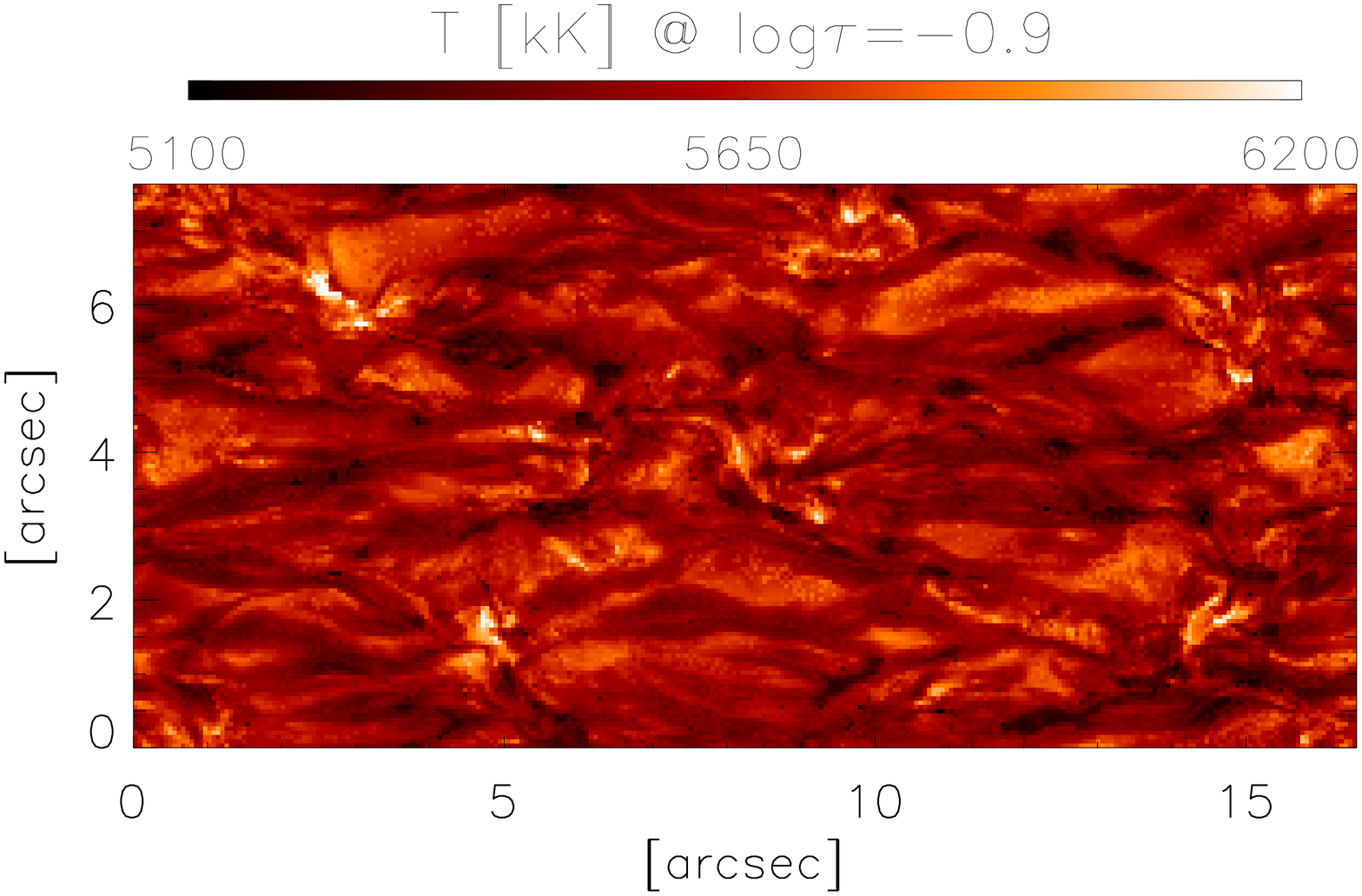}
    \includegraphics[angle=90,width=0.3\linewidth ,trim=3.5cm 0cm 3.5cm 2.5cm,clip=true]{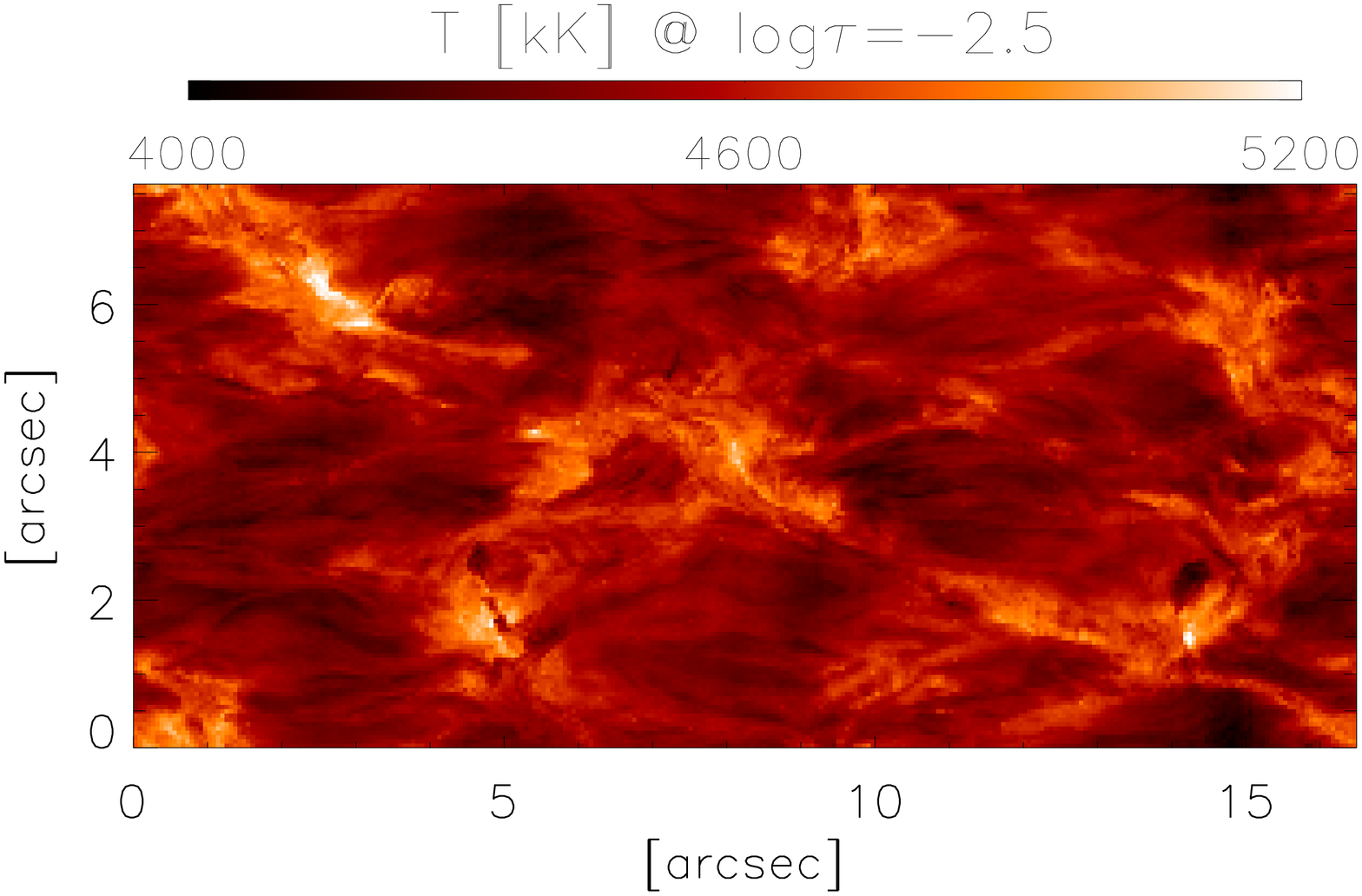} 
     \includegraphics[angle=90,width=0.327\linewidth ,trim=3.5cm 0cm 3.5cm 0cm,clip=true]{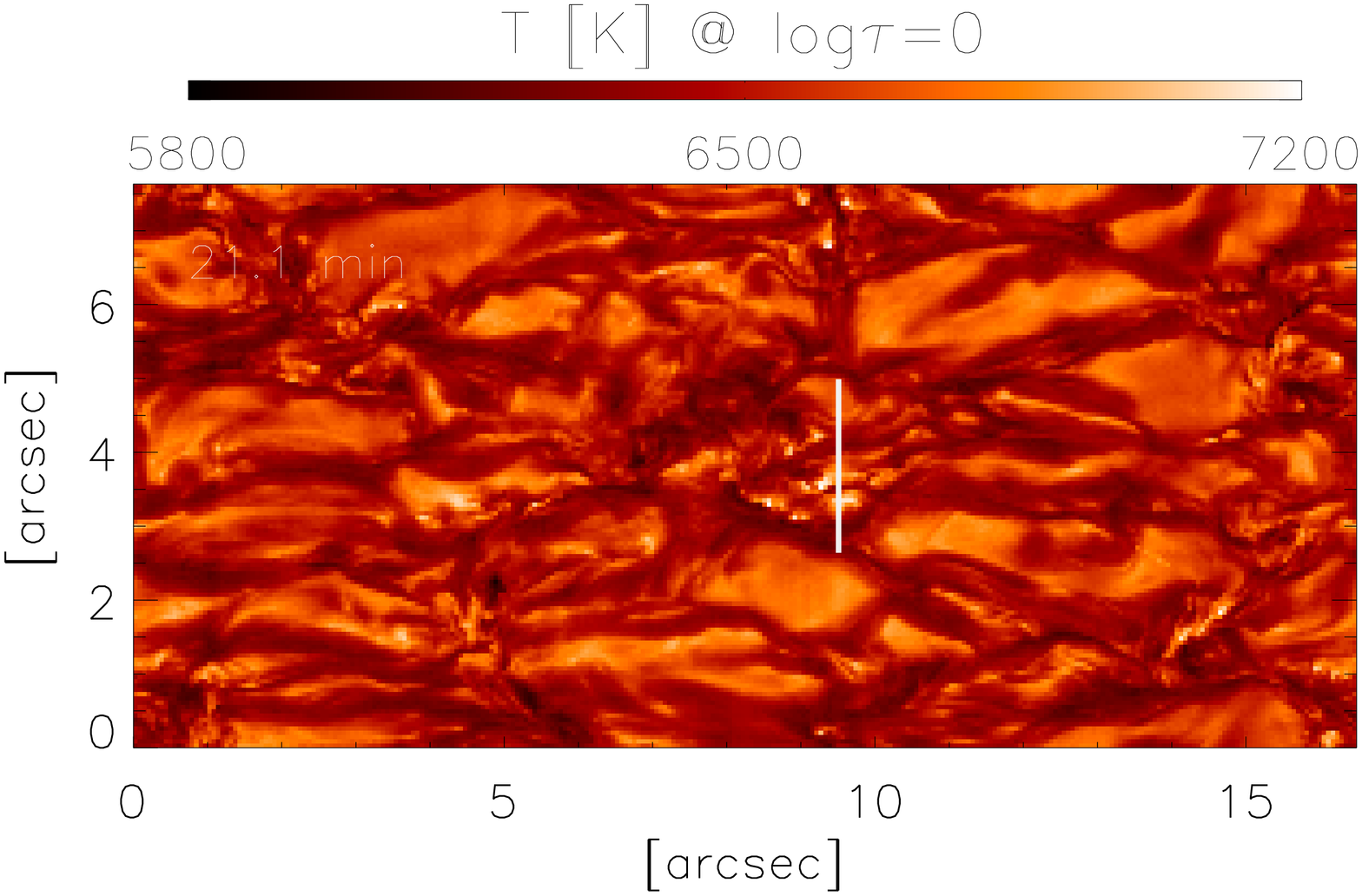}
     \includegraphics[angle=90,width=0.3\linewidth ,trim=3.5cm 0cm 3.5cm 2.5cm,clip=true]{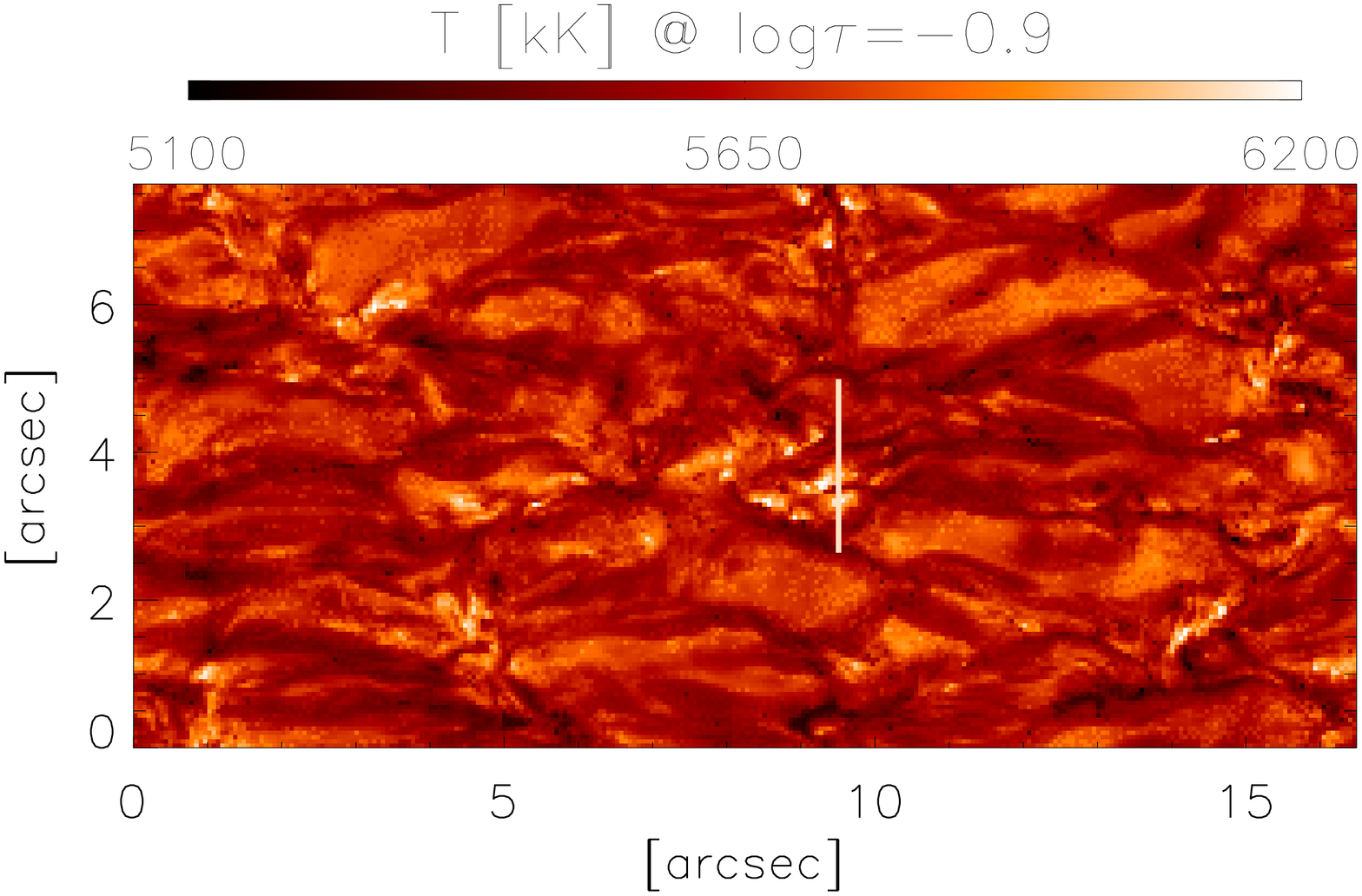}
    \includegraphics[angle=90,width=0.3\linewidth ,trim=3.5cm 0cm 3.5cm 2.5cm,clip=true]{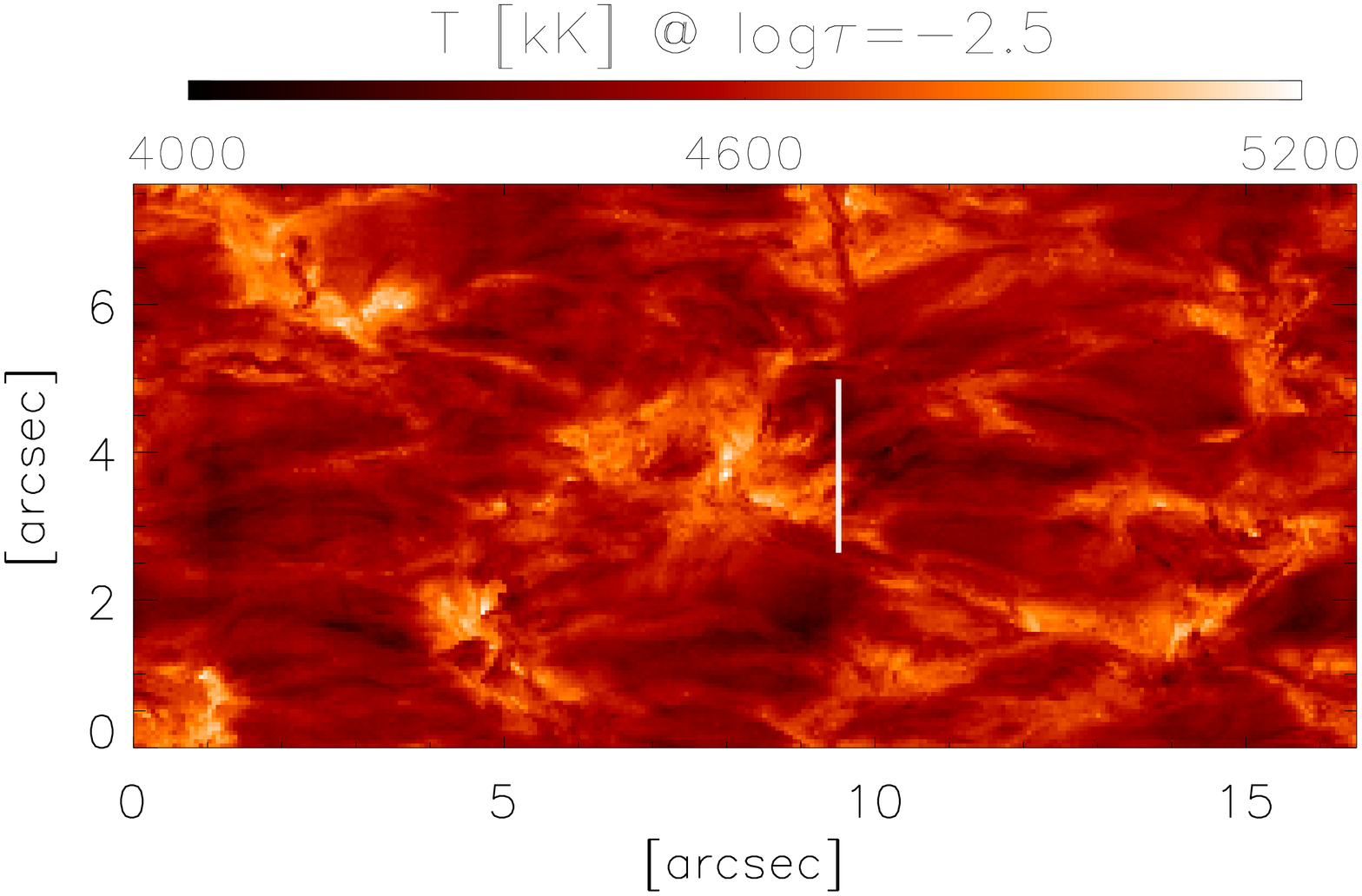} 
   \includegraphics[angle=90,width=0.327\linewidth ,trim=3.5cm 0cm 3.5cm 0cm,clip=true]{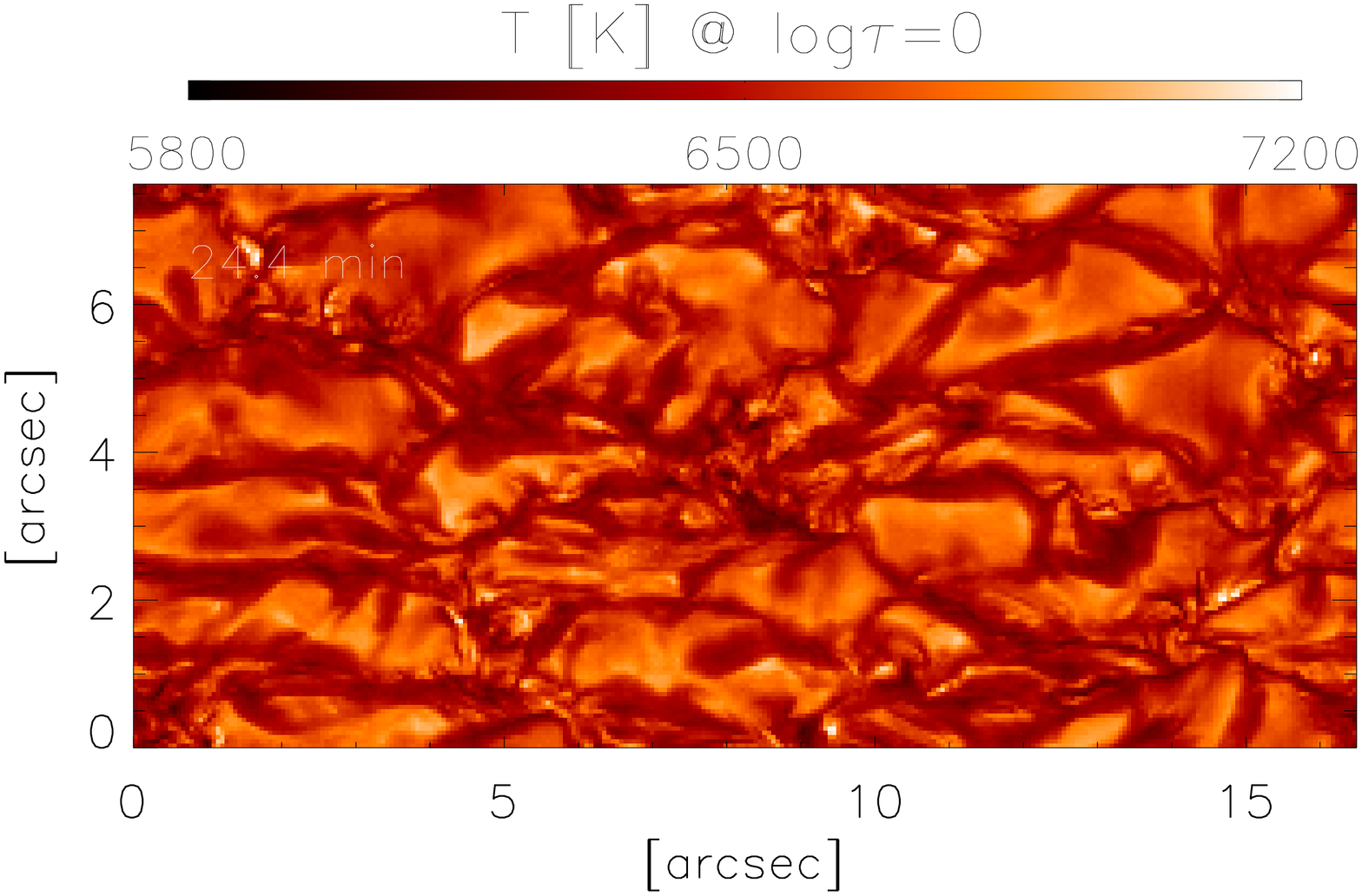}
     \includegraphics[angle=90,width=0.3\linewidth ,trim=3.5cm 0cm 3.5cm 2.5cm,clip=true]{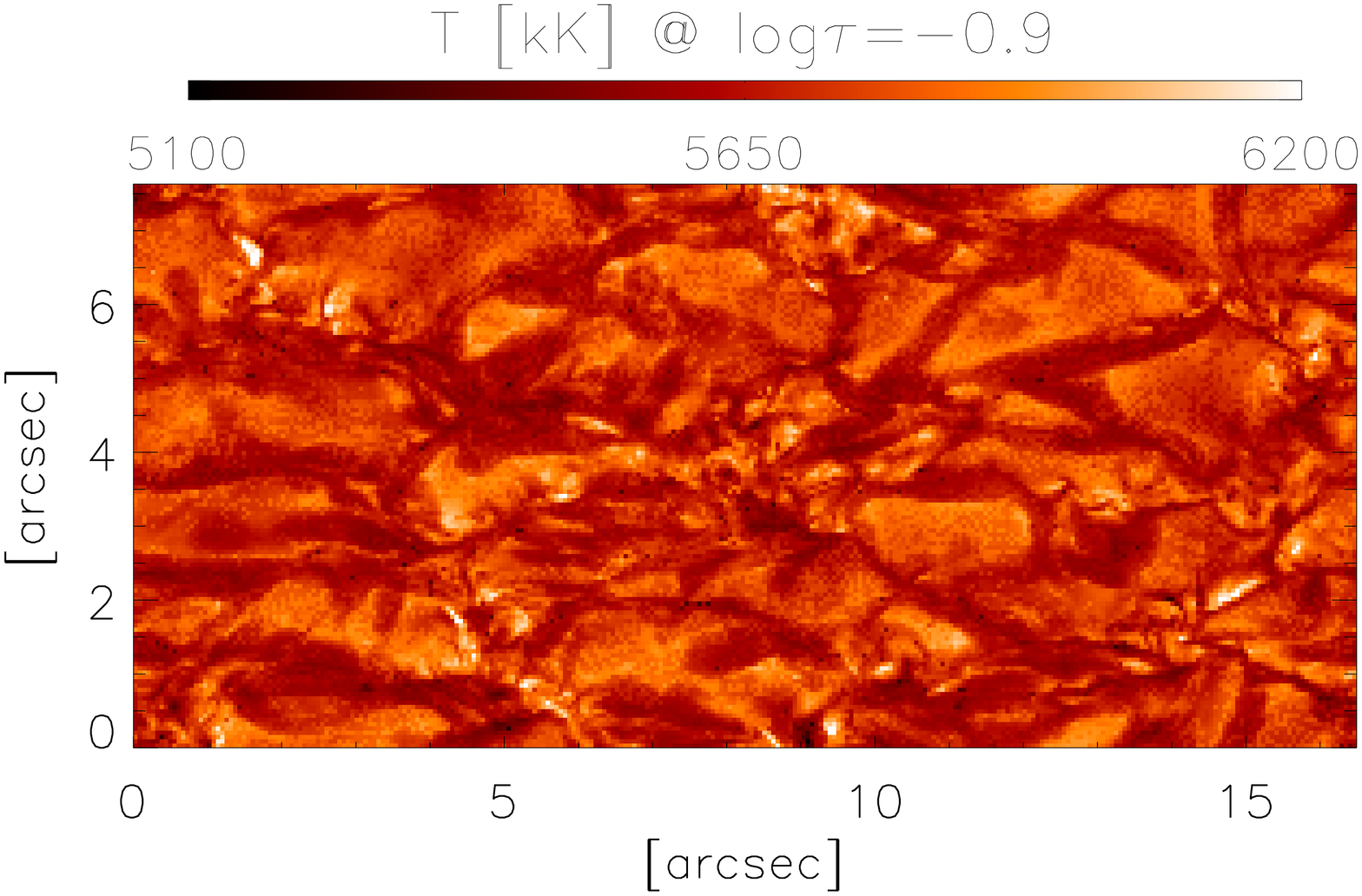}
    \includegraphics[angle=90,width=0.3\linewidth ,trim=3.5cm 0cm 3.5cm 2.5cm,clip=true]{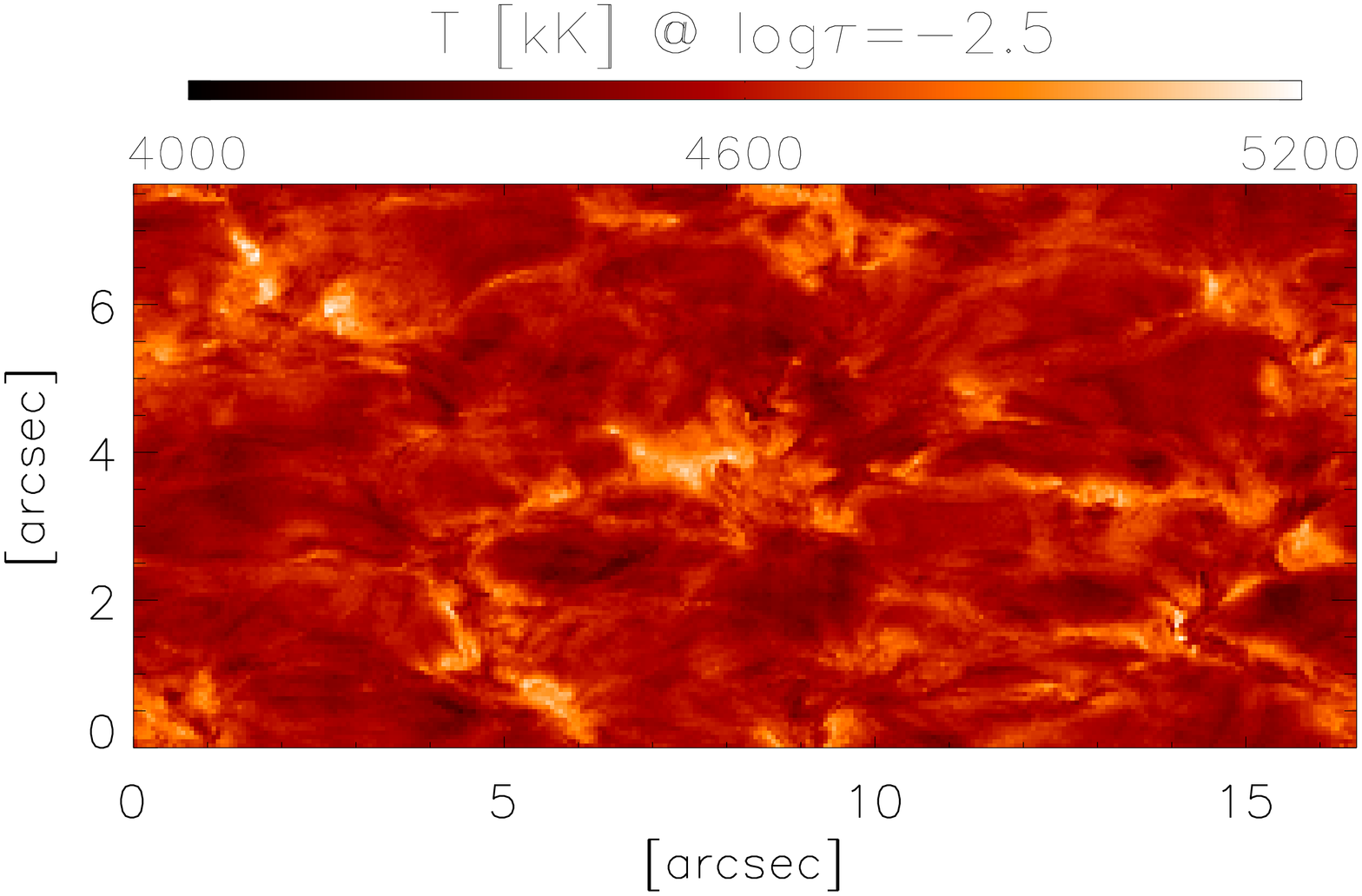} 
   \includegraphics[angle=90,width=0.327\linewidth ,trim=0cm 0cm 3.5cm 0cm,clip=true]{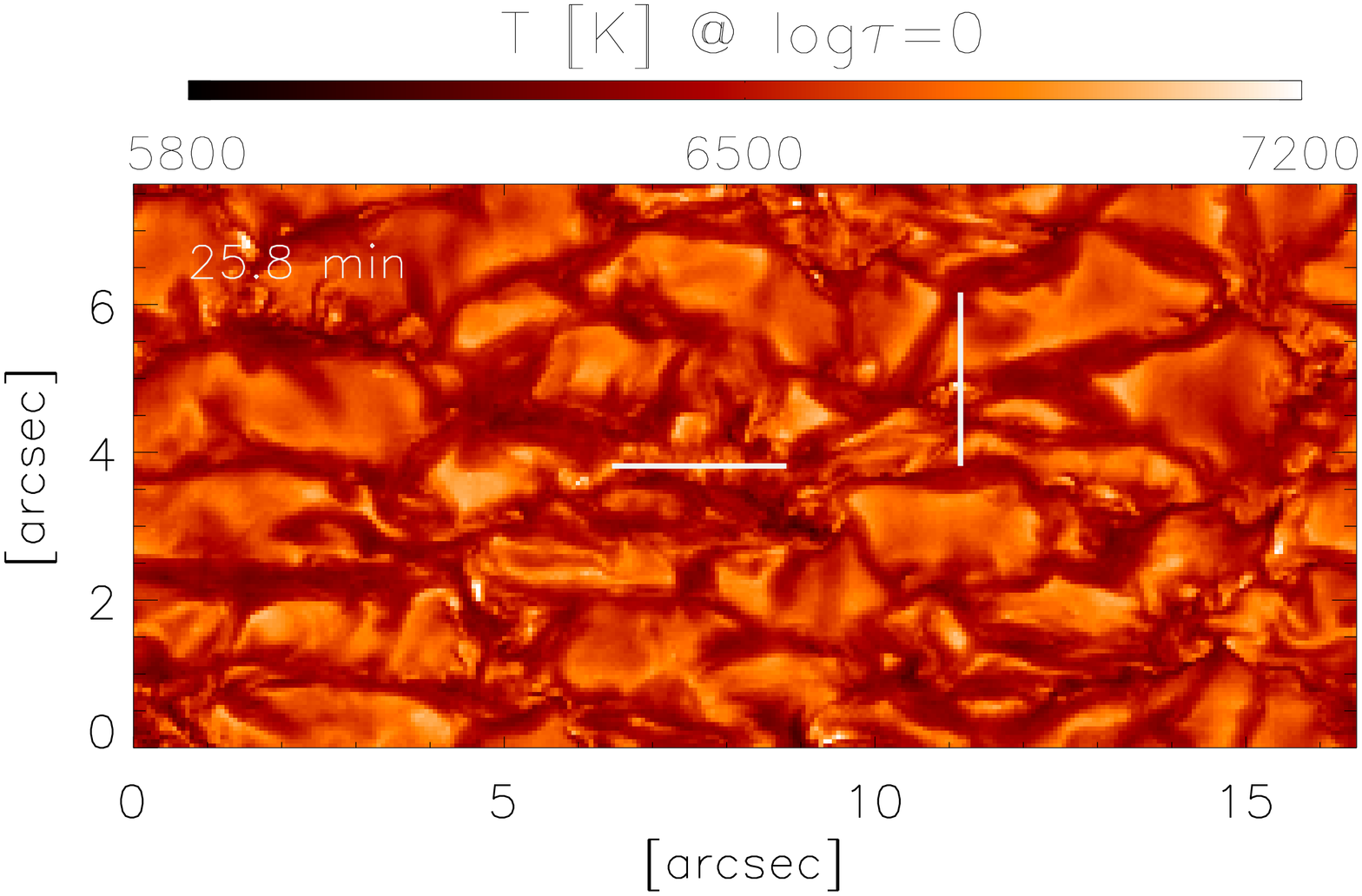}
   \includegraphics[angle=90,width=0.3\linewidth ,trim=0cm 0cm 3.5cm 2.5cm,clip=true]{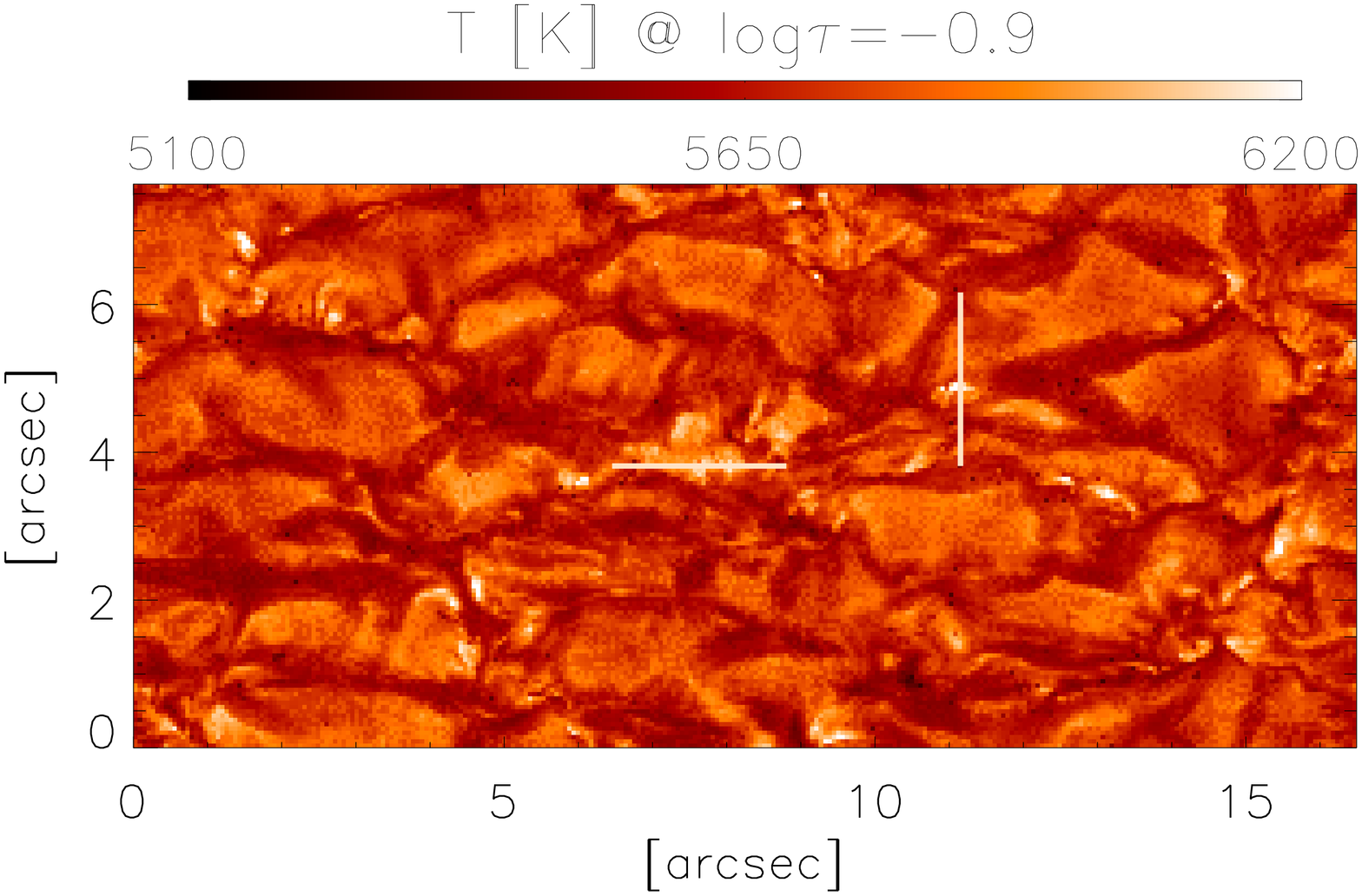}
   \includegraphics[angle=90,width=0.3\linewidth ,trim=0cm 0cm 3.5cm 2.5cm,clip=true]{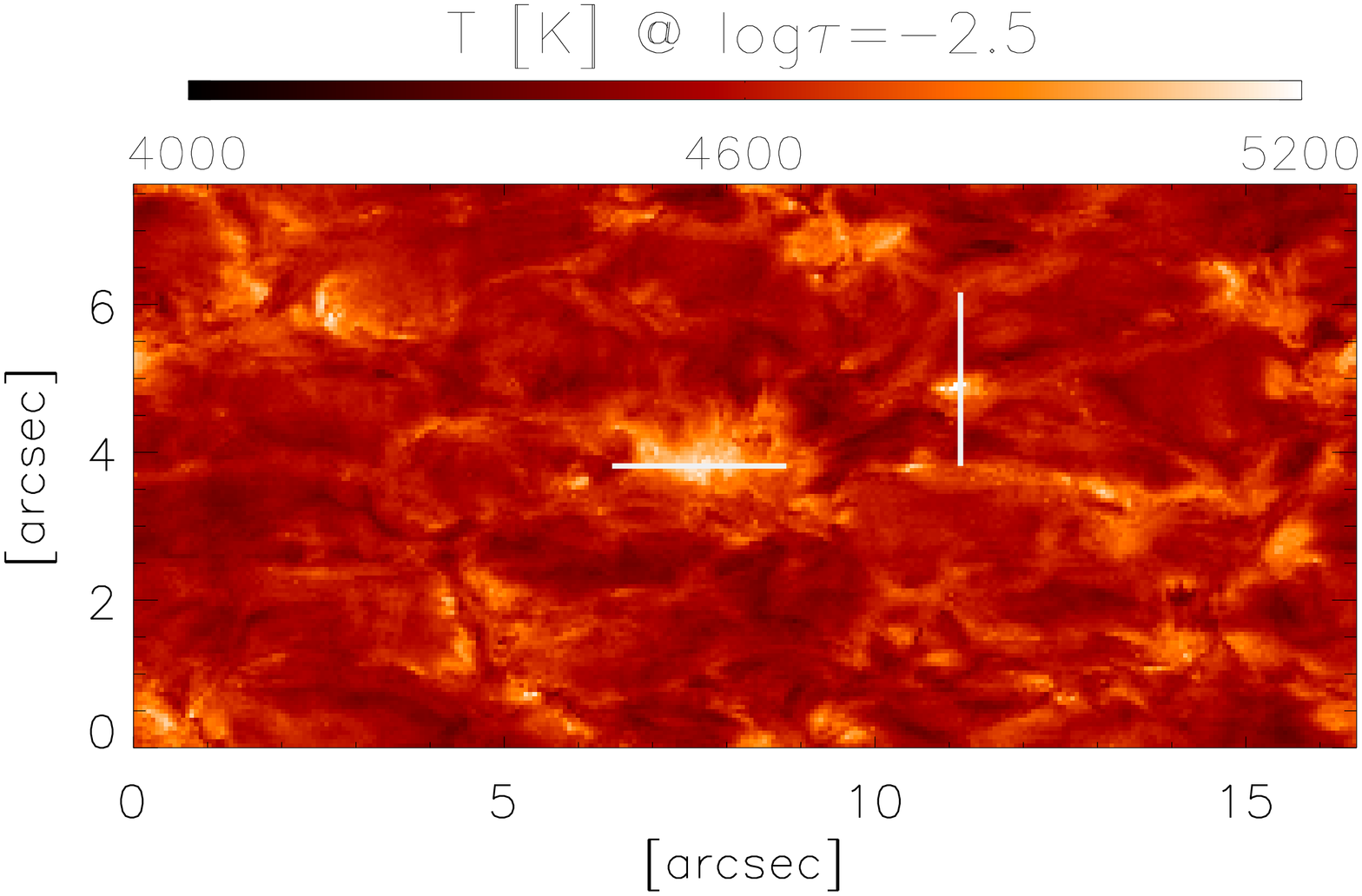}
   \caption{Simulated Sunrise/IMaX observations - Temperatures at $log\tau=0, -0.9$ and $-2.5$ retrieved by the inversions at the same times as in Fig.~\ref{sim_inv_rest}. Vertical/horizontal lines mark the position of the cuts shown in Fig.~\ref{sim_cut_vert}/~\ref{sim_cut_hor}. }
\label{sim_inv}
\end{figure*}

\begin{figure}
  \centering
    \includegraphics[angle=0,width=0.29\linewidth ,trim=0cm 5cm 0cm 0cm,clip=true]{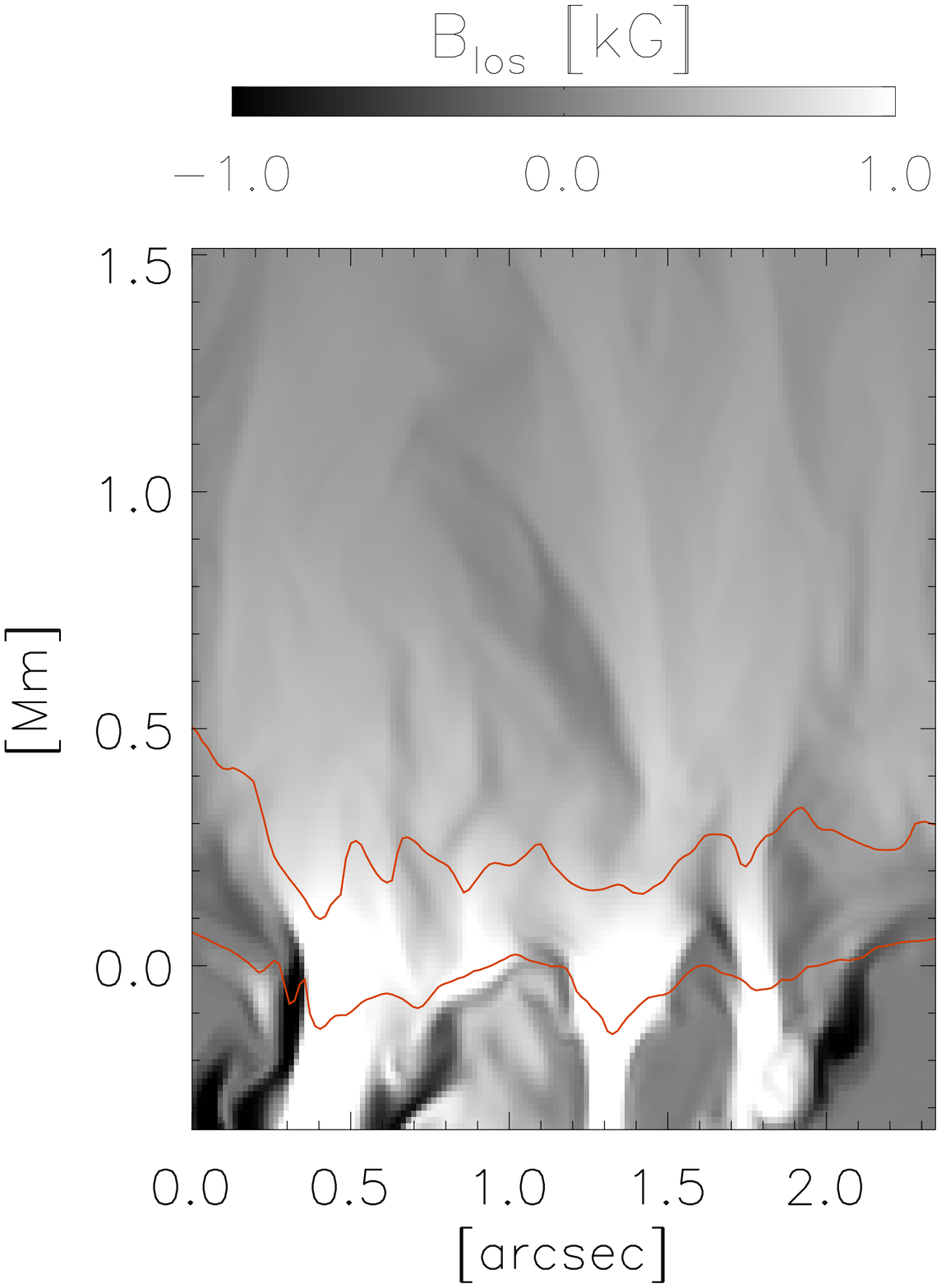}
    \includegraphics[angle=0,width=0.22\linewidth ,trim=4.3cm 5cm 0cm 0cm,clip=true]{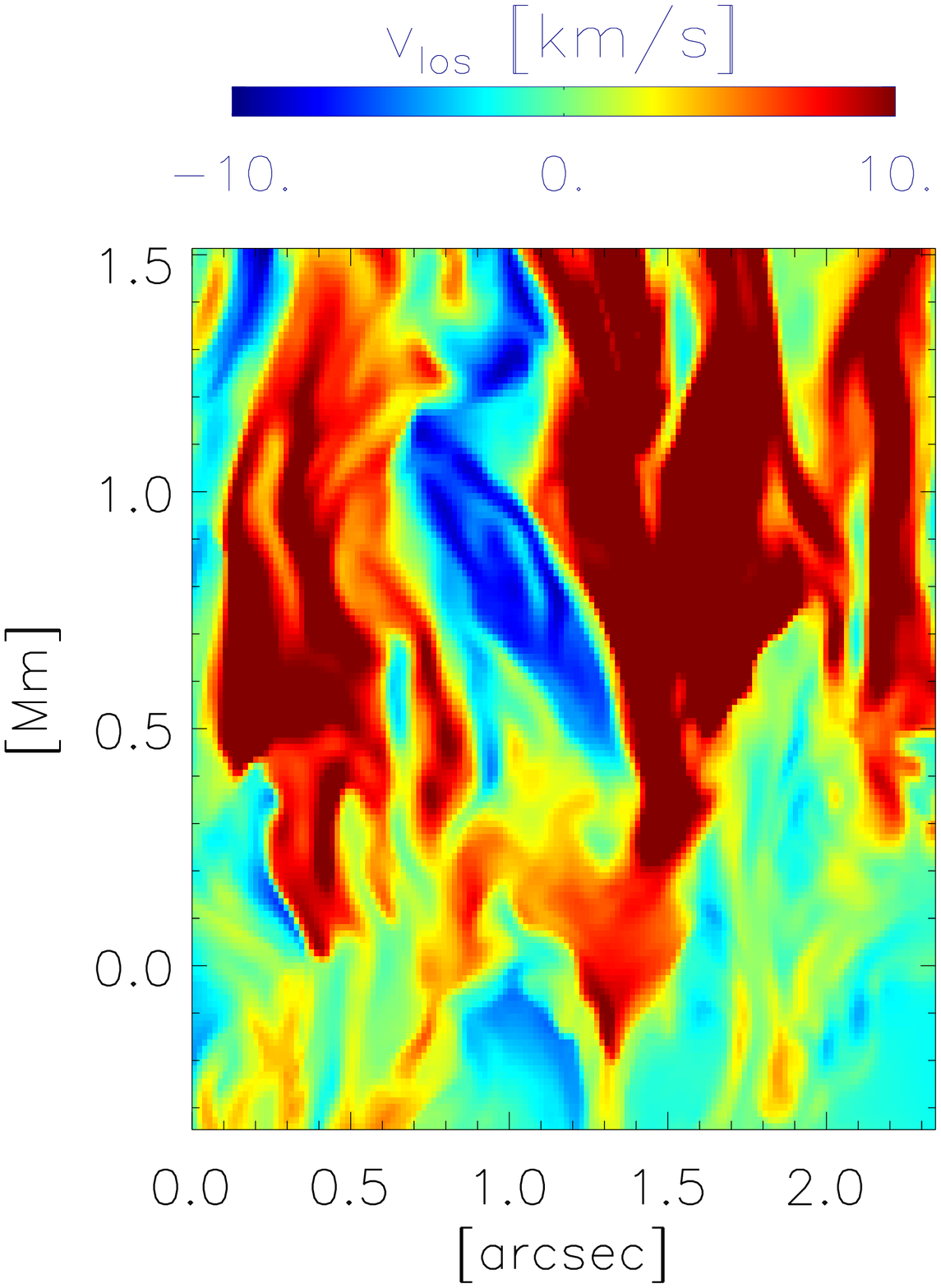} \\
    
   \includegraphics[angle=0,width=0.29\linewidth ,trim=0cm 0cm 0cm 5cm,clip=true]{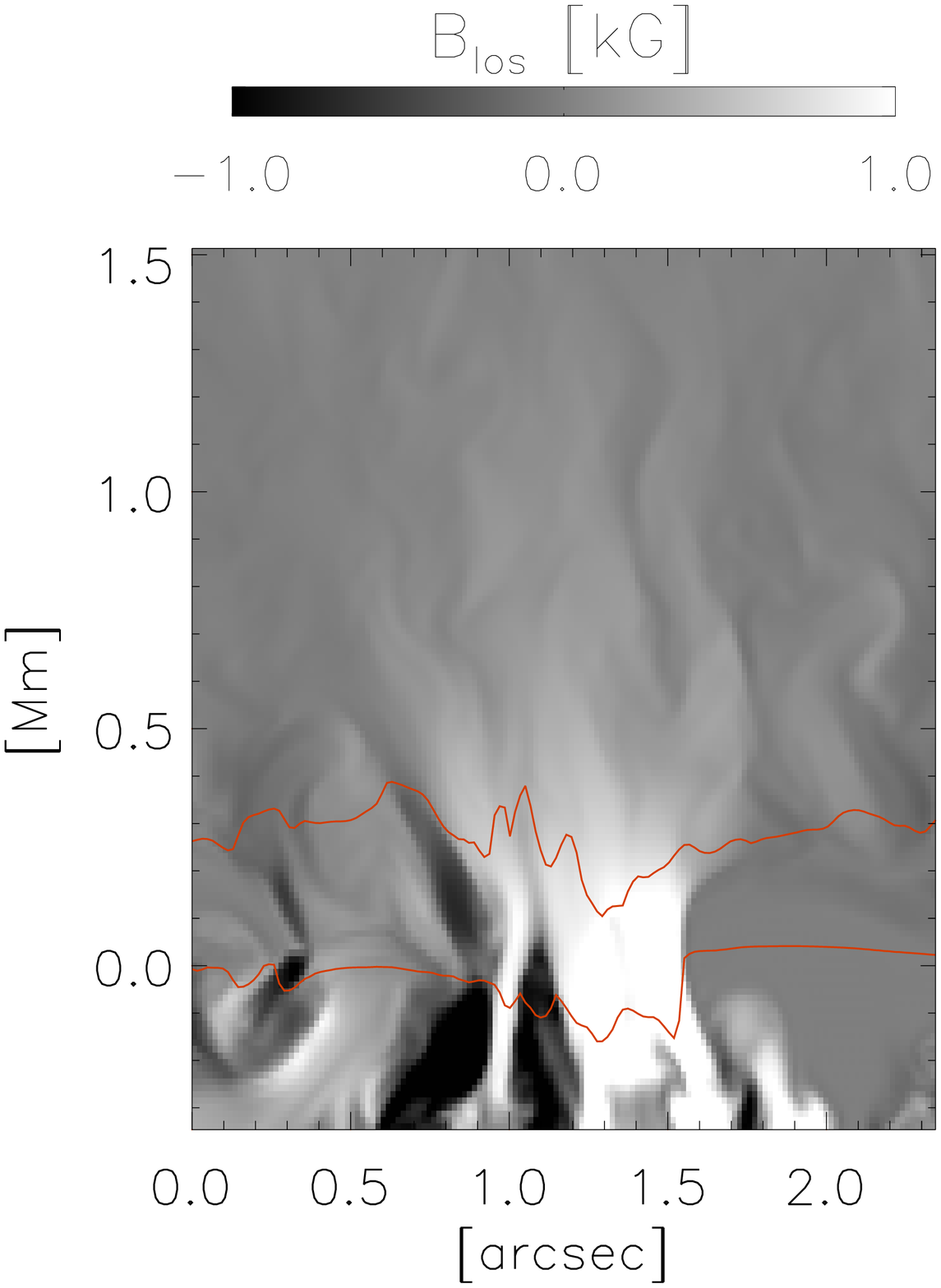}
   \includegraphics[angle=0,width=0.22\linewidth ,trim=4.3cm 0cm 0cm 5cm,clip=true]{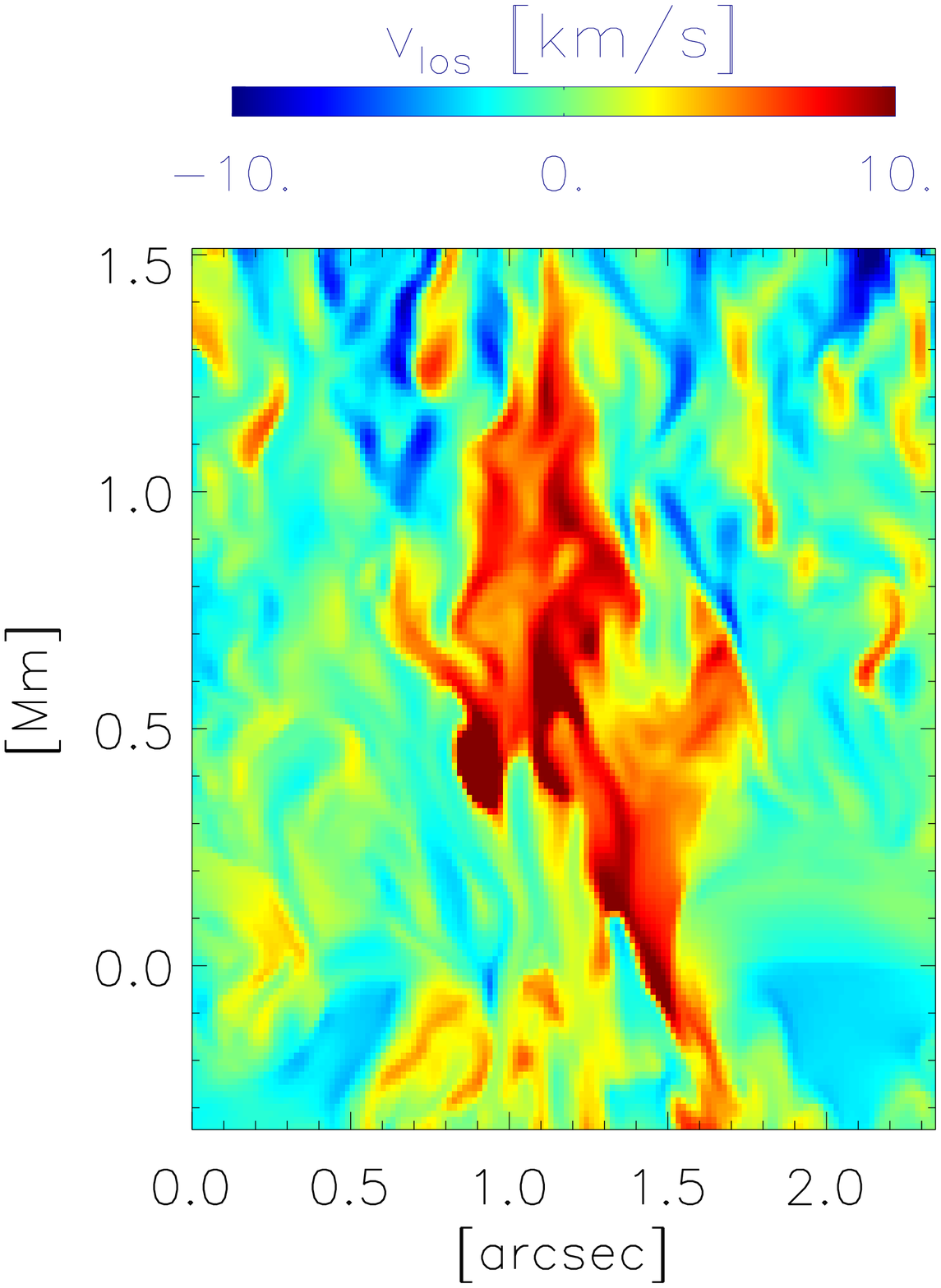}
   \caption{Vertical cuts over locations marked by vertical lines in Fig.~\ref{sim_inv} and Fig.~\ref{sim_inv_rest} show line-of-sight component of the magnetic field (left) and line-of-sight velocity (right; upflows are coloured blue). The $\log \tau = 0$ and  $\log \tau = -2$ levels are outlined by red lines.}
\label{sim_cut_vert}
\end{figure}

\begin{figure}
  \centering
  \includegraphics[angle=0,width=0.29\linewidth ,trim=0cm 5cm 0cm 0cm,clip=true]{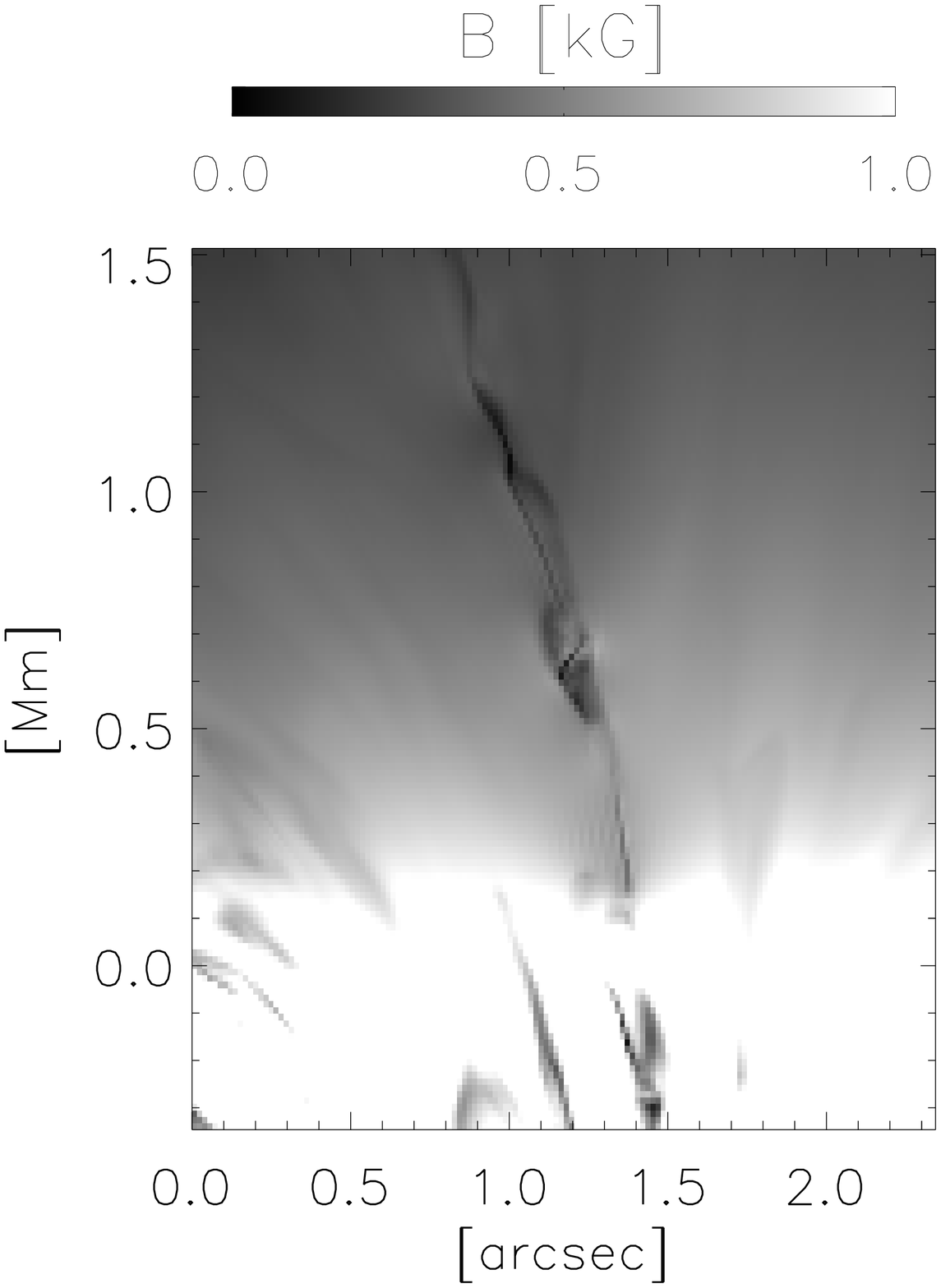} 
   \includegraphics[angle=0,width=0.22\linewidth ,trim=4.3cm 5cm 0cm 0cm,clip=true]{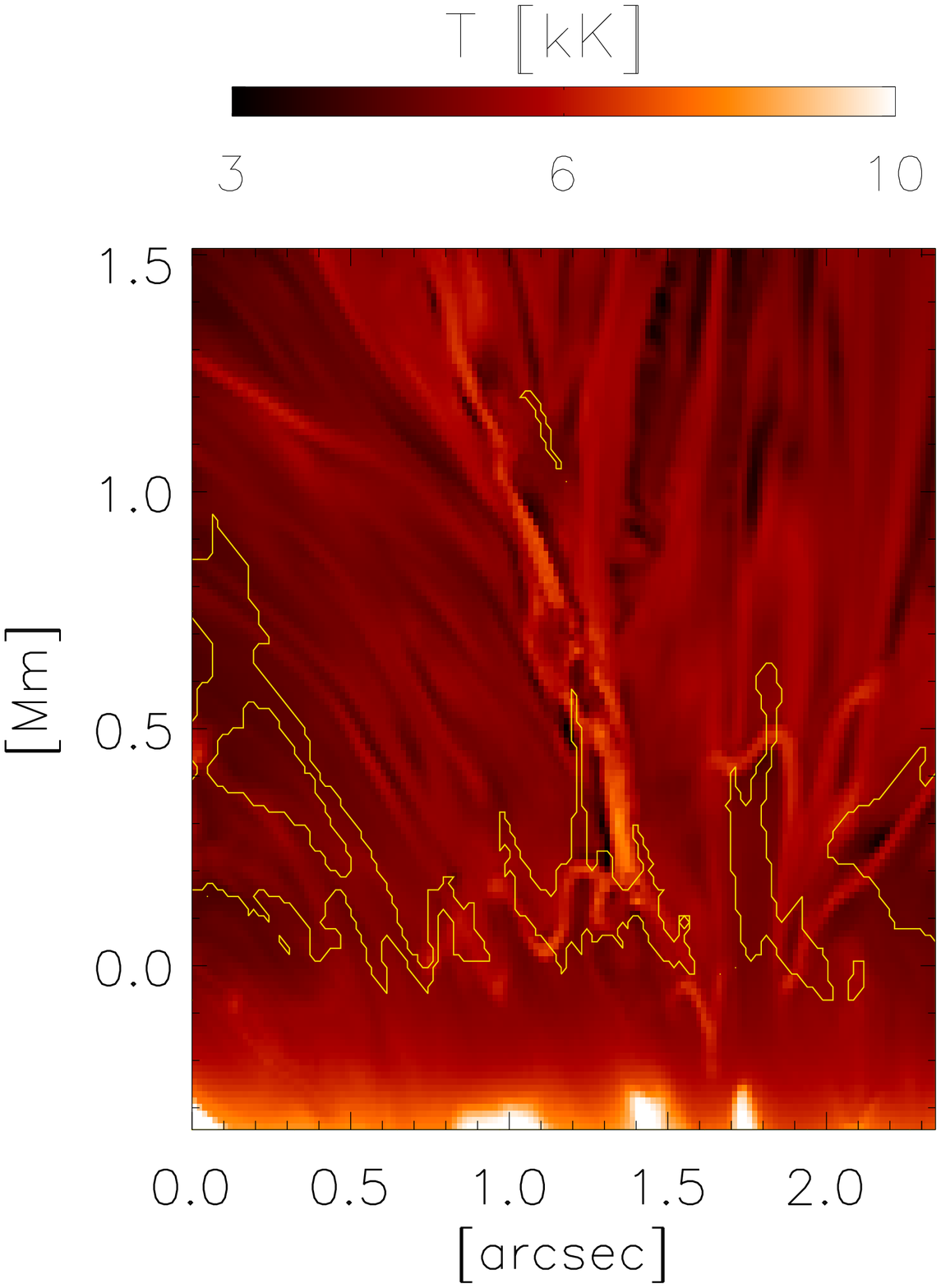} \\
   \includegraphics[angle=0,width=0.29\linewidth ,trim=0cm 0cm 0cm 5cm,clip=true]{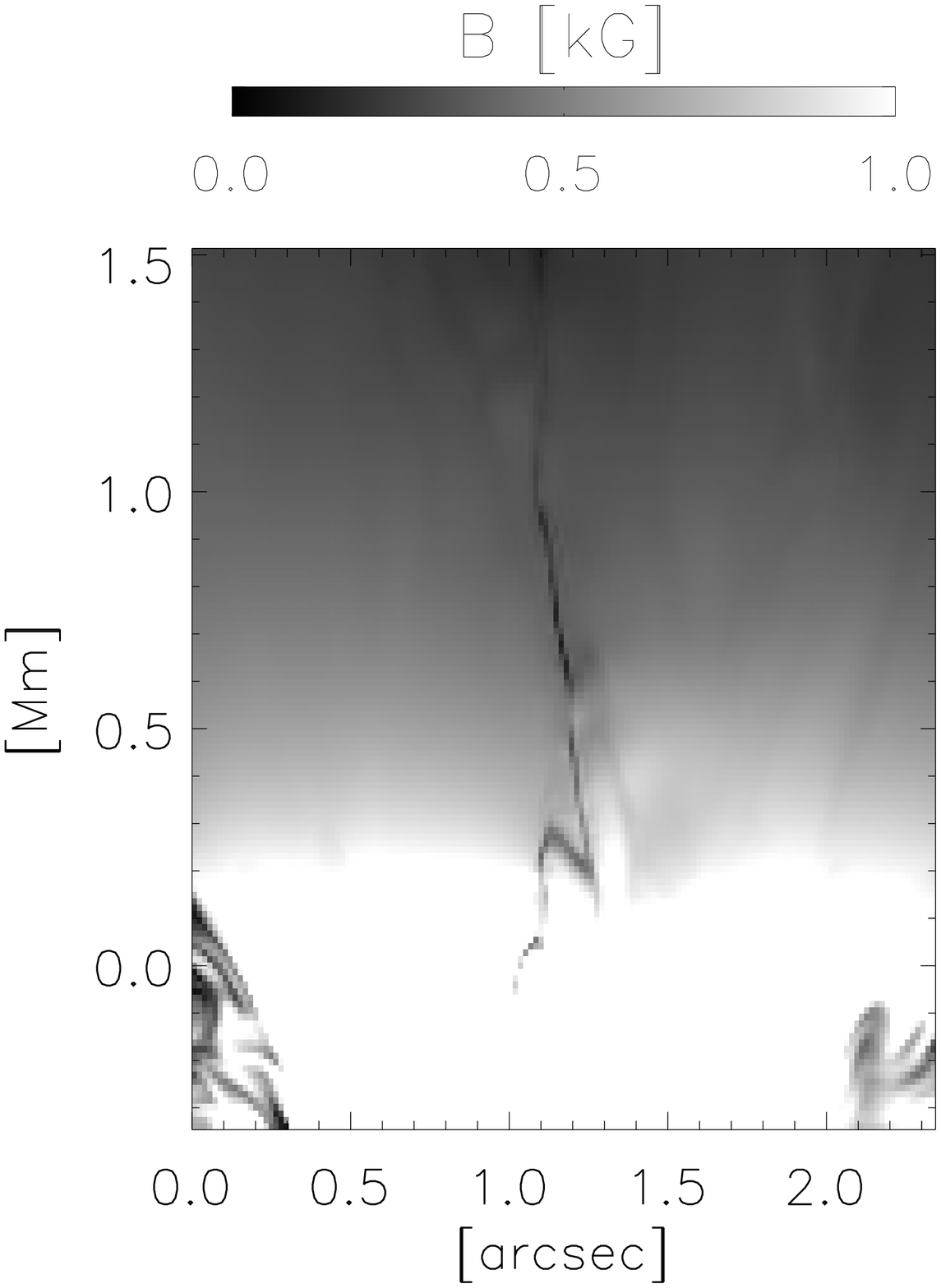}
     \includegraphics[angle=0,width=0.22\linewidth ,trim=4.3cm 0cm 0cm 5cm,clip=true]{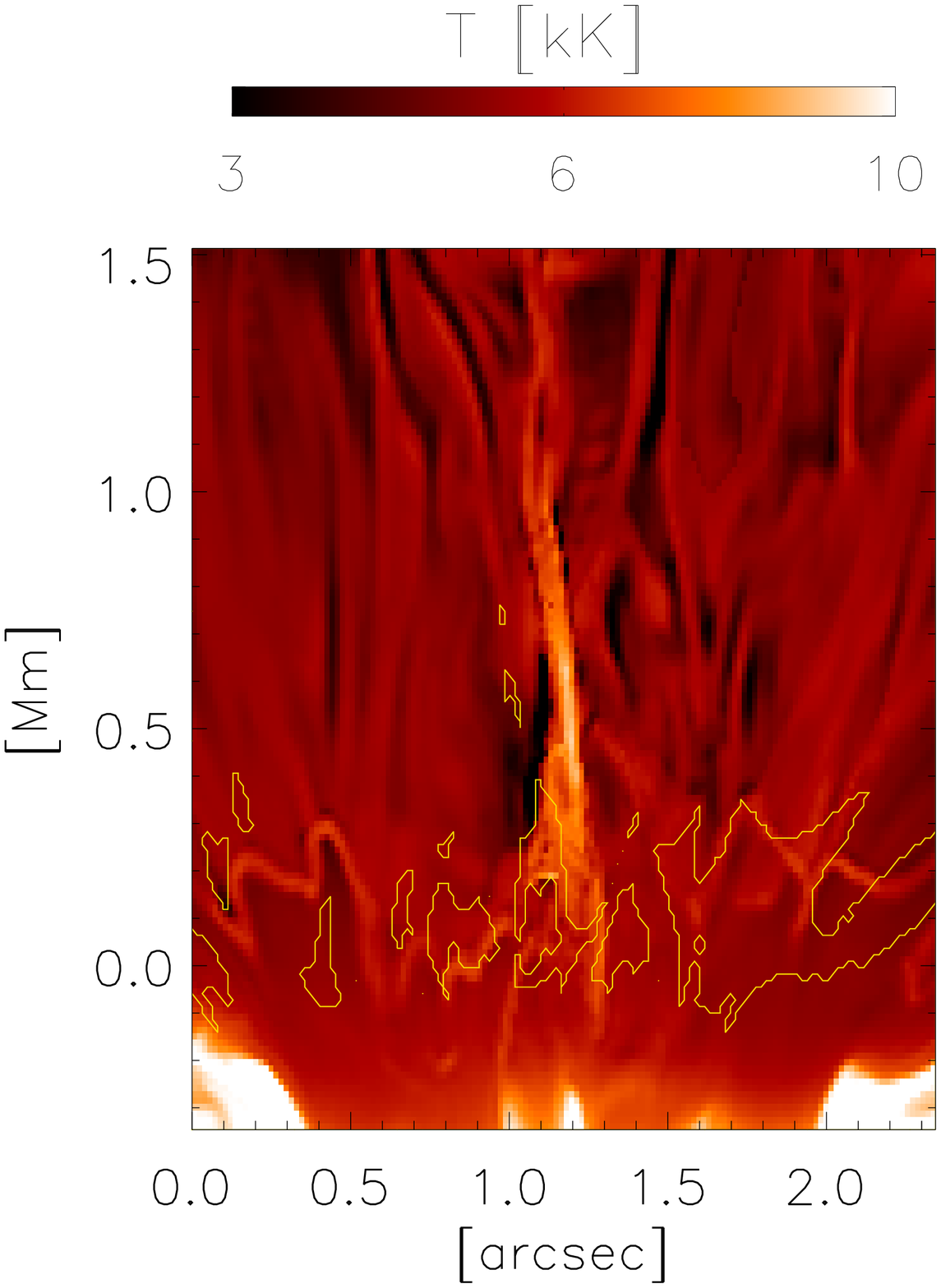}    
   \caption{Vertical cuts over locations marked by the horizontal lines in Fig.~\ref{sim_inv_rest} and Fig.~\ref{sim_inv}  show magnetic field strength and temperature at two instances during the evolution of the simulated EB-like event. Yellow contours show the formation height range of the Fe~I~525.02~nm line. The same snapshots are also shown in Fig.~\ref{topology}.}
\label{sim_cut_hor}
\end{figure}

\section{A simulated event}

We focus on one of the simulated events produced in the numerical experiment performed with the 3D MHD MURaM code \citep{2005A&A...429..335V,Rempel:2009}. The simulation domain extends over $12\times 6 \times 3.5$~Mm of which $2$~Mm is above the surface. Horizontal and vertical grid spacing is 11 and 14~km, respectively. In this run, a thin flux sheet is introduced in fully-developed convecting flow, some 300~km below optical depth unity. The field strength across the sheet changes as a Gaussian function with FWHM = 50~km and reaches a maximum value of 5000~G. The sheet gets undulated due to convection in such a way that crests/troughs are formed where convective uplows/downflows are present. In this way a serpentine type of emergence develops. Fig.~\ref{topology} shows the field topology at two moments during the evolution of the chosen event. The top panel shows the moment when the EB is triggered, i.e. signatures appeared in the blue wing of H$\alpha$ (t=17.5~min). Although the loop system already at that point reached the highest layers of the simulation box, the reconnection happened low down. Extended U-loops expanded from the reconnection site to the top of the box with their troughs moving fast upwards due to magnetic tension. At t=25.8~min (bottom panel in Fig.~\ref{topology}), the current sheet extended upwards as more material arrived at those heights and bald patches were situated far from the photosphere. 

To simulate \sunrise{}/IMaX observations, we synthesize the Fe~I~525.02~nm line by taking into account spectral resolution and sampling of the instrument. Since we compare with reconstructed data which are approximately corrected for stray light, we remove only the spatial frequencies above the diffraction limit of the telescope from the simulated maps and rebin them to pixel size of $0.055$\arcsec. Noise of $7\cdot10^{-3}$~I$_{c}$ is added to the simulated spectropolarimetric signals and then the same inversion strategy is applied as was done to the observations. Results of the inversions are shown in Fig.~\ref{sim_inv_rest} and Fig.~\ref{sim_inv}.\footnote{A movie with the maps of parameters deduced from the inversion of the whole simulated time series is added to on-line material.}

Figure~\ref{sim_inv_rest} also shows the simulated blue wing of H$\alpha$ in the rightmost column, so that the regions where Ellerman bomb-like events appear may be distinguished. Emergent H$\alpha$-wing intensity is calculated with the SPINOR code in LTE, which is an acceptable assumption for the line wings \citep{Leenaarts:etal:2006}. We also add linear Stark broadening as prescribed by \cite{Rutten:2016}. The figures correspond to $\mu = 0.66$ instead of the $\mu = 0.93$ of the observations to create more pronounced flame-like shapes. These images are not spatially degraded. The flame that appears in the middle of the images is the EB-like event we refer to.

Comparison of the inverted maps to the original undegraded parameters directly in the simulations\footnote{Also provided in on-line material.}, reveals that the atmospheric parameters are qualitatively well retrieved. However, a careful quantitative comparison with the corresponding original maps at $\log\tau=-1$, reveals that the retrieved field strength and LOS velocities turn out to be strongly underestimated. This is a consequences of two effects. Firstly, these  inversions assume that these parameters are constant with height, which as a result gives a value that is an average of the height profile of an atmospheric parameter weighted appropriate response function of the spectral line \citep{Borrero:etal:2014}. Secondly, both magnetic field strength and velocity change rapidly with height, especially in this run which starts with a specific magnetic field setup. Also, the retrieved temperature at $\log\tau=-2.5$ resembles more the original temperature at $\log\tau=-1.5$, than the one at the same optical depth. The difference is too large to be attributed simply to spatial degradation. Again, this might simply be a characteristic of this particular run where strong field emergences over the whole simulation domain and atmospheric stratification is somewhat different from the canonical model atmospheres.

Simulations show a similar scenario as the observations. An Ellerman bomb-like event is produced as the footpoints of two newly emerging flux concentrations come into contact. The difference is that this particular simulated EB-like feature changes in brightness, dims and then brightens up again as more flux approaches the cancellation point (at t=17.5~min and then again a minute later), while observations show a constant increase in the AIA 1700 brightness. This is directly related to upflows. They are persistent in observations, but shorter and intermittent in simulations. The simulated upflows develop first at [11\arcsec, 3.5\arcsec] already before t=15.1~min and then again at [6\arcsec, 3.5\arcsec] at t=15.9~min.

Although the observed event seems to be more dynamic and violent as it produces faster flows and higher temperatures, qualitative similarities with simulations can be found. First, as shown in Section 2, in the observations we find one prominent case of a temperature increase at the footpoint of an emerging loop. Since in simulations emergence takes place over the whole simulation domain, similar features can be found in more than one location, as is evident in Fig.~\ref{sim_inv} e.g. at $t=19$~min. These hot spots coincide mostly with edges of elongated granule, but can appear as 'islands' in intergranular lanes, e.g. at [11\arcsec, 5\arcsec] at  t=25.8~min. At these pixels, the inversion code retrieves temperature increase at all three nodes, the same way as in the case of observations. Cuts through two examples marked by vertical lines in Fig.~\ref{sim_inv_rest} and  ~\ref{sim_inv} are displayed in Fig.~\ref{sim_cut_vert}. The cuts show fast downflows and strongly shifted optical depth scale at these locations. In the top case, no opposite polarity is present, as it is in most of these hot spots. Sometimes, as the second example illustrates, opposite polarities do come into contact near the optical depth unity and the temperature is then additionally increased by Ohmic heating. All these hot spots are not connected in any direct way to the EB-like event.

The second similarity with observations can be found at the neutral line between the opposite polarities. Although the onset of the simulated EB-like event starts already at $t=17.5$~min, inverted temperature maps show a temperature increase at this location only in the next snapshot, some 30~s later.  Identically as in observations, after it is retrieved at the top node, similar signatures can be found in the lower nodes too. The largest jump in the temperature is always retrieved at the highest node in both simulations and observations. We choose to examine simulated parameters in more detail in Fig.~\ref{sim_cut_hor}, at two instances where inversions give different stratifications. At the first moment, the inversions give regular temperature profile and in the second, a temperature increase at all nodes is retrieved.

Figure~\ref{sim_cut_hor} shows the neutral line situated vertically in the middle of the plots, where magnetic field strength has a minimum, and the thin signature in temperature outlines where the current density and hence the Ohmic heating is the largest. Contours outline the region where the contribution function to the line depression at the line's nominal wavelength position reaches 99\% of its maximum value. These are the same instances shown also in Fig.~\ref{topology}.

In the first instance, top panel in Fig.~\ref{sim_cut_hor}, the reconnection already started and temperature is increased by more than 2500~K already at 200~km above the surface. Because of this large temperature jump, the formation of the Fe~I~525.02~nm line is cut off at that height, so the line sees only the part of the atmosphere where the temperatures shows no significant changes. With time the jump in the temperature moves downward, so that by $t=25.8$~min temperature reversal happens already close to the the surface. From a relatively normal value there, the temperature gradually increases until it reaches a maximum of 9000~K at a height of around 500~km. In this case the  Fe~I~525.02~nm line is formed very close to the surface, it samples a very shallow layer of less than 50~km in height. The retrieved temperatures are still significantly underestimated even at those heights, which could be a consequence of the spatial degradation.


\section{Conclusions}

During the evolution of NOAA AR 11768, an event at the confluence of two flux emergences produced a significant brightness increase in the SDO/AIA 1700 channel. Given that AIA 1700 brightness was above the $5\sigma$ threshold set by \cite{Gregal:2013}, we claim that this event can be classified as an Ellerman bomb. This claim is further reinforced by the underlying field configuration recovered by \cite{Centeno:etal:this}. They demonstrate that the event was produced during subsequent appearance of two emerging flux regions where the field is aligned in a way that points to a serpentine-like field topology. Observations strongly suggest that this is an environment where EBs inevitably appear \citep{Georgoulis:2002,Pariat:etal:2004,Pariat:2006}.

In this work we analyse in detail photospheric signatures of the event recorded by \sunrise/IMaX. We compare the results with a numerical experiment that reproduces the proposed flux emergence scenario. The simulated event results in a temperature increase which is in agreement with previous EB models 
\citep{Fang:2006,Bello:etal:2013,Berlicki:2014,Fang:2006,Hong:2014} and produces the typically observed morphology of EBs in the wings of H$\alpha$. 

Simulations show the expected field topology at the onset of the EB-like feature. The reconnection seems to start in the photosphere, some 200~km above the surface. As new material and fresh magnetic flux emerges, the current sheet between the two opposite polarities of the two emerging features extends further up and the location where post-reconnection U-loops begin is shifted to above 1~Mm. During the whole evolution, no actual information on the location where the reconnection takes place is contained in the Fe~I~525.02~nm line. As the energy is deposited in the upper photosphere, the temperature rapidly changes first there and then it gets modified also further down by the reconnection aftermath, e.g. downward propagating reconnection jets as they collide with the local plasma. Because of this, the formation height of Fe~I~525.02~nm line is continually  shifted downwards till the moment when the line samples only a very shallow layer near the solar surface. This is in agreement with observations that suggest that these events produce sufficient energy to ionize the neutral metals \citep{Rutten2015} and explains the very low temperature increase found by \cite{Reid2016}. 

Simulations suggest that not all locations where temperature increases are necessarily related to the event itself. Some of those appear at the footpoints of rapidly emerging loops as they expand in the horizontal direction upon reaching the surface. While material is drained, the footpoints are squeezed in between already developed magnetic features and fast-moving emerging material. As a result  fast downflows are generated and footpoints quickly evacuated. These locations might be similar to the high-temperature points presented by \cite{Tortosa2009} and might have similar characteristics from a slanted viewing angles as the flows studied by \cite{Bellot2009} and \cite{Vitas2011}. They also resemble in some ways locations of convective collapse.

Although they show qualitative similarities with the observations, the simulations fail to produce equally high temperatures. Comparison of the field strength and especially the velocities points to the fact that the choice of the initial magnetic field setup is not ideal. To produce an event that is as dynamic and violent as the observation suggest, more magnetic buoyancy is needed that would launch magnetic field into solar atmosphere more efficiently. Also, persistent upflows that last at least three times longer  seem to be essential. For this, instead of an embedded thin flux sheet, one needs to insert into simulation domain more flux or a flux sheet over extended period of time.

   \begin{acknowledgements}
The German contribution to \sunrise{} and its reflight was funded by the
Max Planck Foundation, the Strategic Innovations Fund of the President of the
Max Planck Society (MPG), DLR, and private donations by supporting members of
the Max Planck Society, which is gratefully acknowledged. This work has benefited from the discussions at the meeting 'Solar UV bursts - a new insight to magnetic reconnection' at the International Space Science Institute (ISSI) in Bern. The Spanish contribution was funded by the Ministerio de Economi­a y Competitividad under Projects ESP2013-47349-C6 and ESP2014-56169-C6, partially using European FEDER funds. The HAO contribution was partly funded through NASA grant number NNX13AE95G. The National Solar Observatory (NSO) is operated by the Association of Universities for Research in Astronomy (AURA) Inc. under a cooperative agreement with the National Science Foundation. This work was also partly supported by the BK21 plus program through the National Research Foundation (NRF) funded by the Ministry of Education of Korea. 

\end{acknowledgements}


\end{document}